\documentclass[a4paper,11pt]{article}
\usepackage{jheppub} 
\usepackage{lineno}

\title{Neutrino mass predictions with metaheuristic optimization under $A_4$ modular symmetry}
\author[a]{Muhammad Waheed Aslam,}
\author[a]{Abrar Ahmad Zafar,}
\author[b]{Muhammad Naeem Aslam,}
\author[c]{Amina magdich,}
\author[d]{Yamen El Touati,}
\author[e]{Ahmed A. Hamoud,}
\affiliation[a]{Department of Physics, University of the Punjab, Lahore, Pakistan}
\affiliation[b]{Department of Mathematics, Lahore Garrison University, Lahore, Pakistan}
\affiliation[c]{Department of Computer Science, Applied College, Princess Nourah bint Abdulrahman University, P.O. Box 84428, Riyadh 11671, Saudi Arabia}
\affiliation[d]{Department of Computer Science, Faculty of Computing and Information Technology, Northern Border University, 73213, Arar, Saudi Arabia}
\affiliation[e]{Department of Mathematics,  Taiz University, Taiz 50000, Yemen}
\emailAdd{ahmed.hamoud@taiz.edu.ye}

\abstract{The current research aims to conduct an exhaustive investigation of the neutrino mass model based on $A_4$ discrete non-abelian modular symmetry, specifically in the context of the linear seesaw mechanism, where the conventional type I seesaw mechanism is extended, with particular attention given to the optimization of the parameters using incomprehensible but intelligible-in-time logics optimization algorithm (ILA), a metaheuristic optimization. Unlike other approaches to discrete flavor symmetry, the modular approach to flavor dynamics is found to significantly reduce the number and complexity of flavon fields required to give realistic textures to fermion masses. The key predictions include neutrino masses, $U_\text{PMNS}$ matrices, effective neutrino masses for neutrinoless double beta decay, beta decay, Dirac and Majorana CP violation phases for normal (NO) and inverted mass ordering (IO), offering testable implications. The working efficiency of the ILA optimization technique is also estimated. The optimized neutrino oscillation parameters are well consistent with recent experimental data. Our analysis also aligns with Planck cosmological constraints on the sum of neutrino masses $0.06<\Sigma m<0.12\ \text{eV}$.}
\begin{document}
\maketitle
\flushbottom

\section{Introduction}
\label{sec:intro1}
Standard Model (SM) organizes quarks and leptons into three generations, yet offers no explanation for this family structure, the striking hierarchy of fermion masses, or the markedly different mixing patterns observed between leptons and quarks. Collectively, these open questions comprise the long-standing “flavor problem.”

Neutrinos, though feebly interacting and extraordinarily light, play a significant role in both particle physics and cosmology. They are copiously produced in stellar fusion \cite{raffelt2008neutrinos}, supernovae \cite{burrows1990neutrinos,tamborra2018neutrinos,burrows2000neutrinos,cooperstein1988neutrinos,herant1997neutrinos} and the Big Bang \cite{sarkar2003neutrinos,steigman2005neutrinos} itself, and they underpin some of the greatest puzzles in modern science: the origin of neutrino mass \cite{weinheimer2013neutrino}, the nature of dark matter \cite{blennow2013probing,agarwalla2011neutrino,dodelson1994sterile,bertone2018new,hooper2008strategies}, the Higgs sector \cite{chakraborty2014higgs}, and the cosmic matter–antimatter asymmetry \cite{t2k2020constraint}. The discovery of neutrino oscillations, first hinted at by the solar neutrino deficit and definitively demonstrated by Sudbury Neutrino Observatory (SNO) \cite{boger2000sudbury} and Super-Kamiokande \cite{kamiokande1998evidence}, mandates nonzero neutrino masses and therefore physics beyond the SM. 

Subsequent experiments such as KamLAND \cite{eguchi2003first}
, OPERA \cite{sh2009opera}, Daya Bay \cite{an2012observation,an2015new}, T2K \cite{ahn2006measurement}, MINOS \cite{michael2006observation}, NO$\nu$A, and others) have further confirmed these oscillations and pinned down the two mass‐squared differences and three mixing angles with impressive precision, given in NuFIT 6.0 \cite{esteban2025nufit} as shown in Table \ref{parameters}. However, the mass ordering, the octant of $\theta_{23}$, and the CP-violating phases remain unsettled.
\begin{table}[htbp]
\centering
\begin{tabular}{ccc}
\hline 
Parameters &Best-fit ($3 \sigma$) for NO  &Best-fit ($3 \sigma$) for IO \\ 
\hline 
$sin^2\theta_{12}$ & $0.308$ ($0.275\rightarrow 0.345$)&$0.308$ $(0.275\rightarrow 0.345)$ \\ 
$\theta_{12}/^{\circ}$ & $33.68$ $(31.63\rightarrow 35.95)$ &$33.68$ $(31.63\rightarrow 35.95)$\\ 
$sin^2\theta_{23}$ & $0.470$ $(0.435\rightarrow 0.585)$& $0.562$ $(0.437\rightarrow 0.597)$\\ 
$\theta_{23}/^{\circ}$ &$43.3$ $(41.3\rightarrow 49.9)$&$48.6$ $(41.4\rightarrow 50.6)$ \\ 
$sin^2\theta_{13}$ & $0.02215$ $(0.020230\rightarrow 0.02388)$&$0.02224$ $(0.02053\rightarrow 0.02397)$ \\ 
$\theta_{13}/^{\circ}$ & $8.56$ $(8.19\rightarrow 8.89)$&$8.58$ $(8.24\rightarrow 8.91)$ \\ 
$\delta_\text{CP}/^{\circ}$ &$212$ $(124\rightarrow 364)$&$285$ $(201\rightarrow 348) $ \\ 
$\frac{\Delta m^2_{21}}{10^{-5}\text{eV}^2}$ &7.49 $(6.92\rightarrow 8.05)$&$7.49$ $(6.92\rightarrow 8.05)$ \\ 
$\frac{\Delta m^2_{3l}}{10^{-3}\text{eV}^2}$ & $+2.513$ $(+2.451\rightarrow +2.578)$&$-2.510$ $(-2.584\rightarrow -2.438)$\\
\hline
\end{tabular}
\caption{The table summarizes the Best-fit and $3\sigma$ confidence interval for neutrino oscillation parameters: mixing angles ($\sin^2 \theta_{12}$, $\sin^2 \theta_{23}$, $\sin^2 \theta_{13}$), mass-squared differences ($\Delta m^2_{21}$ and $\Delta m^2_{3l}$), and the $\delta_{\text{CP}}$ phase.}
\label{parameters}
\end{table}

In the three active neutrino flavor, a theoretical framework to describe flavor mixing, provided by the PMNS matrix, a unitary matrix of dimension \( 3 \times 3 \), responsible for mixing the flavor eigenstates \( (\nu_e,\ \nu_\mu,\ \nu_\tau) \) into the corresponding mass eigenstates \( (\nu_1,\ \nu_2,\ \nu_3) \), which possess masses $m_1$, $m_2$, and $m_3$, respectively. The PMNS matrix can be parametrized through successive rotations and complex phases, with three mixing angles $(\theta_{12}, \theta_{23}, \theta_{13})$, a Dirac CP-violating phase $(\delta)$, and two Majorana phases $(\alpha, \beta)$ is given as \cite{vien2016delta}:
\begin{equation}
\begin{aligned}
U_{PMNS} &= U(\theta_{23})\, U(\theta_{13}, \delta)\, U(\theta_{12})\, U_M(\alpha, \beta) \\
&=
\begin{pmatrix}
1 & 0 & 0 \\
0 & c_{23} & s_{23} \\
0 & -s_{23} & c_{23}
\end{pmatrix}
\begin{pmatrix}
c_{13} & 0 & s_{13} e^{-i\delta} \\
0 & 1 & 0 \\
-s_{13} e^{i\delta} & 0 & c_{13}
\end{pmatrix}
\begin{pmatrix}
c_{12} & s_{12} & 0 \\
-s_{12} & c_{12} & 0 \\
0 & 0 & 1
\end{pmatrix}
\begin{pmatrix}
e^{i \alpha} & 0 & 0 \\
0 & e^{i \beta} & 0 \\
0 & 0 & 1
\end{pmatrix}.
\end{aligned}
\end{equation}
Depending on the lightest mass and the sign of the larger mass‐splitting, the spectrum may follow a NO or IO and can range from strongly hierarchical to quasi‐degenerate. Oscillation experiments fix only mass‐squared differences, while searches for neutrinoless double‐beta decay probe the absolute mass scale and Majorana phases.

A compelling theoretical approach to the flavor problem invokes discrete non-Abelian symmetries, such as $S_4$ \cite{Brown:1984dk, Chamoun:2023vnn, Ding:2024inn,Kobayashi:2019xvz,Liu:2020akv,deMedeirosVarzielas:2023crv}, $A_4$ \cite{Ma:2001dn,hirsch2009a4,sruthilaya2018a_4, borah2019linear, R1, R2, R3, R4, R5, R6, R8, R10, R11, R12}, $\Delta(27)$ \cite{Branco:1983tn, Ma:2007wu,Abbas:2014ewa,Chen:2015jta,CentellesChulia:2016fxr,Vien:2020hzy,carcamo2021controlled}, and $T_7$ \cite{T71, T72, T73, T74, T75, T76, T77, T78, T79, T710, T713}, to constrain the neutrino mass matrix and predict mixing angles and CP-violating phases.

More recently, modular symmetries \cite{leontaris1998modular,Kobayashi:2018vbk,Feruglio:2017spp,deAdelhartToorop:2011re}, such as $S_3$ \cite{Mishra:2020gxg,Okada:2019xqk}, $S_4$ \cite{Penedo:2018nmg,Novichkov:2018ovf,Okada:2019lzv}, $A_4$ \cite{Nomura:2022mgf,Abbas:2020vuy,Nomura:2023kwz,Kim:2023jto,Kashav:2022kpk,Nagao:2020snm,Asaka:2020tmo,Nomura:2020opk,Okada:2020dmb,Behera:2020lpd,Ding:2019zxk,Altarelli:2005yx,Kashav:2021zir,Devi:2023vpe,Singh:2023jke,Kikuchi:2023jap,Petcov:2022fjf,Du:2022lij,Ding:2022bzs,Nomura:2021pld,Kuranaga:2021ujd}, $A_5$ \cite{Novichkov:2018nkm,Yao:2020zml}, double covering of $A_4$ \cite{Liu:2019khw,Benes:2022bbg,Okada:2022kee}, $S_4$ \cite{Abe:2023qmr,Abe:2023ilq} and $A_5$ \cite{Wang:2020lxk,Behera:2022wco,Behera:2021eut} have emerged as an elegant alternative where the Yukawa couplings themselves transform as modular forms $\tau$ that perform the role of flavons. When this modulus acquires the vacuum expectation value (VEV), it
breaks the flavor symmetry, obviating the need for flavon fields. Here, $\tau$ is complex variable that is present in the Dedekind eta fucation $\eta(\tau)$ \cite{abbott1991modular}. These frameworks yield highly predictive textures that can be confronted with data from current  and upcoming oscillation experiments such as, JUNO \cite{JUNO:2015zny}, DUNE \cite{DUNE:2020ypp}, Hyper-Kamiokande \cite{Hyper-Kamiokande:2016srs}.

Realizing such models often leads to intricate parameter spaces and non-linear equations. Metaheuristic optimizations have been broadly used across multiple domains \cite{guedria2016improved, kaveh2013engineering, kumar2021design, djemame2019solving, singh2014image, pramanik2015image,azayite2019financial, chiam2009memetic, pan2022design, marinakis2009ant,meissner2006optimized, rauf2018training,esmin2015review, rana2011review,stacey2003particle, eberhart1995new,he2007parameter, alatas2009chaos,babazadeh2009optimization, ibrahim2019hybridization, subbaraj2010hybrid, jiang2020multilayer, yue2019determination, rabady2014global, ruan2016determination, mehmood2019nature, yetis2014forecasting, yassin2016binary, akbar2019novel, Fawzi2019EffectiveMB,AbouOmar2022ObserverbasedIT,Aslam2024NeurocomputingSF,JasimShaban2023NNAAA,Zhang2020HybridTO,Zhang2021NeuralNA,Qadeer2022NeuralNP,Zhang2022MultipleLN,Elsisi2021AnIN,OsornioRos2017IdentificationOP} and are also being utilized in the context of neutrino phenomenology \cite{aslam2025particle, Aslam:2025vvp, doi:10.1142/S0217732325501068,Aslam2025pramana, ASLAM2025964} have proven powerful in identifying best‐fit solutions consistent with experimental constraints.

In this work, we focus our attention to an $A_4$ based modular construction \cite{Behera:2020sfe} (augmented by a Type I seesaw) and use ILA \cite{MIRRASHID2023110305} to optimize its mixing parameters. The choice of this model is motivated by two key aspects. First, in contrast to the canonical seesaw mechanism, which usually requires right-handed neutrinos at an ultra-high scale
($\sim 10^{15}\,\mathrm{GeV}$) and is therefore experimentally inaccessible, the present model is based on the linear seesaw mechanism, which implicitly permits the heavy neutral fermions to lie at the TeV scale. This makes the model more relevant for present and near-future experiments. Second, unlike the conventional discrete flavor symmetry models based on $A_4$, which depends on multiple flavon fields and intricate vacuum-alignment mechanisms, the present framework employs modular $A_4$ symmetry. In this approach, the flavor mixing structure originates solely from the complex modulus $\tau$, where the Yukawa couplings are transformed under modular forms. As a result, the model avoids flavon proliferation and vacuum-alignment ambiguities, significantly improving its theoretical economy and predictability without sacrificing consistency with experimental data. In this study, we have used the ILA algorithm because of the three fundamental differences of ILA with Particle Swarm Optimization (PSO), Genetic Algorithm (GA), Grey Wolf Optimizer (GWO) and others. First, unlike PSO and GWO which are one single continuous population in every iteration, ILA clearly separates the process into three completely independent sequential steps (Exploration, Integration, and Exploitation) with no backward transition, providing users explicit control of each step of the process and ILA inspired from machine learning algorithm k-mean clustering. Second, unlike GA which uses biological operators such as crossover and mutation and GWO which mimics the hierarchy of wolves, ILA offers a new approach to knowledge-based update mechanism based on three mathematically defined parameters: comprehensibility, degree and probability, which dynamically guide each expert's movement in the search space according to their relative distance from both the group best and global best solution. Thirdly, ILA uses a random changing number of clusters with a dynamic k-means clustering strategy, which means that different regions are explored simultaneously with various numbers of clusters — a capability not found in PSO, GA, GWO and others. Finally, the three stages enable user-controlled monitoring and tuning between each stage, which is not possible in other algorithms as they are a continuous loop. The $73$ benchmark problems have been tested using $20$ algorithms, and the average normalized error of ILA is $0.0357$, which is significantly better than the other algorithms, including PSO ($0.1077$), GA ($0.1092$), and GWO ($0.0985$) demonstrate the robustness and accuracy of ILA \cite{MIRRASHID2023110305}. 

For the phenomenological analysis, although the reference paper considered the same theoretical framework, it did not provide predictions for the Dirac CP-violating phase $\delta$ or the Majorana CP phases $\alpha$ and $\beta$. In the present work, the ILA-based optimization enables us to determine these CP-violating phases while satisfying the latest neutrino oscillation constraints. Furthermore, our analysis yields the optimized VEV of the flavon field $v_{\rho}$, which was not determined in the original study.

Section \ref{sub:type-1} introduces the model and its field content; Sections \ref{ILAdetail} review the Incomprehensible but Intelligible-in-time logics: Theory and
optimization algorithm; Section \ref{sec: Numerical Analysis} presents our numerical results for both NO and IO, comparing them with the latest experimental and cosmological bounds; and Section \ref{sec:Conclusion} offers our concluding remarks.
\section{Description of the Model under \texorpdfstring{$A_4$}{TEXT} Modular based Model}
\label{sub:type-1}
We consider a model based on $A_4$ modular symmetry, an extension of SM within a linear seesaw framework, effectively modifying the type-I seesaw structure. In this framework, the Higgs and lepton superfields are transformed according to specific modular representations, as outlined in Table \ref{tab:Model-2}. The particle content is augmented by six singlet heavy superfields, $N_{R_i}$ and $S_{L_i}$ (both $A_4$ triplets), and a singlet weighton $\rho$. Modular Yukawa couplings, also $A_4$ triplets with weight $k_I=2$, replace multiple flavon fields, preventing spectrum overpopulation and enhancing predictability. The VEVs of the Higgs superfields \( \langle H_{u,d} \rangle =\frac{v_{u,d}}{\sqrt{2}}\) are connected to the VEV of the SM Higgs field through the relation
\(
v_H =\sqrt{v_u^2 + v_d^2},
\)
where \( v_u \) and \( v_d \) denote the VEVs of \( H_u \) and \( H_d \), respectively. The ratio of these VEVs is conventionally defined as
\(
\tan\beta = (\frac{v_u}{v_d})\approx 5 
\) \cite{Antusch:2013jca,Okada:2019uoy,Bjorkeroth:2015ora,Mishra:2022egy}.
\begin{center} 
\begin{table}[htpb]
\centering
\begin{tabular}{c c c c c c c c c c}
\hline  
Fields & ~$e_R^c$~& ~$\mu_R^c$~  & ~$\tau_R^c$~& ~$L_{\wp L}$~& ~$N_R$~& ~$S_L^c$~& ~$H_{u,d}$~&~$\rho$ & $\textbf{
Y}=(y_1, y_2, y_3)$\\ \hline 
$SU(2)_L$ & $1$ & $1$ & $1$ & $2$ & $1$ & $1$ &$2$&$1$&$-$ \\ \hline
$U(1)_Y$ & $1$ & $1$ & $1$ & $-\frac{1}{2}$ & $0$ &$0$& $\frac{1}{2}, -\frac{1}{2}$ &$0$&$-$  \\ \hline
$U(1)_X$ & $1$ & $1$ & $1$ & $-1$ & $1$ & $-2$ & 0 &$1$ & $-$  \\ \hline
$A_4$ & $1$ & $1'$ & $1''$ & $1, 1^{\prime \prime}, 1^{\prime }$ & $3$ & $3$ & $1$ & $1$  & $3$ \\ \hline

$k_I$ & $1$ & $1$ & $1$ & $-1$ & $-1$ & $-1$ & $0$  & $0$ & $2$\\ 
\hline
\end{tabular}
\caption{Particle content and Yukawa couplings under $SU(2)_L \times U(1)_Y \times U(1)_{X} \times A_4 $ modular symmetry, where, $k_I$ being the modular weight.}
\label{tab:Model-2}
\end{table}
\end{center}
A global $U(1)_X$ symmetry~\cite{Behera:2020sfe} forbids unwanted superpotential terms. The $A_4$ symmetry is assumed to break above the electroweak scale~\cite{Dawson:2017ksx}, with $\langle\rho\rangle\neq 0$ generating masses for the new superfields. Suitable $A_4$-singlet charge assignments for charged leptons yield a diagonal mass matrix matching observed values. Dirac $M_D$ and pseudo-Dirac $M_{LS}$ mass matrices arise from contractions of $N_{R_i}$ and $S_{L_i}$ with the triplet Yukawa coupling $\mathbf{Y}$ respectively, while $M_{RS}$ results from $N_{R_i}$ and $S_{L_i}$ mixing with $\alpha_{NS} \gg \beta_{NS}$. The super-potential of the model can be expressed as
\begin{equation}
    \begin{aligned}
        \label{NRc}
        \mathcal{W} &=  y_{\ell_{}}^{\wp \wp}  {L}_{\wp_L} H_d ~\wp_R^c +\alpha_D   {L}_{\wp_L} H_u~ (\textbf{Y} N_R)_{1, 1^{\prime \prime}, 1^\prime} +  \beta_D \left[   {L}_{\wp_L} H_u~ (\textbf{Y} S^c_L)_{1, 1^\prime, 1^{\prime \prime}} \right]\frac{\rho^3}{\Lambda^3}  \\
&+ [\alpha_{NS} \textbf{Y} ({S^c_L} N_R)_{\rm sym} + \beta_{NS} \textbf{Y} ({S^c_L} N_R)_{\rm Anti-sym} ]\rho\;,
    \end{aligned}
\end{equation}
The superpotential in Equation \ref{NRc} involves diagonal matrices $\alpha_D$ and $\beta_D$ with six free parameters, plus $\alpha_{NS}$ and $\beta_{NS}$. The Yukawa couplings $Y = (y_1, y_2, y_3)$, transforming as an $A_4$ triplet with modular weight 2, can be formulated using the Dedekind eta-function $\eta(\tau) =\sqrt[24]{q} \prod _{k=1}^{\infty } \left(1-q^k\right)$ and its derivative with $ q=\exp (2i \pi  \tau )$, where $q$ is a complex, are given as
\begin{equation}
\begin{aligned}
     \label{eq:yukawacoup4}
   y_1&=\frac{i }{2 \pi } \left(\frac{\eta '\left(\frac{\tau +1}{3}\right)}{\eta \left(\frac{\tau +1}{3}\right)}+\frac{\eta '\left(\frac{\tau +2}{3}\right)}{\eta \left(\frac{\tau +2}{3}\right)}+\frac{\eta '\left(\frac{\tau }{3}\right)}{\eta \left(\frac{\tau }{3}\right)}-\frac{27 \eta '(3 \tau )}{\eta (3 \tau )}\right),\\
  y_2&=-\frac{i}{\pi } \left(\omega ^2\frac{ \eta '\left(\frac{\tau +1}{3}\right)}{\eta \left(\frac{\tau +1}{3}\right)}+\omega\frac{  \eta '\left(\frac{\tau +2}{3}\right)}{\eta \left(\frac{\tau +2}{3}\right)}+\frac{\eta '\left(\frac{\tau }{3}\right)}{\eta \left(\frac{\tau }{3}\right)}\right),\\
   y_3&=-\frac{i}{\pi }  \left(\omega ^2\frac{ \eta '\left(\frac{\tau +2}{3}\right)}{\eta \left(\frac{\tau +2}{3}\right)}+\omega\frac{  \eta '\left(\frac{\tau +1}{3}\right)}{\eta \left(\frac{\tau +1}{3}\right)}+\frac{\eta '\left(\frac{\tau }{3}\right)}{\eta \left(\frac{\tau }{3}\right)}\right).
\end{aligned}
\end{equation}
The charged lepton mass matrix is inherently diagonal, arising from the transformation properties of the left-handed and right-handed charged leptons, which belong to singlet representations. Owing to the tensor product structure of the $A_4$ symmetry, the mass matrix of charged leptons takes the form, $M_\ell=diag(y_{\ell_{}}^{ee} v_d/\sqrt{2}, \ y_{\ell_{}}^{\mu \mu} v_d/\sqrt{2}, \ y_{\ell_{}}^{\tau \tau} v_d/\sqrt{2})=diag(m_e,\ m_\mu,\ m_\tau )$, with $m_e$, $m_\mu$ and $m_\tau$ denote the physical masses of the charged leptons~\cite{particle2022review}. The light neutrino mass matrix is
\begin{equation}
    m_\nu = -M_D M_{RS}^{-1} M_{LS}^T ~+~ {\rm transpose} = 
\begin{pmatrix}
M_{11} & M_{12} & M_{13} \\
M_{12} & M_{22} & M_{23} \\
M_{13} & M_{23} & M_{33} 
\end{pmatrix},
\end{equation}
with,
\begin{equation}
\begin{aligned}
\label{eq:eee111}
M_{11}=&(3 \alpha _1 \beta _1 v_u^2 v_{\rho }^2 (y_1^4 (\alpha _{NS}+3 \beta _{NS}){}^2-12 y_2 y_3 y_1^2 \alpha _{NS} (\alpha _{NS}+2 \beta _{NS})-8 (y_2^3+y_3^3) y_1 \alpha _{NS}^2\\&-36 y_2^2 y_3^2 \beta _{NS}^2))/(4 \Lambda ^3 (y_1^3 \alpha _{NS} (\alpha _{NS}+3 \beta _{NS}){}^2-3 y_2 y_3 y_1 (\alpha _{NS}-3 \beta _{NS}) (2 \alpha _{NS} \beta _{NS}\\&+\alpha _{NS}^2+3 \beta _{NS}^2)+(y_2^3+y_3^3) \alpha _{NS} (\alpha _{NS}^2-9 \beta _{NS}^2))),
\end{aligned}
\end{equation}
\begin{equation}
\begin{aligned}
\label{eq:eee222}
M_{12}=&  -(3 v_u^2 v_{\rho }^2 (2 y_2 y_1^3 (\alpha _{NS}+3 \beta _{NS}) (\alpha _1 \beta _2 \alpha _{NS}+\alpha _2 \beta _1 (\alpha _{NS}-3 \beta _{NS}))+9 y_3^2 y_1^2 (\alpha _{NS}\\&+\beta _{NS}) (\alpha _1 \beta _2 (\alpha _{NS}-\beta _{NS})+\alpha _2 \beta _1 (\alpha _{NS}+\beta _{NS}))+6 y_2^2 y_3 y_1 (\alpha _2 \beta _1+\alpha _1 \beta _2) (\alpha _{NS} \beta _{NS}\\&+2 \alpha _{NS}^2+3 \beta _{NS}^2)+2 y_2 (y_2^3 \alpha _{NS} (\alpha _2 \beta _1 (\alpha _{NS}+3 \beta _{NS})+\alpha _1 \beta _2 (\alpha _{NS}-3 \beta _{NS}))\\&+y_3^3 (\alpha _1 \beta _2 (\alpha _{NS}+3 \beta _{NS}){}^2+\alpha _2 \beta _1 (\alpha _{NS}-3 \beta _{NS}){}^2))))/(8 \Lambda ^3 (y_1^3 \alpha _{NS} (\alpha _{NS}+3 \beta _{NS}){}^2\\&-3 y_2 y_3 y_1 (\alpha _{NS}-3 \beta _{NS}) (2 \alpha _{NS} \beta _{NS}+\alpha _{NS}^2+3 \beta _{NS}^2)\\&+(y_2^3+y_3^3) \alpha _{NS} (\alpha _{NS}^2-9 \beta _{NS}^2))),
\end{aligned}
\end{equation}
\begin{equation}
\begin{aligned}
\label{eq:eee333}
 M_{13}=& -(3 v_u^2 v_{\rho }^2 (2 y_3 y_1^3 (\alpha _{NS}+3 \beta _{NS}) (\alpha _3 \beta _1 \alpha _{NS}+\alpha _1 \beta _3 (\alpha _{NS}-3 \beta _{NS}))+9 y_2^2 y_1^2 (\alpha _{NS}\\&+\beta _{NS}) (\alpha _3 \beta _1 (\alpha _{NS}-\beta _{NS})+\alpha _1 \beta _3 (\alpha _{NS}+\beta _{NS}))+6 y_2 y_3^2 y_1 (\alpha _3 \beta _1+\alpha _1 \beta _3) \\&(\alpha _{NS} \beta _{NS}+2 \alpha _{NS}^2+3 \beta _{NS}^2)+2 y_3 (y_2^3 (\alpha _3 \beta _1 (\alpha _{NS}+3 \beta _{NS}){}^2+\alpha _1 \beta _3 (\alpha _{NS}-3 \beta _{NS}){}^2)\\&+y_3^3 \alpha _{NS} (\alpha _1 \beta _3 (\alpha _{NS}+3 \beta _{NS})+\alpha _3 \beta _1 (\alpha _{NS}-3 \beta _{NS})))))/(8 \Lambda ^3 (y_1^3 \alpha _{NS} (\alpha _{NS}+3 \beta _{NS}){}^2\\&-3 y_2 y_3 y_1 (\alpha _{NS}-3 \beta _{NS}) (2 \alpha _{NS} \beta _{NS}+\alpha _{NS}^2+3 \beta _{NS}^2)\\&+(y_2^3+y_3^3) \alpha _{NS} (\alpha _{NS}^2-9 \beta _{NS}^2))),
 \end{aligned}
\end{equation}
\begin{equation}
\begin{aligned}
\label{eq:eee444}
 M_{22}=&-(3 \alpha _2 \beta _2 y_3 v_u^2 v_{\rho }^2 (4 y_1^3 \alpha _{NS} (2 \alpha _{NS}+3 \beta _{NS})+6 y_2 y_3 y_1 (\alpha _{NS} \beta _{NS}+2 \alpha _{NS}^2+3 \beta _{NS}^2)\\&-y_3^3 (\alpha _{NS}^2-9 \beta _{NS}^2)+8 y_2^3 \alpha _{NS}^2))/(4 \Lambda ^3 (y_1^3 \alpha _{NS} (\alpha _{NS}+3 \beta _{NS}){}^2-3 y_2 y_3 y_1 (\alpha _{NS}\\&-3 \beta _{NS}) (2 \alpha _{NS} \beta _{NS}+\alpha _{NS}^2+3 \beta _{NS}^2)+(y_2^3+y_3^3) \alpha _{NS} (\alpha _{NS}^2-9 \beta _{NS}^2))),
\end{aligned}
\end{equation}
\begin{equation}
\begin{aligned}
\label{eq:eee555}
 M_{23}=&-(3 v_u^2 v_{\rho }^2 (2 y_1^4 (\alpha _3 \beta _2+\alpha _2 \beta _3) \alpha _{NS} (\alpha _{NS}+3 \beta _{NS})+6 y_2 y_3 y_1^2 (\alpha _2 \beta _3 (\alpha _{NS}-\beta _{NS}) (2 \alpha _{NS}\\&+3 \beta _{NS})+\alpha _3 \beta _2 (\alpha _{NS} \beta _{NS}+2 \alpha _{NS}^2+3 \beta _{NS}^2))+2 (y_2^3+y_3^3) y_1 (\alpha _2 \beta _3 (\alpha _{NS}+3 \beta _{NS}){}^2\\&+\alpha _3 \beta _2 \alpha _{NS} (\alpha _{NS}-3 \beta _{NS}))+9 y_2^2 y_3^2 (\alpha _2 \beta _3 (\alpha _{NS}-\beta _{NS}){}^2+\alpha _3 \beta _2 (\alpha _{NS}+\beta _{NS}){}^2)))\\&/(8 \Lambda ^3 (y_1^3 \alpha _{NS} (\alpha _{NS}+3 \beta _{NS}){}^2-3 y_2 y_3 y_1 (\alpha _{NS}-3 \beta _{NS}) (2 \alpha _{NS} \beta _{NS}+\alpha _{NS}^2+3 \beta _{NS}^2)\\&+(y_2^3+y_3^3) \alpha _{NS} (\alpha _{NS}^2-9 \beta _{NS}^2))),
 \end{aligned}
\end{equation}
\begin{equation}
\begin{aligned}
\label{eq:eee666}
 M_{33}=&-(3 \alpha _3 \beta _3 y_2 v_u^2 v_{\rho }^2 (4 y_1^3 \alpha _{NS} (2 \alpha _{NS}+3 \beta _{NS})+6 y_2 y_3 y_1 (\alpha _{NS} \beta _{NS}+2 \alpha _{NS}^2+3 \beta _{NS}^2)\\&-y_2^3 (\alpha _{NS}^2-9 \beta _{NS}^2)+8 y_3^3 \alpha _{NS}^2))/(4 \Lambda ^3 (y_1^3 \alpha _{NS} (\alpha _{NS}\\&+3 \beta _{NS}){}^2-3 y_2 y_3 y_1 (\alpha _{NS}-3 \beta _{NS}) (2 \alpha _{NS} \beta _{NS}+\alpha _{NS}^2+3 \beta _{NS}^2)\\&+(y_2^3+y_3^3) \alpha _{NS} (\alpha _{NS}^2-9 \beta _{NS}^2))).
\end{aligned}
\end{equation}
Here, the structure of the mass matrices of $M_D$, $M_{LS}$, and $M_{RS}$ are given as:
\begin{align}
M_D&=\frac{v_u}{\sqrt2}
\left(\begin{array}{ccc}
\alpha_1 & 0 & 0 \\ 
0 & \alpha_2 & 0 \\ 
0 & 0 & \alpha_3 \\ 
\end{array}\right)
\left(\begin{array}{ccc}
y_1 &y_3 &y_2 \\ 
y_2 &y_1 &y_3 \\ 
y_3 &y_2 &y_1 \\ 
\end{array}\right)_{LR},              
\label{Eq:Mell} 
\end{align}
\begin{align}
M_{LS}&=\frac{v_u}{\sqrt2}\left(\frac{v_{\rho}}{\sqrt{2}\Lambda}\right)^3
\left(\begin{array}{ccc}
\beta_1 & 0 & 0 \\ 
0 & \beta_2 & 0 \\ 
0 & 0 & \beta_3 \\ 
\end{array}\right)
\left(\begin{array}{ccc}
y_1 &y_3 &y_2 \\ 
y_2 &y_1 &y_3 \\ 
y_3 &y_2 &y_1 \\ 
\end{array}\right)_{LR},              
\label{Eq:Mell1} 
\end{align}
\begin{align}
M_{RS}&=\frac{v_{\rho}}{\sqrt2}
 \left[
 \frac{\alpha_{NS}}{3}\left(\begin{array}{ccc}
2y_1 & -y_3 & -y_2 \\ 
-y_3 & 2y_2 & -y_1 \\ 
-y_2 & -y_1 & 2y_3 \\ 
\end{array}\right)_{sym}
+
\beta_{NS}
\left(\begin{array}{ccc}
0 &y_3 & -y_2 \\ 
-y_3 & 0 & y_1 \\ 
y_2 & -y_1 &0 \\ 
\end{array}\right)_{asym}
\right]. \label{yuk:MRS}
\end{align}
\section{IbI Logic Optimization Framework}
\label{ILAdetail}

In this section, we present a novel optimization framework based on the principles of IbI logic theory, aimed at achieving efficient and reliable performance across a wide class of optimization problems. This method, referred to as the IbI Logic Algorithm (ILA), combines structured search strategies with logical refinement to improve solution quality and convergence.

\subsection{Overall Structure of ILA}

\begin{figure}
    \centering
    \includegraphics[width=0.7\linewidth]{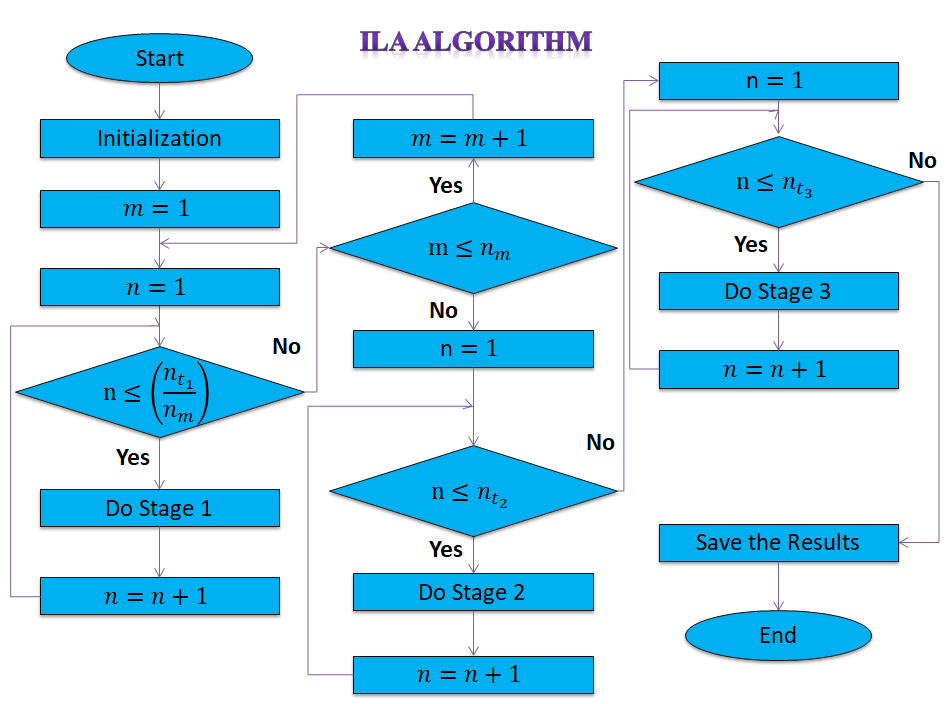}
    \caption{Schematic representation of the ILA workflow}
    \label{fig:ILAAlgorithm}
\end{figure}

The ILA procedure begins with an initialization stage, followed by three sequential and mutually independent phases: \textit{Exploration}, \textit{Integration}, and \textit{Exploitation}. Each phase serves a distinct role in the optimization process, and once completed, the algorithm progresses forward without revisiting earlier stages. As illustrated in Figure~\ref{fig:ILAAlgorithm}, the exploration phase is responsible for performing a diversified, group-based search over the solution space. The integration phase then aggregates and processes the global information obtained during exploration. Finally, the exploitation phase applies IbI logic-based refinement to enhance convergence and improve the accuracy of the obtained solutions. If the number of iterations assigned to these phases are denoted by $n_{t_1}$, $n_{t_2}$, and $n_{t_3}$, respectively, then the total number of iterations is given by $N_T = n_{t_1} + n_{t_2} + n_{t_3}$.
Further details regarding the functionality and implementation of each phase are provided in the subsequent subsections.

\subsection{ILA Parameters}

The effectiveness of the ILA framework mainly depends on two primary parameters: the population size and the total number of iterations, both of which must be specified prior to execution. In addition, several auxiliary hyperparameters (see Table~\ref{hyeperparameters}) can be adjusted depending on the nature and complexity of the optimization problem. While careful tuning of these parameters can enhance convergence speed and solution accuracy, the algorithm maintains robust performance even when default parameter values are used.
\subsection{The ILA stages}
This section provides a comprehensive description of the initialization phase along with the three fundamental stages of the proposed algorithm.
\begin{table}[htbp]
\centering
\begin{tabular}{ccc}
\hline 
Hyperparameters & Abbrs. &Default Values \\ 
\hline 
Number of experts \\ (each is a non-logic expert) & $n_{NL}$& 200\\ 
Number of iterations & $N_T$& 600 \\ 
Number of models & $n_m$&$5$\\ 
The minimum permissible values\\ for the IbI parameters & $B_{min}$&$0.4$ \\ 
The maximum permissible values\\ for the IbI parameters &$B_{max}$&$0.6$ \\ 
Maximum number of groups \\ in each model &$n_{g, max}$&$n_{NL}/2$ \\ 
Stage 1 maximum \\ iteration percentage &$P_{S1}$&$33\%$ \\ 
Number of replications \\for clustering & $n_{rep}$&$10$\\
Stage 2 maximum \\ iteration percentage & $P_{S2}$ &$33\%$ \\ 
Attempts to reach classification\\ convergence over $n_{\text{rep}}$ replications
& $t_{clus}$& $100$\\
\hline
\end{tabular}
\caption{The parameters of ILA}
\label{hyeperparameters}
\end{table}
\subsubsection{Preparation Phase}

Following the initialization of the ILA parameters (Table~\ref{hyeperparameters}), the algorithm enters the preparation phase, which precedes the execution of Stage~1. In this phase, the total number of iterations assigned to Stage~1 is first determined as
\begin{equation}
\label{equation2}
    n_{t_1} = p_{s_1}\, N_T,
\end{equation}
where $p_{s_1}$ denotes the proportion of the total iterations allocated to this stage. The iteration budget is then equally distributed among the $n_m$ models, such that each model is assigned
\begin{equation}
\label{equation1}
    t_m = \frac{n_{t_1}}{n_m}.
\end{equation}
The $n_m$ models operate independently, each acting as a distinct search agent within the optimization framework. Every model generates its own population of candidate solutions (referred to as experts), processes knowledge updates under different configurations, and evolves according to its internal dynamics. This multi-model structure introduces diversity into the search process, improves exploration of the solution space, and increases the likelihood of avoiding local minima. The models are executed sequentially, with each model running for $t_m$ iterations. The output population of one model is used as the input for the subsequent model. Upon completion of all $n_m$ models, Stage~1 concludes, and the refined population is passed to Stage~2.

\begin{figure}
    \centering
    \includegraphics[width=0.8\linewidth]{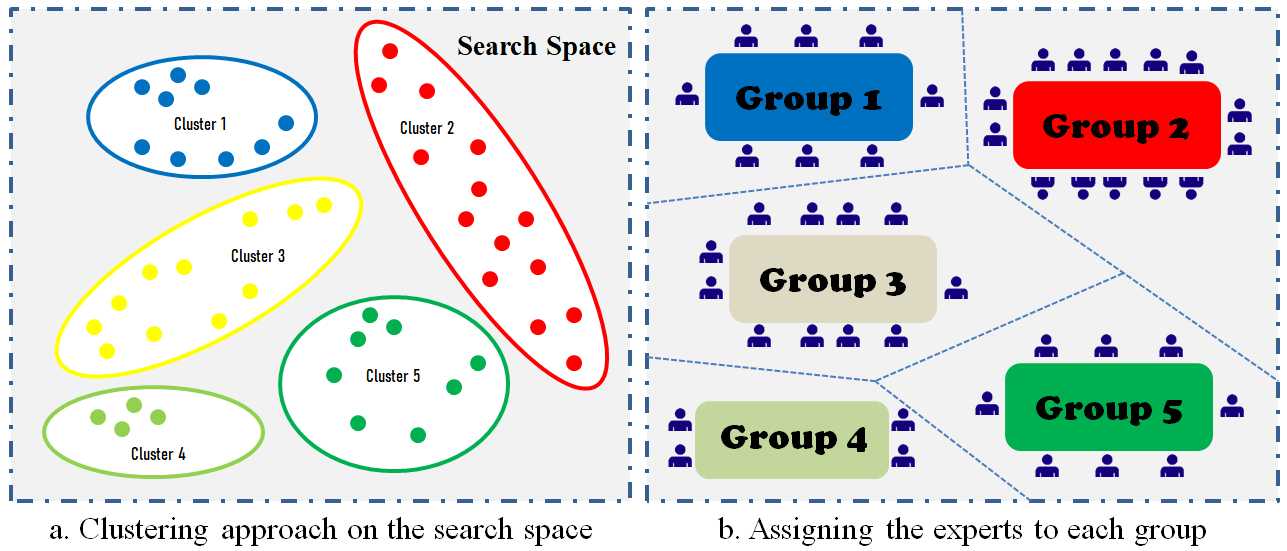}
    \caption{Grouping of experts into clusters}
    \label{fig:figure2}
\end{figure}

\vspace{2mm}
\noindent\textbf{Clustering:}
Prior to the optimization of each model, the population (experts) is partitioned into groups using the $k$-means clustering algorithm, as illustrated in Figure~\ref{fig:figure2}.
Each cluster corresponds to a localized region of the search space, within which experts collaboratively refine candidate solutions based on IbI logic. For the $m$-th model, the number of clusters is randomly selected as
\begin{equation}
\label{equation3}
    n_{g,m} = \mathrm{randi}(n_{g,\max}), \qquad m = 1, \dots, n_m.
\end{equation}
In cases where the desired number of clusters cannot be generated, the clustering structure from the previous model is retained. Since clustering is performed using the $k$-means algorithm with the Squared Euclidean Distance (SED) metric, the resulting cluster sizes may not be uniform. However, alternative distance measures can also be incorporated if required. Once the clustering process is completed, each expert is assigned to its respective group, and the Stage~1 optimization proceeds based on this grouped structure.
\subsubsection{Stage 1: Group-Based Exploration}

This stage constitutes the global exploration phase of the algorithm, where each group focuses on a specific region of the search space and seeks to identify an optimal non-logic (\textit{NL}) value within that region. During this process, all experts $E$ within a group iteratively improve their fitness over $t_m$ iterations by leveraging locally shared information.

At the beginning of each iteration, three fundamental quantities are defined. The first is the \emph{Logic} $L$, representing the globally best expert (in terms of \textit{NL}) obtained from the previous iteration across all groups. The second is the best-performing expert within the current group, denoted by $E_{I,g}$. The third is the group average $A_g$, computed from the experts’ values in the previous iteration.

Based on these quantities, three IbI-inspired measures are evaluated for each expert $i$: the comprehensibility $C_i$, the degree $D_i$, and the probability $P_i$. These are defined as
\begin{equation}
\label{equation4}
    C_i = \sqrt{\sum_{j=1}^{n_{NL}} (E_j - L)^2},
\end{equation}
\begin{equation}
\label{equation5}
    D_i = \sqrt{\sum_{j=1}^{n_{NL}} (E_j - E_{j,p})^2},
\end{equation}
\begin{equation}
\label{equation6}
    P_i = \sqrt{\sum_{j=1}^{n_{NL}} (E_j - E_{I,g})^2}, \qquad g = 1, \dots, n_{g,m},
\end{equation}
where $E_{i,p}$ denotes the value of expert $i$ in the previous iteration.

To ensure consistency, these quantities are normalized into the interval $[0,1]$, yielding the ratios $R_{C,i}$, $R_{D,i}$, and $R_{P,i}$:
\begin{equation}
\label{equation7}
    R_{C,i} = \frac{C_i - C_{\min}}{C_{\max} - C_{\min}},
\end{equation}
\begin{equation}
\label{equation8}
    R_{D,i} = \frac{D_i - D_{\min}}{D_{\max} - D_{\min}},
\end{equation}
\begin{equation}
\label{equation9}
    R_{P,i} = \frac{P_i - P_{\min}}{P_{\max} - P_{\min}},
\end{equation}
where the subscripts ``min'' and ``max'' correspond to the minimum and maximum values within the population.

\paragraph*{Knowledge Construction:}

At each iteration, the knowledge of expert $i$ is constructed through two components, namely $K_{0,S1}$ and $K_{1,S1}$. The thresholds $B_C$, $B_P$, and $B_D$ are randomly sampled from the interval $[B_{\min}, B_{\max}]$. The first component, $K_{0,S1}$, is determined based on the normalized comprehensibility and probability:
\begin{equation}
\label{equation10}
    K_{0,i,S1} = R_{P,i}\,\frac{E_i + E_r}{2}, \quad \text{if } R_{C,i} \le B_C,\ R_{P,i} \le B_P,
\end{equation}
\begin{equation}
\label{equation11}
    K_{0,i,S1} = R_{P,i}\,\frac{E_i + E_{a,g}}{2}, \quad \text{if } R_{C,i} \le B_C,\ R_{P,i} > B_P,
\end{equation}
\begin{equation}
\label{equation12}
    K_{0,i,S1} = R_{P,i}\,\frac{E_{I,g} + E_r}{2}, \quad \text{if } R_{C,i} > B_C,\ R_{P,i} \le B_P,
\end{equation}
\begin{equation}
\label{equation13}
    K_{0,i,S1} = R_{P,i}\,\frac{E_{I,g} + E_{a,g}}{2}, \quad \text{if } R_{C,i} > B_C,\ R_{P,i} > B_P.
\end{equation}

\begin{figure}
    \centering
    \includegraphics[width=0.8\linewidth]{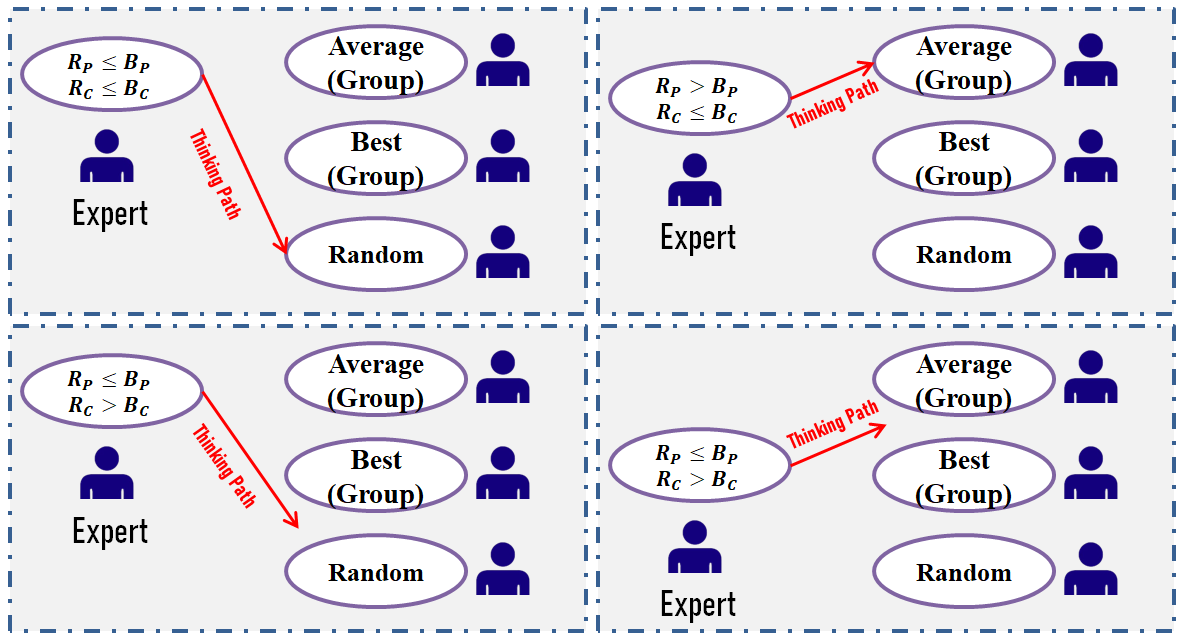}
    \caption{Computation process of $K_0$.}
    \label{fig:figure3}
\end{figure}
These rules reflect adaptive behavior: experts closer to the global logic tend to exploit their own positions, while those farther away rely more on the group’s best solution. Similarly, higher probability values promote exploration, whereas lower values encourage convergence toward the group mean.

The second component, $K_{1,S1}$, is governed by the degree ratio:
\begin{equation}
\label{equation14}
    K_{1,i,S1} = c_1 E_{a,g}, \quad \text{if } R_{D,i} \le B_D,
\end{equation}
\begin{equation}
\label{equation15}
    K_{1,i,S1} = c_1 E_u, \quad \text{if } R_{D,i} > B_D,
\end{equation}
where $E_r$ is a randomly selected expert from the group, $E_{a,g}$ denotes the group average, $E_u$ is a randomly generated point in the search space, and $c_1 \in [0,1]$.

\begin{figure}
    \centering
    \includegraphics[width=0.8\linewidth]{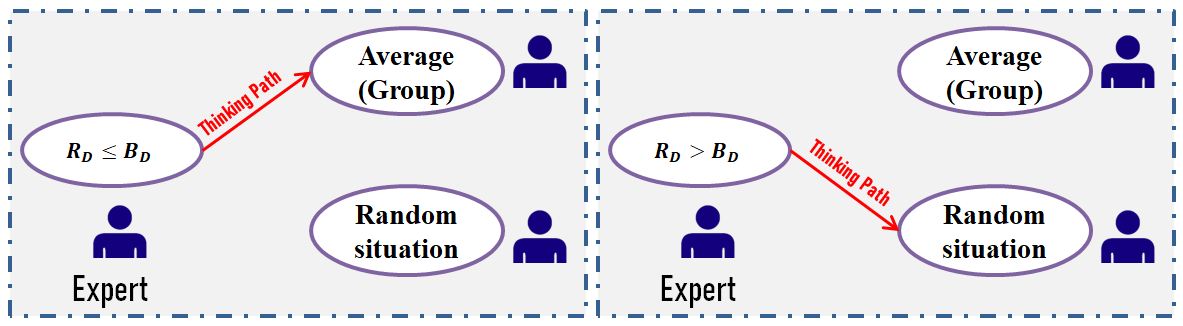}
    \caption{Computation process of $K_1$.}
    \label{fig:figure4}
\end{figure}

The overall knowledge update is then defined as
\begin{equation}
\label{equation16}
    K_{S1} = \frac{\left| K_{0,S1} + K_{1,S1} \right|}{2}.
\end{equation}

\paragraph*{Expert Update:}

The position of each expert is updated through two candidate solutions. The first update is given by
\begin{equation}
\label{equation17}
    E_{S1,new1} = E_i + c_2\, K_{S1},
\end{equation}
where $c_2 \in [-1.5, 1.5]$. A refined candidate is then constructed by incorporating the best expert in the group:
\begin{equation}
\label{equation18}
    E_{S1,new2} = c_3\, E_{S1,new1} + c_4\, E_{I,g},
\end{equation}
with $c_3 \in [-1.5, 1.5]$ and $c_4 \in [0,1]$.

\begin{figure}
    \centering
    \includegraphics[width=0.8\linewidth]{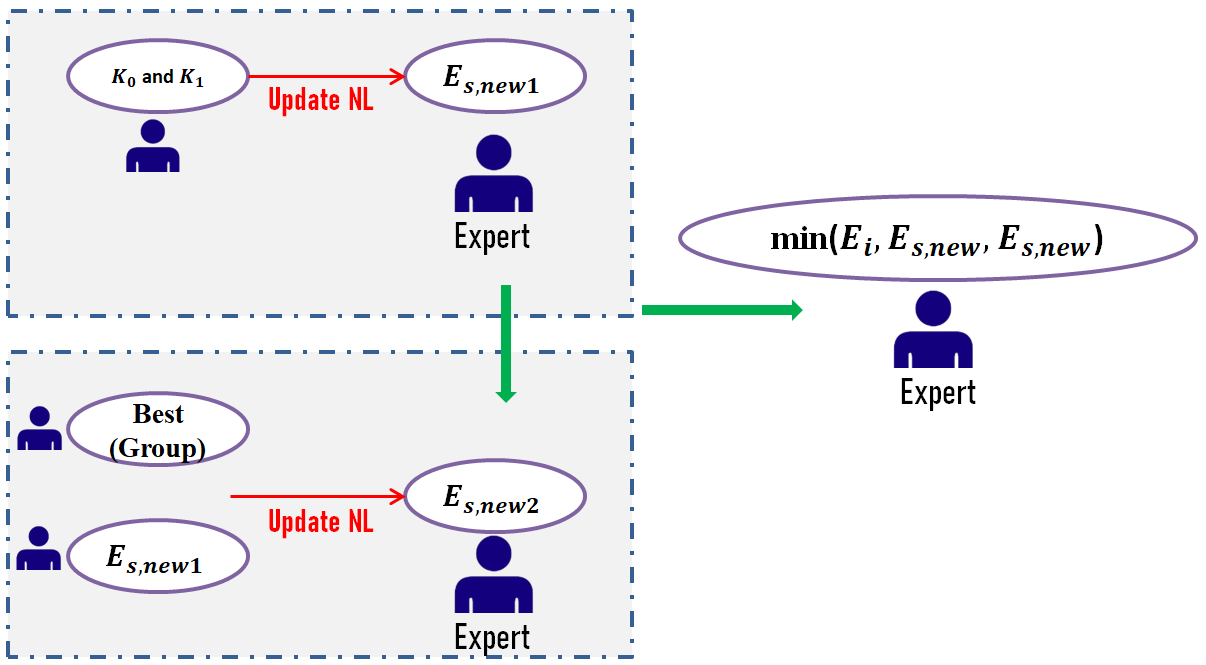}
    \caption{Determination of the final \textit{NL} value in Stage~1.}
    \label{fig:figure5}
\end{figure}

\begin{figure}
    \centering
    \includegraphics[width=0.8\linewidth]{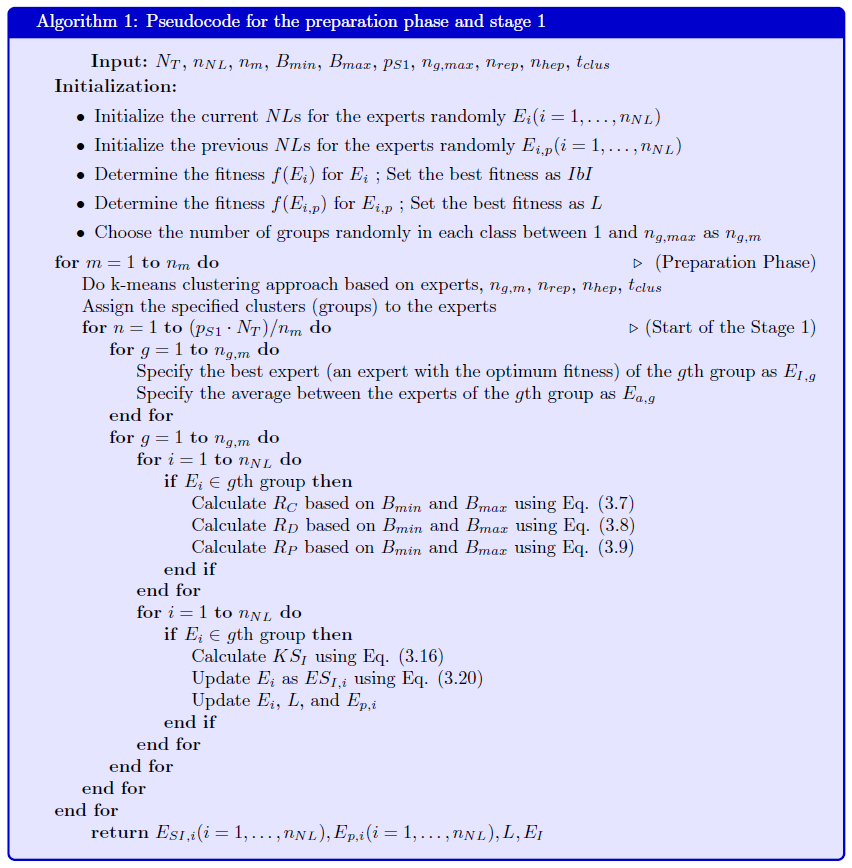}
    \caption{Preparation phase and Stage~1 (Group-based exploration).}
    \label{fig:algorithm1}
\end{figure}

The better candidate is selected according to the objective function $f$:
\begin{equation}
\label{equation19}
    E_{S1,new} = \min \left( E_{S1,new1},\ E_{S1,new2} \right),
\end{equation}
and compared with the current solution:
\begin{equation}
\label{equation20}
    E_{i,S1} = \min \left( E_i,\ E_{S1,new} \right).
\end{equation}
Algorithm~1 (Figure~\ref{fig:algorithm1}) summarizes the overall workflow of Stage~1. Upon completion of all iterations in this stage, the updated population is transferred to Stage~2 without revisiting previous steps. The parameter $E_I$ represents the globally best expert among all candidates, as illustrated in Figure~\ref{fig:figure5}.

\subsubsection{Stage 2: Global Integration}

In this stage, all experts are merged into a single population, enabling the algorithm to exploit the collective knowledge accumulated during Stage~1. The unified population is then evolved over $n_{t_2}$ iterations to further refine the non-logic (\textit{NL}) values.

At the beginning of this stage, the normalized ratios $R_{C,i}$ and $R_{D,i}$ are computed using Equations~(\ref{equation7}) and~(\ref{equation8}). Since all experts now belong to a single group, the probability measure is redefined with respect to the global best expert $E_I$. Accordingly, Equation~(\ref{equation9}) is replaced by
\begin{equation}
\label{equation21}
P_i = \sqrt{\sum_{j=1}^{n_{NL}} (E_j - E_I)^2}.
\end{equation}

At each iteration, the ratios $R_{P,i}$, $R_{C,i}$, and $R_{D,i}$ are utilized to construct an updated knowledge representation $K_{S2}$ for each expert. The first component, $K_{0,S2}$, is defined as
\begin{equation}
\label{equation22}
K_{0,i,S2} = \frac{R_{P,i}}{2}(E_i + E_R), \quad \text{if } R_{C,i} \leq B_C,\ R_{P,i} \leq B_P,
\end{equation}
\begin{equation}
\label{equation23}
K_{0,i,S2} = \frac{R_{P,i}}{2}(E_i + E_A), \quad \text{if } R_{C,i} \leq B_C,\ R_{P,i} > B_P,
\end{equation}
\begin{equation}
\label{equation24}
K_{0,i,S2} = \frac{R_{P,i}}{2}(E_I + E_R), \quad \text{if } R_{C,i} > B_C,\ R_{P,i} \leq B_P,
\end{equation}
\begin{equation}
\label{equation25}
K_{0,i,S2} = \frac{R_{P,i}}{2}(E_I + E_A), \quad \text{if } R_{C,i} > B_C,\ R_{P,i} > B_P,
\end{equation}
where $E_R$ denotes a randomly selected expert from the population and $E_A$ represents the population mean.

The second component, $K_{1,S2}$, is governed by the degree ratio:
\begin{equation}
\label{equation26}
K_{1,i,S2} = c_5 E_A, \quad \text{if } R_{D,i} \leq B_D,
\end{equation}
\begin{equation}
\label{equation27}
K_{1,i,S2} = c_5 E_u, \quad \text{if } R_{D,i} > B_D,
\end{equation}
where $E_u$ is a randomly generated point in the search space and $c_5 \in [0,1]$.

The combined knowledge update is then expressed as
\begin{equation}
\label{equation28}
K_{S2} = \frac{\left| K_{0,S2} + K_{1,S2} \right|}{2}.
\end{equation}

Using this knowledge, the first candidate update for each expert is given by
\begin{equation}
\label{equation29}
E_{S2,new1} = E_i + c_6 K_{S2},
\end{equation}
where $c_6 \in [-0.75, 0.75]$. A refined candidate is then obtained by incorporating the global best solution:
\begin{equation}
\label{equation30}
E_{S2,new2} = c_7 E_{S2,new1} + c_8 E_I,
\end{equation}
where $c_7 \in [-0.75, 0.75]$ and $c_8 \in [0,1]$.

The better candidate is selected based on the objective function:
\begin{equation}
\label{equation31}
E_{S2,new} = \min \left( E_{S2,new1},\ E_{S2,new2} \right),
\end{equation}
and the expert is updated accordingly:
\begin{equation}
\label{equation32}
E_{i,S2} = \min \left( E_i,\ E_{S2,new} \right).
\end{equation}

\begin{figure}
    \centering
    \includegraphics[width=0.8\linewidth]{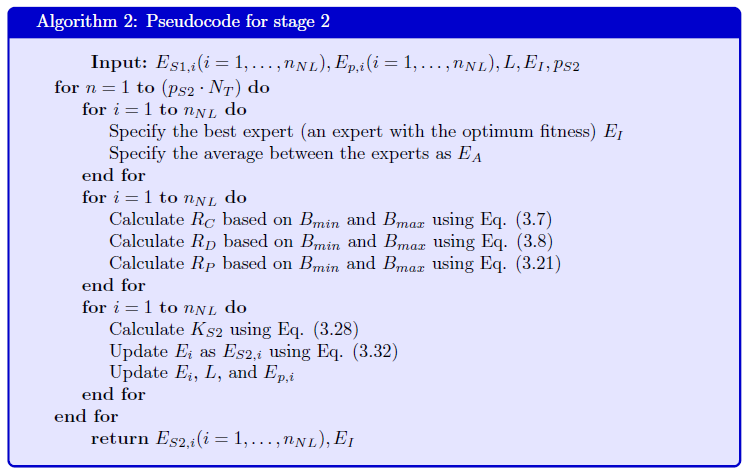}
    \caption{Stage~2 (Global integration phase).}
    \label{fig:algorithm2}
\end{figure}

Algorithm~2 (Figure~\ref{fig:algorithm2}) outlines the complete workflow of the integration stage.

\subsubsection{Stage 3: IbI Logic-Based Refinement}

In the final stage, the algorithm performs a fine-grained refinement of the solutions by applying incremental updates guided by the accumulated global knowledge. This stage emphasizes exploitation, where each expert adjusts its position based on both collective information and the influence of the best-performing solution. The process is iterated for $n_{t_3}$ steps or until a predefined convergence criterion is satisfied.

At each iteration, the knowledge for expert $i$, denoted by $K_{i,S3}$, is constructed using one of two strategies depending on the selected knowledge factor:
\begin{equation}
\label{equation33}
K_{i,S3} = \left| E_A - E_R \right|, \quad \text{if knowledge factor} = 1,
\end{equation}
\begin{equation}
\label{equation34}
K_{i,S3} = \left| E_A - E_I \right|, \quad \text{if knowledge factor} = 2,
\end{equation}
where $E_A$ denotes the population mean, $E_R$ is a randomly selected expert, and $E_I$ represents the global best expert.

Using this knowledge, two candidate updates are generated. The first update is given by
\begin{equation}
\label{equation35}
E_{S3,new1} = E_i + c_9\, K_{S3},
\end{equation}
followed by a refined update that incorporates the influence of the global best solution:
\begin{equation}
\label{equation36}
E_{S3,new2} = c_{10}\, E_{S3,new1} + c_{11}\, E_I.
\end{equation}

The optimal candidate is selected according to the objective function:
\begin{equation}
\label{equation37}
E_{S3,new} = \min \left( E_{S3,new1},\ E_{S3,new2} \right),
\end{equation}
and the expert is updated as
\begin{equation}
\label{equation38}
E_{i,S3} = \min \left( E_i,\ E_{S3,new} \right).
\end{equation}

Here, $c_9$ and $c_{10}$ are random coefficients sampled from the interval $[-0.25, 0.25]$, while $c_{11} \in [0,1]$. These parameters ensure controlled, small-scale adjustments that enhance convergence stability during the final refinement stage.

\begin{figure}
    \centering
    \includegraphics[width=0.8\linewidth]{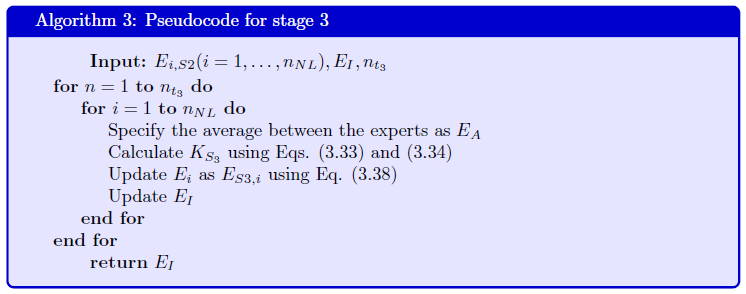}
    \caption{Stage~3 (IbI logic-based refinement phase).}
    \label{fig:algorithm3}
\end{figure}

Algorithm~3 (Figure~\ref{fig:algorithm3}) summarizes the complete optimization procedure of Stage~3.

\section{Numerical Analysis and Discussion}
\label{sec: Numerical Analysis}
In the framework where the charged lepton mass matrix is assumed to be diagonal, the effective light neutrino mass matrix \( m_\nu \), as derived previously, must be brought to a diagonal form through a unitary transformation governed by the, \( U_{\text{PMNS}} \). This transformation renders the neutrino mass matrix diagonal, ensuring the relation:
\begin{equation}\label{eq:e5}
m_\nu = U^* m_{\text{diag}} U^\dagger,
\end{equation}
where \( m_{\text{diag}} = \text{diag}(m_1, m_2, m_3) \) denotes the diagonal matrix composed of the physical masses of the three light neutrino mass eigenstates. Consequently, the neutrino mass eigenvalues are encapsulated within \( m_{\text{diag}} \), and the matrix \( U \) encodes the mixing parameters that relate flavor and mass eigenstates.

 \begin{align}
 \label{em}
m_{\text{diag}}=\begin{cases}
\text{diag}(m_{1},\sqrt{m_{1}^{2}+\Delta m_{21}^{2}},\sqrt{m_{1}^{2}+\Delta m_{31}^{2}}),~~ &\text{in NO} \\
\text{diag}( \sqrt{m_{3}^{2}+\Delta m_{31}^{2}}, \sqrt{m_{3}^{2}-\Delta m_{31}^{2}+\Delta m_{21}^{2}},m_{3}),~~ &\text{in IO}
\end{cases}
 \end{align}

Incorporating the most recent global neutrino oscillation data from NuFIT 6.0~\cite{esteban2025nufit}, the objective (or fitness) function, denoted as \( \Delta^{\prime}\), is formulated by embedding equation~\ref{eq:e5} within its structure. This formulation ensures consistency with contemporary experimental observations and facilitates a data-driven analysis of the neutrino mass matrix.

\begin{equation}
\label{eq:sum}
\begin{aligned}
\Delta^{\prime}=|\Delta^{\prime}_{1}|+|\Delta^{\prime}_{2}|+|\Delta^{\prime}_{3}|+|\Delta^{\prime}_{4}|+|\Delta^{\prime}_{5}|+|\Delta^{\prime}_{6}|,
\end{aligned}
\end{equation}
with,
\begin{equation}
\begin{aligned}
\Delta^{\prime}_{1}=\Bigl[&\Bigl(c_{13}^{2} ( e^{2 i \delta} m_{1} c_{12}^{2} + e^{2 i \alpha} m_{2} s_{12}^{2}) + e^{2 i \beta} m_{3} s_{13}^{2}\Bigl)
-\Bigl((3 \alpha _1 \beta _1 v_u^2 v_{\rho }^2 (y_1^4 (\alpha _{NS}+3 \beta _{NS}){}^2\\&-12 y_2 y_3 y_1^2 \alpha _{NS} (\alpha _{NS}+2 \beta _{NS})-8 (y_2^3+y_3^3) y_1 \alpha _{NS}^2-36 y_2^2 y_3^2 \beta _{NS}^2))\\&/(4 \Lambda ^3 (y_1^3 \alpha _{NS} (\alpha _{NS}+3 \beta _{NS}){}^2-3 y_2 y_3 y_1 (\alpha _{NS}-3 \beta _{NS}) (2 \alpha _{NS} \beta _{NS}\\&+\alpha _{NS}^2+3 \beta _{NS}^2)+(y_2^3+y_3^3) \alpha _{NS} (\alpha _{NS}^2-9 \beta _{NS}^2)))\Bigl)\Bigl]^2,
\end{aligned}
\end{equation}
\begin{equation}
\begin{aligned}
\Delta^{\prime}_{2}=\Bigl[&\Bigl(c_{13} \Big[ e^{i \beta} m_{3} s_{13} s_{23} 
- e^{2 i \delta} m_{1} c_{12} \big( c_{23} s_{12} + e^{-i \beta} c_{12} s_{13} s_{23} \big) 
+ e^{2 i \alpha} m_{2} s_{12} \big( c_{12} c_{23}\\& - e^{-i \beta} s_{12} s_{13} s_{23} \big) \Big]\Bigl)
+\Bigl((3 v_u^2 v_{\rho }^2 (2 y_2 y_1^3 (\alpha _{NS}+3 \beta _{NS}) (\alpha _1 \beta _2 \alpha _{NS}+\alpha _2 \beta _1 (\alpha _{NS}-3 \beta _{NS}))\\&+9 y_3^2 y_1^2 (\alpha _{NS}+\beta _{NS}) (\alpha _1 \beta _2 (\alpha _{NS}-\beta _{NS})+\alpha _2 \beta _1 (\alpha _{NS}+\beta _{NS}))\\&+6 y_2^2 y_3 y_1 (\alpha _2 \beta _1+\alpha _1 \beta _2) (\alpha _{NS} \beta _{NS}+2 \alpha _{NS}^2+3 \beta _{NS}^2)+2 y_2 (y_2^3 \alpha _{NS} (\alpha _2 \beta _1 (\alpha _{NS}\\&+3 \beta _{NS})+\alpha _1 \beta _2 (\alpha _{NS}-3 \beta _{NS}))+y_3^3 (\alpha _1 \beta _2 (\alpha _{NS}+3 \beta _{NS}){}^2+\alpha _2 \beta _1 (\alpha _{NS}\\&-3 \beta _{NS}){}^2))))/(8 \Lambda ^3 (y_1^3 \alpha _{NS} (\alpha _{NS}+3 \beta _{NS}){}^2-3 y_2 y_3 y_1 (\alpha _{NS}-3 \beta _{NS}) (2 \alpha _{NS} \beta _{NS}\\&+\alpha _{NS}^2+3 \beta _{NS}^2)+(y_2^3+y_3^3) \alpha _{NS} (\alpha _{NS}^2-9 \beta _{NS}^2)))\Bigl)\Bigl]^2,
\end{aligned}
\end{equation}
\begin{equation}
\begin{aligned}
\Delta^{\prime}_{3}&=\Bigl[\Bigl(c_{13} \bigl( e^{-i \beta} c_{23}( e^{2 i \beta} m_{3} - e^{2 i \delta} m_{1} c_{12}^{2} - e^{2 i \alpha} m_{2} s_{12}^{2} ) s_{13} 
+ ( e^{2 i \delta} m_{1} - e^{2 i \alpha} m_{2} ) c_{12} s_{12} s_{23} \bigl)\Bigl)
\\&+\Bigl((3 v_u^2 v_{\rho }^2 (2 y_3 y_1^3 (\alpha _{NS}+3 \beta _{NS}) (\alpha _3 \beta _1 \alpha _{NS}+\alpha _1 \beta _3 (\alpha _{NS}-3 \beta _{NS}))+9 y_2^2 y_1^2 (\alpha _{NS}+\beta _{NS})\\& (\alpha _3 \beta _1 (\alpha _{NS}-\beta _{NS})+\alpha _1 \beta _3 (\alpha _{NS}+\beta _{NS}))+6 y_2 y_3^2 y_1 (\alpha _3 \beta _1+\alpha _1 \beta _3) (\alpha _{NS} \beta _{NS}+2 \alpha _{NS}^2\\&+3 \beta _{NS}^2)+2 y_3 (y_2^3 (\alpha _3 \beta _1 (\alpha _{NS}+3 \beta _{NS}){}^2+\alpha _1 \beta _3 (\alpha _{NS}-3 \beta _{NS}){}^2)+y_3^3 \alpha _{NS} (\alpha _1 \beta _3 (\alpha _{NS}\\&+3 \beta _{NS})+\alpha _3 \beta _1 (\alpha _{NS}-3 \beta _{NS})))))/(8 \Lambda ^3 (y_1^3 \alpha _{NS} (\alpha _{NS}+3 \beta _{NS}){}^2-3 y_2 y_3 y_1 (\alpha _{NS}\\&-3 \beta _{NS}) (2 \alpha _{NS} \beta _{NS}+\alpha _{NS}^2+3 \beta _{NS}^2)+(y_2^3+y_3^3) \alpha _{NS} (\alpha _{NS}^2-9 \beta _{NS}^2)))\Bigl)\Bigl]^2,
\end{aligned}
\end{equation}
\begin{equation}
\begin{aligned}
\Delta^{\prime}_4=\Bigl[&\Bigl(m_{3} c_{13}^{2} s_{23}^{2} 
+ e^{2 i \delta} m_{1}( c_{23} s_{12} + e^{-i \beta} c_{12} s_{13} s_{23} )^{2} 
+ e^{2 i \alpha} m_{2} ( c_{12} c_{23} - e^{-i \beta} s_{12} s_{13} s_{23} )^{2}\Bigl)
\\&+\Bigl((3 \alpha _2 \beta _2 y_3 v_u^2 v_{\rho }^2 (4 y_1^3 \alpha _{NS} (2 \alpha _{NS}+3 \beta _{NS})+6 y_2 y_3 y_1 (\alpha _{NS} \beta _{NS}+2 \alpha _{NS}^2+3 \beta _{NS}^2)\\&-y_3^3 (\alpha _{NS}^2-9 \beta _{NS}^2)+8 y_2^3 \alpha _{NS}^2))/(4 \Lambda ^3 (y_1^3 \alpha _{NS} (\alpha _{NS}+3 \beta _{NS}){}^2-3 y_2 y_3 y_1 (\alpha _{NS}\\&-3 \beta _{NS}) (2 \alpha _{NS} \beta _{NS}+\alpha _{NS}^2+3 \beta _{NS}^2)+(y_2^3+y_3^3) \alpha _{NS} (\alpha _{NS}^2-9 \beta _{NS}^2)))\Bigl)\Bigl]^2,
\end{aligned}
\end{equation}
\begin{equation}
\begin{aligned}
\Delta^{\prime}_5&=\Bigl[\Bigl(m_{3} c_{13}^{2} c_{23} s_{23} 
- e^{2 i \delta} m_{1} ( - e^{-i \beta} c_{12} c_{23} s_{13} + s_{12} s_{23} ) 
( c_{23} s_{12} + e^{-i \beta} c_{12} s_{13} s_{23} ) \\&
- e^{2 i \alpha} m_{2} ( e^{-i \beta} c_{23} s_{12} s_{13} + c_{12} s_{23} ) 
( c_{12} c_{23} - e^{-i \beta} s_{12} s_{13} s_{23} )\Bigl)
\\&+\Bigl((3 v_u^2 v_{\rho }^2 (2 y_1^4 (\alpha _3 \beta _2+\alpha _2 \beta _3) \alpha _{NS} (\alpha _{NS}+3 \beta _{NS})+6 y_2 y_3 y_1^2 (\alpha _2 \beta _3 (\alpha _{NS}-\beta _{NS}) (2 \alpha _{NS}\\&+3 \beta _{NS})+\alpha _3 \beta _2 (\alpha _{NS} \beta _{NS}+2 \alpha _{NS}^2+3 \beta _{NS}^2))+2 (y_2^3+y_3^3) y_1 (\alpha _2 \beta _3 (\alpha _{NS}+3 \beta _{NS}){}^2\\&+\alpha _3 \beta _2 \alpha _{NS} (\alpha _{NS}-3 \beta _{NS}))+9 y_2^2 y_3^2 (\alpha _2 \beta _3 (\alpha _{NS}-\beta _{NS}){}^2+\alpha _3 \beta _2 (\alpha _{NS}+\beta _{NS}){}^2)))\\&/(8 \Lambda ^3 (y_1^3 \alpha _{NS} (\alpha _{NS}+3 \beta _{NS}){}^2-3 y_2 y_3 y_1 (\alpha _{NS}-3 \beta _{NS}) (2 \alpha _{NS} \beta _{NS}+\alpha _{NS}^2+3 \beta _{NS}^2)\\&+(y_2^3+y_3^3) \alpha _{NS} (\alpha _{NS}^2-9 \beta _{NS}^2)))\Bigl)\Bigl]^2,
\end{aligned}
\end{equation}
\begin{equation}
\begin{aligned}
\Delta^{\prime}_6=\Bigl[&\Bigl(m_{3} c_{13}^{2} c_{23}^{2} 
+ e^{2 i \alpha} m_{2} ( e^{-i \beta} c_{23} s_{12} s_{13} + c_{12} s_{23} )^{2} 
+ e^{2 i \delta} m_{1} ( e^{-i \beta} c_{12} c_{23} s_{13} - s_{12} s_{23} )^{2}\Bigl)\\&
+\Bigl((3 \alpha _3 \beta _3 y_2 v_u^2 v_{\rho }^2 (4 y_1^3 \alpha _{NS} (2 \alpha _{NS}+3 \beta _{NS})+6 y_2 y_3 y_1 (\alpha _{NS} \beta _{NS}+2 \alpha _{NS}^2+3 \beta _{NS}^2)\\&-y_2^3 (\alpha _{NS}^2-9 \beta _{NS}^2)+8 y_3^3 \alpha _{NS}^2))/(4 \Lambda ^3 (y_1^3 \alpha _{NS} (\alpha _{NS}\\&+3 \beta _{NS}){}^2-3 y_2 y_3 y_1 (\alpha _{NS}-3 \beta _{NS}) (2 \alpha _{NS} \beta _{NS}+\alpha _{NS}^2+3 \beta _{NS}^2)\\&+(y_2^3+y_3^3) \alpha _{NS} (\alpha _{NS}^2-9 \beta _{NS}^2)))\Bigl)\Bigl]^2,
\end{aligned}
\end{equation}

In Equation~\ref{eq:sum}, the $\Delta'$ is constructed from the full reconstruction condition ($m_\nu = U^* m_{\text{diag}} U^\dagger$), which leads to six independent matrix relations corresponding to the symmetric entries of the neutrino mass matrix. Explicitly, the deviations are defined as ($\Delta'_r = (m_\nu)_{ij} - (U^* m_{\text{diag}} U^\dagger)_{ij}$) for the six independent components. We further clarify that the rephasing freedom of the PMNS matrix has been properly taken into account, since the reconstruction is performed using the standard physical parameterization of $U_{\text{PMNS}}$, which ensures rephasing-invariant observables. In addition, the analysis is carried out using the absolute values of these deviations. 

\begin{figure}
    \centering
    \includegraphics[width=1\linewidth]{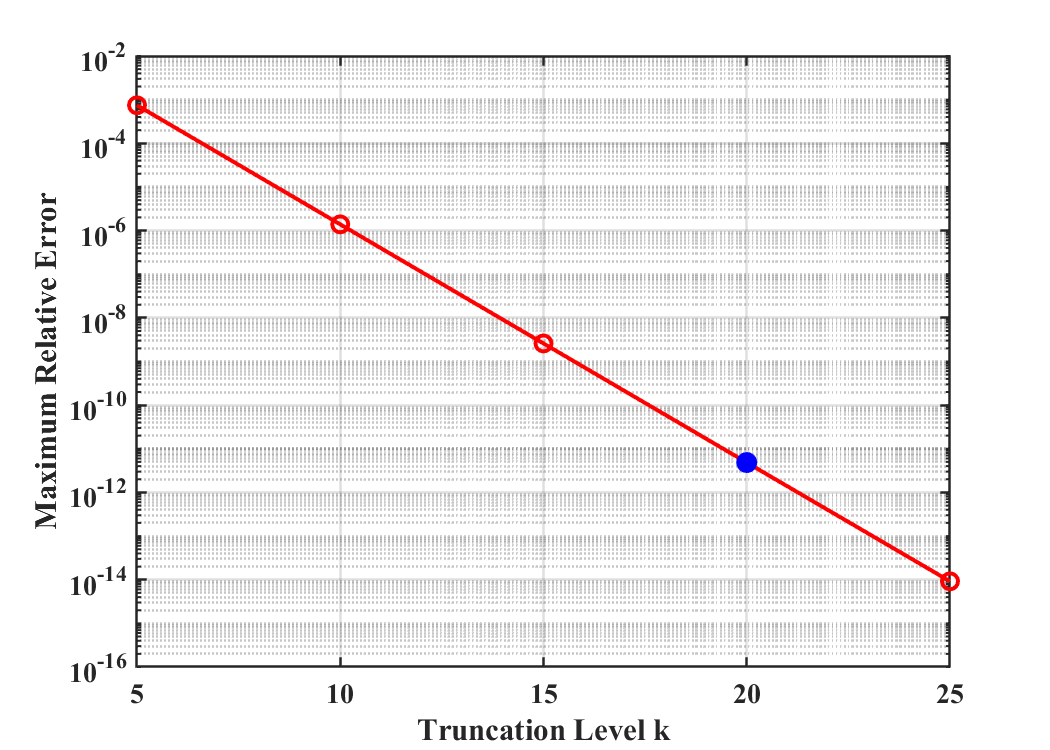}
    \caption{Convergence analysis of the truncated Dedekind eta function. The maximum relative error is computed with respect to a high-accuracy reference solution obtained using (N=200). The rapid decay of the error demonstrates that the truncation level (N=20) provides sufficient numerical accuracy for the considered parameter range.}
    \label{fig:convergencecheck}
\end{figure}

In our analysis, the Dedekind eta function is evaluated by truncating the infinite product to \( k = 20 \), which provides sufficient numerical precision for the range of parameters considered. To verify the numerical accuracy of the Dedekind eta function evaluation, a convergence analysis was performed by comparing truncated products with a high-accuracy reference solution computed using (k=200). The maximum relative error rapidly decreased with increasing truncation level and became negligible for ($k\geq20$), confirming that the selected truncation level ((k=20)) is sufficient for the parameter range considered as given in \ref{fig:convergencecheck}.

The best-fit values of the model parameters are optimized using ILA for NO and IO. The optimization process was carried out for 600 iterations, which constituted one independent run. Based on this iterative process, we obtained the optimal values for the fundamental model parameters, along with the neutrino mass eigenvalues given in Table \ref{tab:optimal values for Normal hierarchy} (\ref{tab:optimal values for Inverted hierarchy}) for NO (IO). In our analysis, we employ the ILA, which does not rely on a traditional $\chi^{2}$--minimization scheme. Instead, ILA performs optimization within pre-defined, experimentally motivated initial parameter ranges, including mass-squared splittings ($\Delta m^{2}_{21}$ and $\Delta m^{2}_{3l}$), the mixing angles 
($\sin^{2}\theta_{12}$, $\sin^{2}\theta_{23}$, $\sin^{2}\theta_{13}$) and the leptonic CP-violating phase $\delta_{\text{CP}}$ ($3\sigma$ confidence
interval).  The ILA updates the parameters only within these experimentally consistent ranges. Given these allowed intervals (The allowed ranges of all model parameters are given as 
$
-0.5 \leq Re(\tau) \leq 0.5, \
1 \leq Im(\tau) \leq 2, \
10^{-6} \leq \{\alpha_D, \beta_D, \gamma_D\} \leq 10^{-5},\ 10^{-3} \leq \{\alpha_D', \beta_D', \gamma_D'\} \leq 10^{-2}, \ 0 \leq \alpha_{NS} \leq 0.5, \ 0 \leq \beta_{NS} \leq 10^{-4}, \
10~\text{TeV} \leq v_\rho \leq 100~\text{TeV}, \
100~\text{TeV} \leq \Lambda \leq 1000~\text{TeV}, \ \text{and}\ 0 \leq m_1 \leq 0.01~\text{eV} \text{ for NO}, \;
\text{and } 0 \leq m_3 \leq 0.01~\text{eV} \text{ for IO}$), the ILA proceeds to estimate all mixing parameters and, along with the remaining model parameters, simultaneously; the corresponding optimized best-fit values are summarized in Tables \ref{tab:optimal values for Normal hierarchy} and \ref{tab:optimal values for Inverted hierarchy}. Since our main purpose is to compare the phenomenological measures with modern metaheuristic optimization algorithm, we choose the same parameters ranges as given in previous work. Moreover, The range $\tau$ corresponds to a representative region inside the standard modular fundamental domain, where $|Re(\tau)| \leq 1/2$ and $Im(\tau) > 0$. The optimization was performed using positive ranges for the couplings primarily to reduce the dimensionality of the scan and improve numerical convergence. Since physical observables depend on combinations of these parameters and phases can be absorbed through field redefinitions or effective coefficients, restricting the scan to positive magnitudes does not reduce the generality of the phenomenological analysis. The small values of the Yukawa-like parameters allow the model to reproduce neutrino masses in the sub-eV range while keeping the heavy states near the TeV scale. These are motivated by the linear seesaw structure where the light neutrino masses are naturally suppressed by small Dirac-type with the heavy scale $v_\rho$ and cutoff scale $\Lambda$.

In the optimization procedure, only $\tau$ is treated as an independent variable, while $y_{1,2,3}$ are computed from the value of $\tau$.
\begin{table}[htbp]
\centering
\begin{tabular}{llll}
\hline
 Parameters& Optimal Value & Parameters &Optimal Value
 \\
\hline
\hline
$\theta_{12}$ & $32.123^{\circ} $ &$\theta_{13}$ & $8.829^{\circ}$ \\
\hline
$\theta_{23}$ & $ 41.395 ^{\circ}$ &$\delta$ &$308.388^{\circ}$ \\
\hline
$\alpha$ & $49.970 ^{\circ}$ &$\beta$ & $-20.215^{\circ}$ \\
\hline
$v_\rho$ & $51.945\ \text{TeV}$ & $\Lambda$&  $988.346\ \text{TeV}$ \\
\hline
$\alpha_1$ & $1\times 10^{-6}$ & $\alpha_2$&  $1\times 10^{-5}$ \\
\hline
$\alpha_3$ & $8.475\times10^{-6}$ & $\beta_1$&  $3.860\times10^{-3}$ \\
\hline
$\beta_2$ & $9.449\times10^{-3}$&  $\beta_3$& $3.585\times10^{-3}$\\
\hline
$\alpha_{NS}$ & $0.460$&  $\beta_{NS}$& $6.752\times10^{-5}$\\
\hline
$\tau$ & $0.307+1.199i $ & $y_1$& $0.998 + 5.991\times10^{-3} i$\\
\hline
$y_2$ & $-0.388 - 0.293 i$ & $y_3$& $-3.309\times10^{-2} - 0.114 i$\\
\hline
$m_1$ & $3.863\ \text{meV}$ &$m_2$ &$9.478\ \text{meV}$ \\
\hline
$m_3$ & $50.278\ \text{meV}$ & &  \\
\hline
\end{tabular}
\caption{The optimized parameters include $\alpha_1$, $\alpha_2$, $\alpha_3$, $\beta_1$, $\beta_2$, $\beta_3$, $\alpha_{NS}$, $\beta_{NS}$, $y_1$, $y_2$, $y_3$, $\tau$, VEVs of $v_\rho$, cut-off scale $\Lambda$, mixing angles ($\theta_{12}$, $\theta_{13}$, $\theta_{23}$), phases ($\alpha$, $\beta$, $\delta$), and neutrino masses ($m_1$, $m_2$, $m_3$) using ILA for NO, achieving an objective function minimum of $6.012\times10^{-8}$. \label{tab:optimal values for Normal hierarchy}}
\end{table}
\begin{table}[htbp]
\centering
\begin{tabular}{llll}
\hline
 Parameters& Optimal Value & Parameters &Optimal Value
 \\
\hline
\hline
$\theta_{12}$ & $34.115^{\circ} $ &$\theta_{13}$ & $8.252^{\circ}$ \\
\hline
$\theta_{23}$ & $ 44.108^{\circ}$ &$\delta$ &$271.239^{\circ}$ \\
\hline
$\alpha$ & $28.237 ^{\circ}$ &$\beta$ & $57.296^{\circ}$ \\
\hline
$v_\rho$ & $90.906\ \text{TeV}$ & $\Lambda$&  $796.869\ \text{TeV}$ \\
\hline
$\alpha_1$ & $5.900\times10^{-6}$ & $\alpha_2$&  $1.720\times10^{-6}$ \\
\hline
$\alpha_3$ & $1.588\times10^{-6}$ & $\beta_1$&  $5.423\times10^{-3}$ \\
\hline
$\beta_2$ & $7.739\times 10^{-3}$&  $\beta_3$& $2.296\times 10^{-3}$\\
\hline
$\alpha_{NS}$ & $0.442$&  $\beta_{NS}$& $3.190\times10^{-5}$\\
\hline
$\tau$ & $0.5+1.168i $ & $y_1$& $0.992+ 1.054\times 10^{-16} i$\\
\hline
$y_2$ & $-0.259 - 0.448 i$ & $y_3$& $6.747\times10^{-2} - 0.117 i$\\
\hline
$m_1$ & $50.628\ \text{meV}$ &$m_2$ &$51.362\ \text{meV}$ \\
\hline
$m_3$ & $7.294\ \text{meV}$ & &  \\
\hline
\end{tabular}
\caption{The optimized parameters include $\alpha_1$, $\alpha_2$, $\alpha_3$, $\beta_1$, $\beta_2$, $\beta_3$, $\alpha_{NS}$, $\beta_{NS}$, $y_1$, $y_2$, $y_3$, $\tau$, VEVs of $v_\rho$, cut-off scale $\Lambda$, mixing angles ($\theta_{12}$, $\theta_{13}$, $\theta_{23}$), phases ($\alpha$, $\beta$, $\delta$), and neutrino masses ($m_1$, $m_2$, $m_3$) using ILA for IO, achieving an objective function minimum of $1.023\times10^{-7}$.
\label{tab:optimal values for Inverted hierarchy}}
\end{table}
The lepton mixing matrices for both NO and IO are determined as follows:
\begin{equation}
\label{eq:UPMNS Normal}
\begin{aligned}
|U_{PMNS}|=\begin{cases}\left(
\begin{array}{ccc}
 0.837 & 0.525 & 0.153 \\
 0.480 & 0.585 & 0.653 \\
 0.262 & 0.618 & 0.741 \\
\end{array}
\right),
~~ &\text{in NO} \\
\left(
\begin{array}{ccc}
 0.819 & 0.555 & 0.144 \\
 0.453 & 0.566 & 0.689 \\
 0.352 & 0.609 & 0.711 \\
\end{array}
\right), ~~ &\text{in IO}
\end{cases}
\end{aligned}
\end{equation}

The effective neutrino mass parameters \cite{rodejohann2011neutrino, mitra2012neutrinoless, bilenky2012neutrinoless, rodejohann2012neutrinolessssw,vergados2012theory}  relevant to both beta decay $\langle m_{\beta} \rangle$ and neutrinoless double beta decay $\langle m_{ee} \rangle$ offer critical insights into the fundamental nature of neutrinos, including their mass scale and whether they are Majorana particles. These parameters are derived from the elements of the lepton mixing matrix $U_{ei}$ (where $i=1,\ 2,\ 3$) and the corresponding light neutrino mass eigenstates $m_i$. The mathematical expressions for $\langle m_{\beta} \rangle$ and $\langle m_{ee} \rangle$ are given below:
\begin{equation}
\label{effective majorana parameter1}
\langle m_{ee} \rangle = \left| \sum_{i=1}^{3} U_{ei}^2 m_i \right|, \quad
\langle m_{\beta} \rangle=\sqrt{\sum_{i=1}^{3} |U_{ei}|^2m_i^2}.
\end{equation}
These expressions capture both the modulus-squared contributions to observable masses in beta decay experiments and the complex interference effects arising from CP-violating and Majorana phases in the context of neutrinoless double beta decay. Using the best-fit parameter values obtained through the ILA technique, we have computed the values of $\langle m_{ee} \rangle$ and $\langle m_{\beta} \rangle$ for both the NO and IO mass ordering scenarios. These values are based on the detailed structure of the PMNS matrix elements and the predicted light neutrino masses within our model framework.
\begin{equation}
\label{effective majorana parametervaluesPSO}
\begin{aligned}
\langle {m_{ee}} \rangle=\begin{cases} 0.842\ \text{meV},
~~ &\text{in NO} \\
27.922\ \text{meV},~~ &\text{in IO}
\end{cases},\qquad
\langle {m_{\beta}} \rangle=\begin{cases} 13.988\ \text{meV},
~~ &\text{in NO} \\
50.344\ \text{meV},~~ &\text{in IO}
\end{cases}.
\end{aligned}
\end{equation}
Our analysis demonstrates that the resulting effective mass parameters are well within the sensitivities of upcoming and planned experimental searches, such as those conducted by the LEGEND, nEXO, and KATRIN collaborations. Specifically, the predicted value of $\langle m_{ee} \rangle$ provides a testable signature in neutrinoless double beta decay, offering a pathway to distinguish between different mass hierarchies and to probe the CP-violating nature of neutrinos. 

In our study, we efficiently tuned the parameter space using ILA to get the optimal fit over 100 independent runs, each run having 600 iterations for both the hierarchy cases. ILA optimization has been demonstrated as being capable of navigating the complex, high-dimensional parameter space. It is well-suited to global optimization problems in which the objective function is nonlinear and possibly multi-modal. ILA was able to identify optimal parameter sets obeying experimental constraints on neutrino mixing angles, mass-squared differences, and CP-violating phases. The convergence accuracy of ILA was seen to reach around the benchmark of about $6.012\times10^{-8}$ ($1.023\times10^{-7}$) for NO (IO).

\begin{figure}
    \centering
\includegraphics[width=0.6\linewidth]{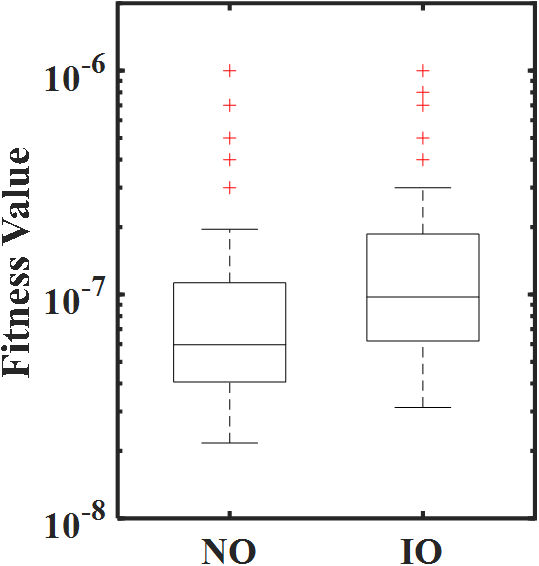}
    \caption{Boxplot of the fitness values obtained for the NO and IO case.}
    \label{fig:fitnessNO}
\end{figure}

Figure \ref{fig:fitnessNO} presents the box plots of the fitness values for the NO and IO cases. For NO, the vertical axis is displayed on a logarithmic scale, spanning approximately from $\sim 1\times10^{-8}$ to $\sim 5\times10^{-7}$. The red horizontal line inside the blue box represents the median fitness value ($\sim 6\times10^{-8}$), while the box boundaries indicate the interquartile range (IQR), enclosing the central $50\%$ of the data. The whiskers span the full range of values within 1.5 times the IQR, and the red crosses denote outliers associated with relatively poorer fits. The above distribution clearly shows that most of the values are between $\sim2\times10^{-8}$ and $\sim2\times10^{-7}$, while a few outliers are higher up to $\sim10^{-6}$. In the case of IO, it can be seen that fitness values are between $\sim1\times10^{-8}$ and $\sim7\times10^{-7}$. The median in this case is slightly higher and is $\sim1\times10^{-7}$. This value for IO is slightly higher than that for the NO case. The distribution for most of the values in the IO case is between $\sim3\times10^{-8}$ and $\sim3\times10^{-7}$, which is slightly higher than that for the NO case. A small number of outliers appear at higher values, reaching up to $\sim 10^{-6}$, illustrating less optimal fits.
    
\begin{figure}
    \centering
\includegraphics[width=1\linewidth]{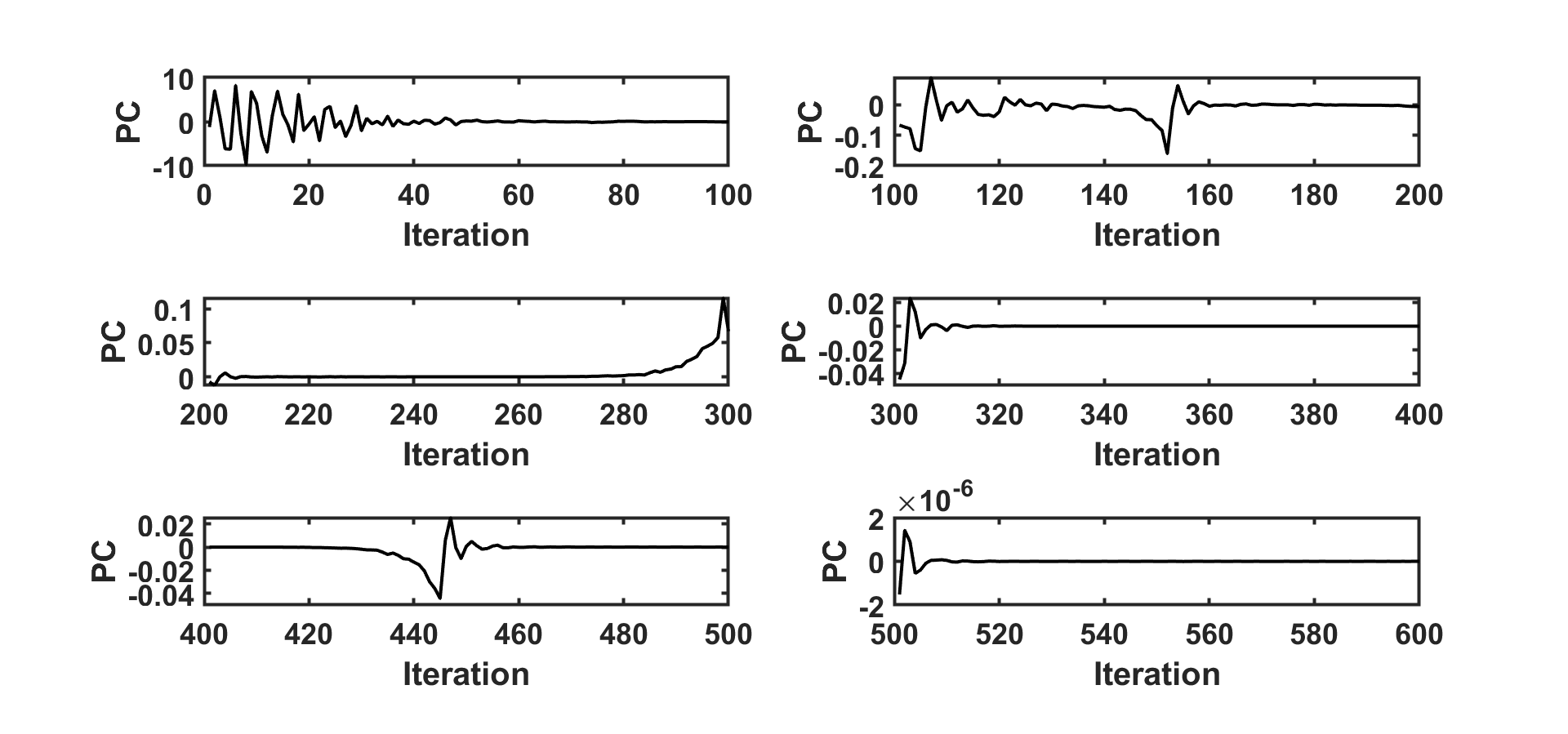}
    \caption{Best population convergence (PC) performance of ILA through iteration wise for NO}
    \label{fig:NH600}
\end{figure}

\begin{figure}
    \centering
\includegraphics[width=1\linewidth]{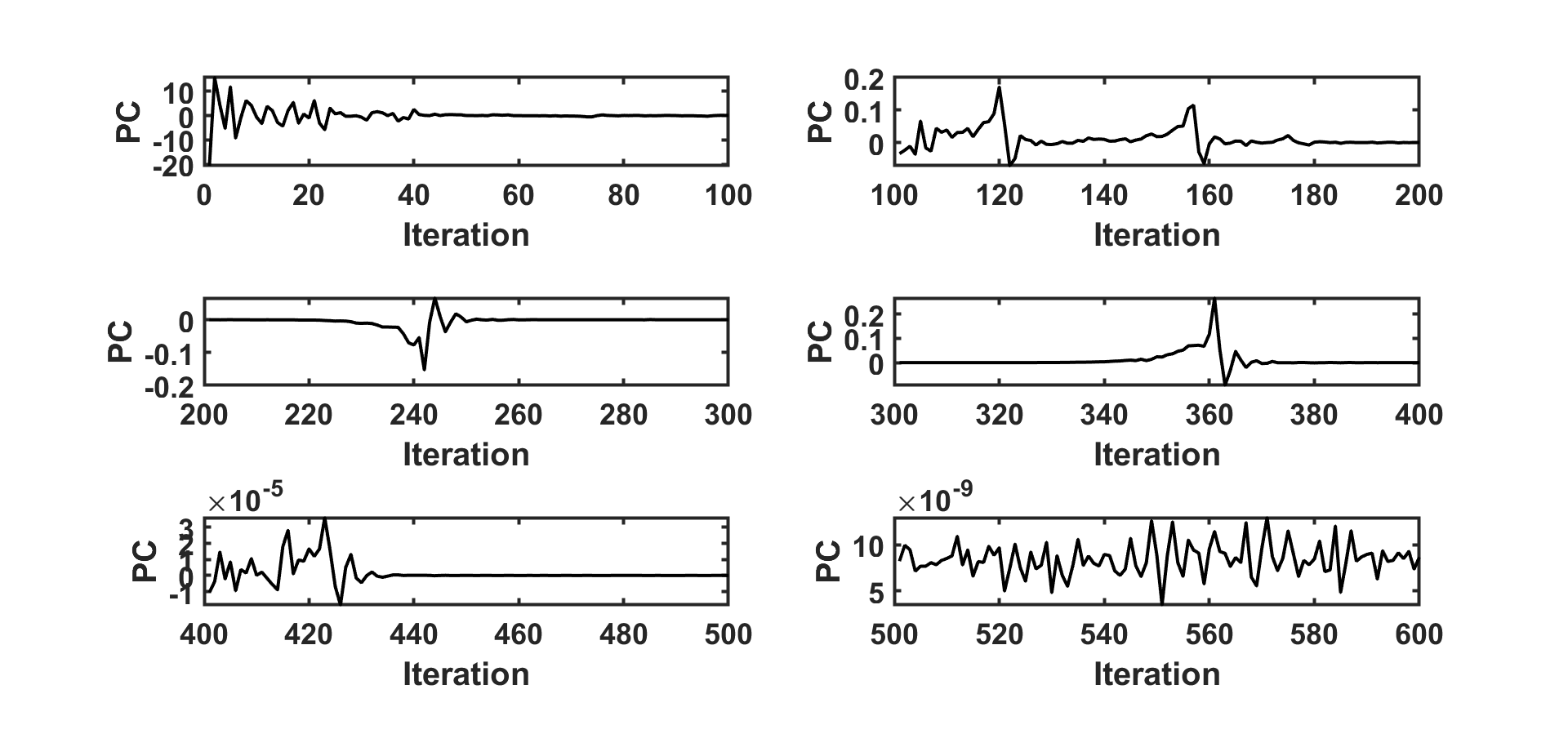}
    \caption{Best PC performance of ILA through iteration wise for IO}
    \label{fig:IH600}
\end{figure}

One key measure to evaluate the effectiveness and convergence behavior of the ILA algorithm is the change in the population distance over iterations, as shown in Figures \ref{fig:NH600} and \ref{fig:IH600}. Note: Figures \ref{fig:NH600} and \ref{fig:IH600} have six panels in total. Each panel shows a segment of the same optimization process. The optimization process consists of 600 iterations in total. Each panel in the figures shows 100 iterations of the same population distance evolution. 
The results show that there is a steady decline in population distance, which eventually tends to zero. This shows that the entire population is converging towards a common solution in the search space. Thus, such a convergence indicates that the algorithm is performing well, implying a correct implementation of the algorithm. This not only shows the stability of the optimization process but also indicates that the algorithm is likely to converge to a global or near-global optimum. Consequently, the observed convergence behavior affirms both the robustness and reliability of the ILA methodology.

\begin{figure}
    \centering
    \includegraphics[width=0.7\linewidth]{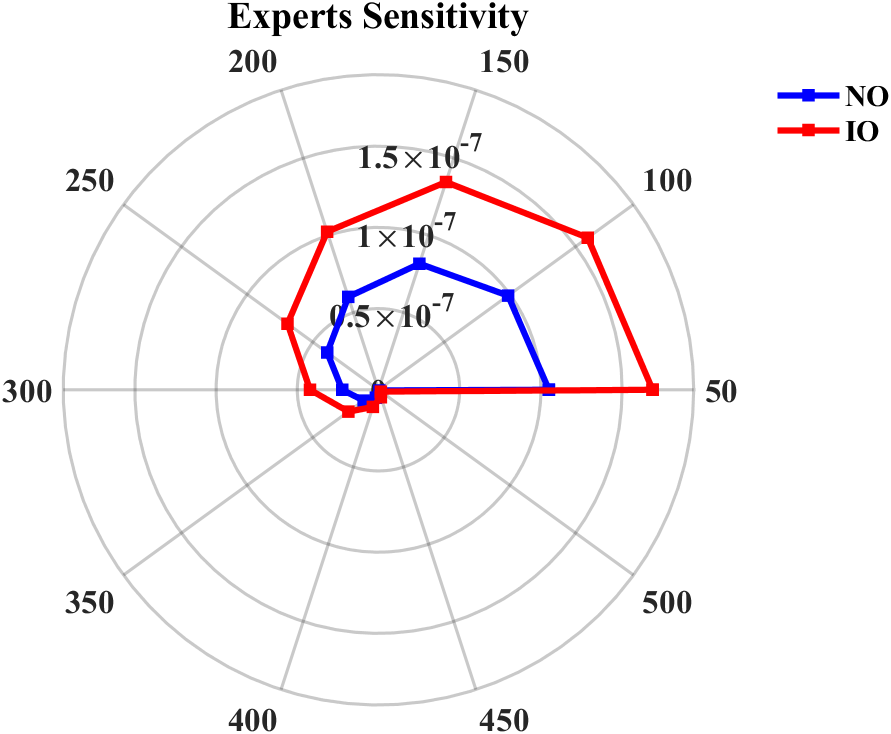}
    \caption{Sensitivity analysis of the ILA with respect to the number of experts. The optimization performance is evaluated for different expert populations while keeping the remaining algorithmic parameters fixed. The results indicate that the selected number of experts achieves a favorable balance between solution quality and computational cost.}
    \label{fig:placeholder1}
\end{figure}

\begin{figure}
    \centering
    \includegraphics[width=0.7\linewidth]{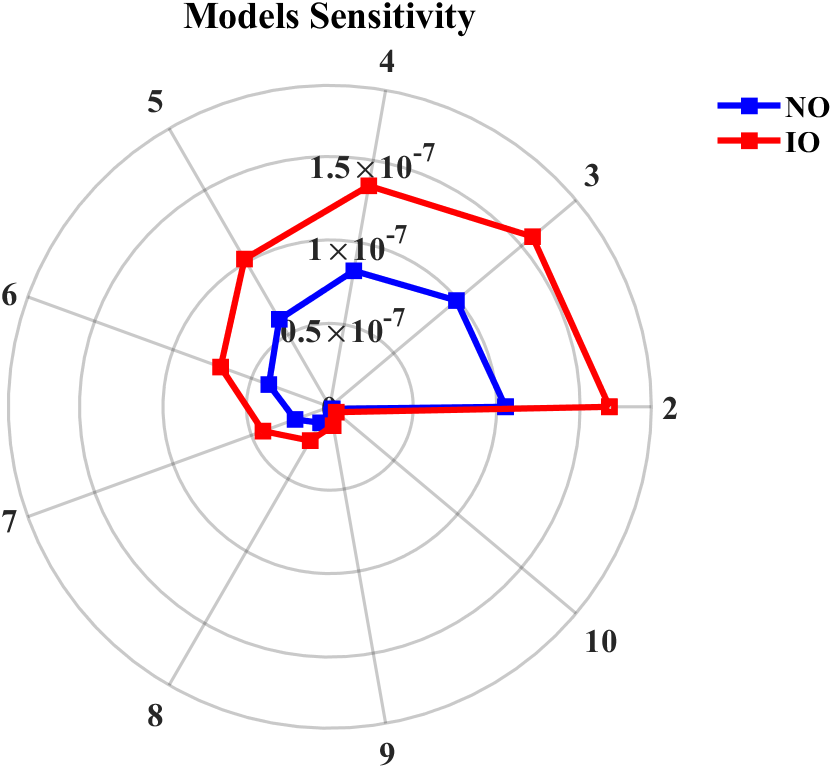}
    \caption{Sensitivity analysis of the ILA with respect to the number of models. The convergence behavior and objective function value are examined for different model populations. The selected configuration lies within a stable performance region, indicating robustness of the optimization process.}
    \label{fig:placeholder2}
\end{figure}

\begin{figure}
    \centering
    \includegraphics[width=0.7\linewidth]{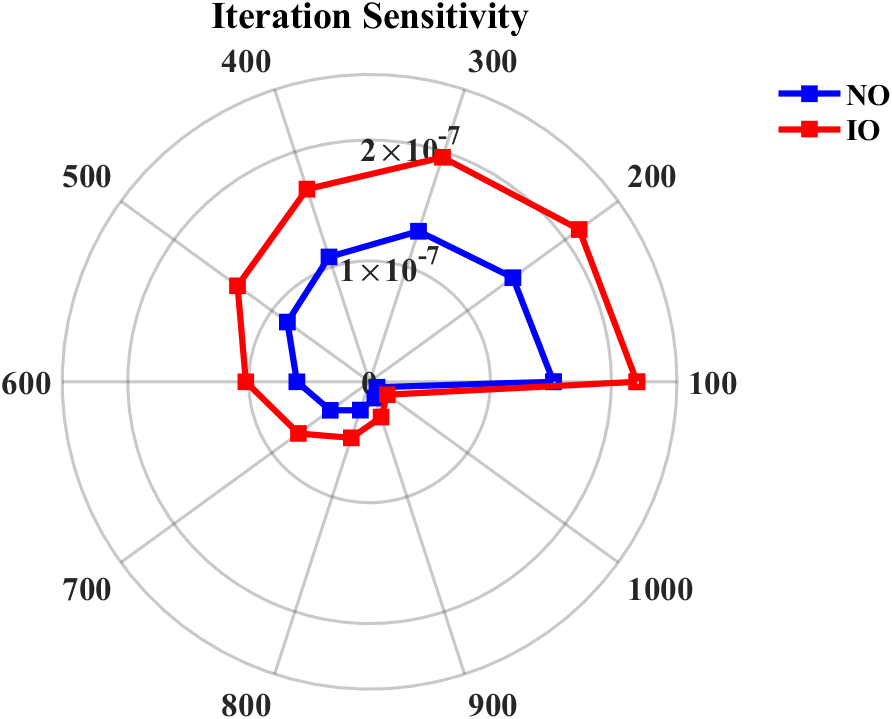}
    \caption{Sensitivity analysis of the ILA with respect to the maximum number of iterations. The objective function value is monitored as the iteration budget increases. The results show that the algorithm reaches a stable convergence region beyond the selected iteration count, confirming the adequacy of the adopted stopping criterion.}
    \label{fig:placeholder3}
\end{figure}

In addition, a sensitivity study was carried out for the principal ILA hyperparameters, including the number of experts, number of models, and maximum number of iterations as shown in Figures \ref{fig:placeholder1}, \ref{fig:placeholder2} and \ref{fig:placeholder3} respectively. The results demonstrated stable algorithmic behavior within the selected parameter ranges, indicating that the adopted settings provide a good balance between optimization performance and computational efficiency.

\begin{figure*}[t]
\centering

\includegraphics[width=0.53\linewidth]{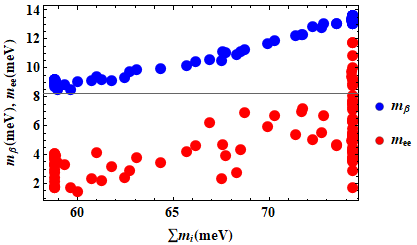}
\includegraphics[width=0.46\linewidth]{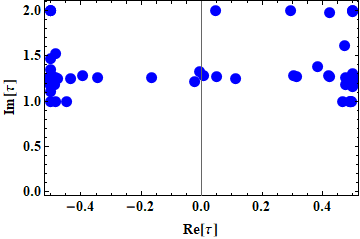}
\includegraphics[width=0.50\linewidth]{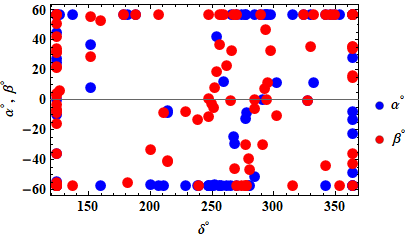}
\includegraphics[width=0.48\linewidth]{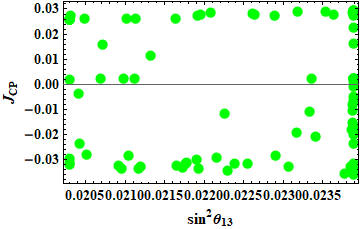}
\includegraphics[width=0.65\linewidth]{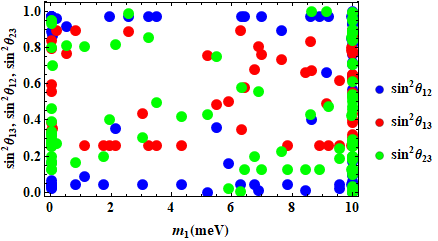}
\caption{The panels show the correlation plots among the relevant physical observables obtained from 100 independent optimization runs for the NO.}
\label{fig:NHcorrelation}
\end{figure*}

\begin{figure*}[t]
\centering
\includegraphics[width=0.53\linewidth]{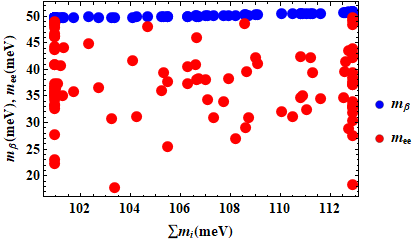}
\includegraphics[width=0.46\linewidth]{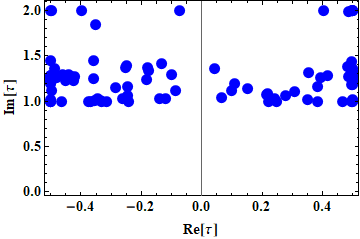}
\includegraphics[width=0.50\linewidth]{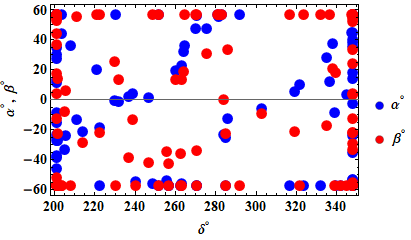}
\includegraphics[width=0.48\linewidth]{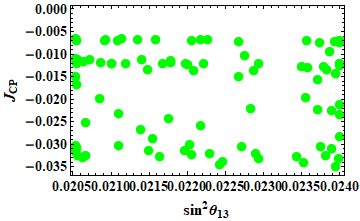}
\includegraphics[width=0.65\linewidth]{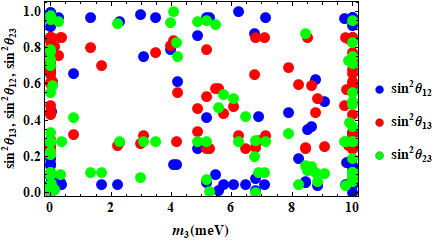}
\caption{The panels show the correlation plots among the relevant physical observables obtained from 100 independent optimization runs for the IO.}
\label{fig:IHcorrelation}
\end{figure*}

Due to the relatively large number of free parameters in the model, different parameter combinations may yield similar phenomenological predictions. To examine the robustness and uniqueness of the obtained solutions, the optimization procedure was repeated through 100 independent runs across the entire parameter space for both the NO and IO scenarios. The resulting best-fit solutions were further analyzed by constructing correlation plots among the relevant physical observables. These correlation plots, presented in the corresponding Figure~\ref{fig:NHcorrelation} (Figure~\ref{fig:IHcorrelation}) for the NO (IO), provide valuable insight into the allowed parameter regions, parameter interdependencies, and possible degeneracies of the model. The observed clustering of solutions and the consistency of the best-fit points demonstrate that the optimization algorithm repeatedly converges toward phenomenologically viable regions of the parameter space, thereby confirming the stability and reliability of the obtained results.

From the above analysis, it appears that all values of the observed parameters within their experimentally allowed regions are treated as equally accessible, with no explicit penalty applied. To address this, we have additionally performed an analysis in which the observed parameters are fixed at their central (best-fit) values and only the remaining model parameters are optimized, thereby effectively introducing a penalty and constraining the fit more tightly. In this case, the objective function converges to values of the order $10^{-7}$ for both normal and inverted hierarchies and the corresponding optimized best-fit values are summarized in Tables \ref{tab:optimal values for Normal hierarchy2} and \ref{tab:optimal values for Inverted hierarchy2}.
\begin{table}[htbp]
\centering
\begin{tabular}{llll}
\hline
 Parameters& Optimal Value & Parameters &Optimal Value
 \\
\hline
\hline
$\alpha$ & $-58.020 ^{\circ}$ &$\beta$ & $-9.163^{\circ}$ \\
\hline
$v_\rho$ & $17.126\ \text{TeV}$ & $\Lambda$&  $581.994\ \text{TeV}$ \\
\hline
$\alpha_1$ & $2.300\times 10^{-6}$ & $\alpha_2$&  $1.000\times 10^{-6}$ \\
\hline
$\alpha_3$ & $8.517\times10^{-6}$ & $\beta_1$&  $2.904\times10^{-3}$ \\
\hline
$\beta_2$ & $9.919\times10^{-3}$&  $\beta_3$& $3.3438\times10^{-3}$\\
\hline
$\alpha_{NS}$ & $0.200$&  $\beta_{NS}$& $9.635\times10^{-9}$\\
\hline
$\tau$ & $0.449+1.228i $ & $y_1$& $0.995 + 1.686\times 10^{-3}i$\\
\hline
$y_2$ & $-0.269 - 0.369 i$ & $y_3$& $0.032 - 0.100i$\\
\hline
$m_1$ & $3.272\ \text{meV}$ &$m_2$ &$9.252\ \text{meV}$ \\
\hline
$m_3$ & $50.236\ \text{meV}$ & &  \\
\hline
\end{tabular}
\caption{The optimized parameters include $\alpha_1$, $\alpha_2$, $\alpha_3$, $\beta_1$, $\beta_2$, $\beta_3$, $\alpha_{NS}$, $\beta_{NS}$, $y_1$, $y_2$, $y_3$, $\tau$, VEVs of $v_\rho$, cut-off scale $\Lambda$, phases ($\alpha$, $\beta$), and neutrino masses ($m_1$, $m_2$, $m_3$) using the best-fit values given in Table \ref{parameters} for NO, achieving an objective function minimum of $2.1231 \times 10^{-7}$. \label{tab:optimal values for Normal hierarchy2}}
\end{table}
\begin{table}[htbp]
\centering
\begin{tabular}{llll}
\hline
 Parameters& Optimal Value & Parameters &Optimal Value
 \\
\hline
\hline
$\alpha$ & $52.586^{\circ}$ &$\beta$ & $52.184^{\circ}$ \\
\hline
$v_\rho$ & $35.697\ \text{TeV}$ & $\Lambda$&  $810.399\ \text{TeV}$ \\
\hline
$\alpha_1$ & $7.907\times10^{-6}$ & $\alpha_2$&  $6.168\times10^{-6}$ \\
\hline
$\alpha_3$ & $1.705\times10^{-6}$ & $\beta_1$&  $8.189\times10^{-3}$ \\
\hline
$\beta_2$ & $4.649\times 10^{-3}$&  $\beta_3$& $4.750\times 10^{-3}$\\
\hline
$\alpha_{NS}$ & $0.206$&  $\beta_{NS}$& $2.475\times10^{-5}$\\
\hline
$\tau$ & $ 0.234+1.156i $ & $y_1$& $1.000 + 8.374\times 10^{-3} i$\\
\hline
$y_2$ & $-0.469 - 0.253 i$ & $y_3$& $-0.08 - 0.118 i$\\
\hline
$m_1$ & $50.418\ \text{meV}$ &$m_2$ &$51.155\ \text{meV}$ \\
\hline
$m_3$ & $5.651\ \text{meV}$ & &  \\
\hline
\end{tabular}
\caption{The optimized parameters include $\alpha_1$, $\alpha_2$, $\alpha_3$, $\beta_1$, $\beta_2$, $\beta_3$, $\alpha_{NS}$, $\beta_{NS}$, $y_1$, $y_2$, $y_3$, $\tau$, VEVs of $v_\rho$, cut-off scale $\Lambda$, phases ($\alpha$, $\beta$), and neutrino masses ($m_1$, $m_2$, $m_3$) using the best-fit values
given in Table \ref{parameters} for IO, achieving an objective function minimum of $1.002 \times 10^{-7}$.
\label{tab:optimal values for Inverted hierarchy2}}
\end{table}

Since the lower and upper bound ranges adopted here
are the same as those used in the original study where numerical analyses and predictions were
already performed. In our case, the observed parameters are fixed to their best-fit (central) values, and ILA optimizes the remaining model parameters within these established ranges. Consequently, the optimized solutions naturally lie within the same phenomenologically allowed region, which
cannot be dismissed since they correspond to best-fit values consistent with experimental data. 

\section{Conclusion}
\label{sec:Conclusion}
In this work, we have analyzed a comprehensive exploration of the neutrino mass model based on $A_4$ discrete non-Abelian modular symmetry formulated within a linear seesaw framework that modifies the conventional type-I seesaw structure with focus on optimizing the model parameters using ILA.  In this setup, the Higgs and lepton superfields transform under well-defined modular representations along with a singlet weighton \(\rho\), allowing the flavor structure to be generated with a reduced number of flavon fields compared to conventional flavor models. This economy of fields avoids unnecessary spectrum proliferation while retaining the possibility of incorporating flavons, thereby improving both predictability and model-building flexibility. The effectiveness of ILA in computing the $U_{PMNS}$ matrix and neutrino mass parameters is demonstrated for both NO and IO. For NO (IO), ILA produces neutrino masses as $m_1 = 3.8630 \ \text{meV}$ ($m_1 = 50.628\ \text{meV}$), $m_2 = 9.4775\ \text{meV}$ ($m_2 = 51.3624 \ \text{meV}$), $m_3 =50.2784 \ \text{meV}$ ($m_3 = 7.2937\ \text{meV}$), with effective neutrino mass parameters as $m_{\beta} = 13.9880\ \text{meV}$ ($m_{\beta} = 50.3445 \ \text{meV}$), $m_{ee} = 0.8416 \ \text{meV}$ ($m_{ee} = 27.9219 \ \text{meV}$), demonstrating the robustness of ILA for precision modeling in neutrino mass scenarios. We also performed an additional analysis in which the ILA algorithm was executed using the best-fit NuFIT 6.0 values for the observed parameters, while optimizing the remaining model parameters. The model complies with the latest cosmological limit on the total neutrino mass, \(\Sigma m_i \leq 0.12~\mathrm{eV}\), and also yields predictions for the mixing angles \((\theta_{ab},\; a < b,\; a,b \in \{1,2,3\})\), the Dirac phase \(\delta\), and the Majorana phases \(\alpha\) and \(\beta\), all in good agreement with current experimental data.
\section*{Data availability}
The datasets used and/or analysed during the current study are available from the corresponding author on reasonable request.
\section*{Funding:}
No Funding
\bibliographystyle{JHEP}
\bibliography{biblio}

@article{kamiokande1998evidence,
author = {{Fukuda}, Y. and {Hayakawa}, T. and {Ichihara}, E. and {Inoue}, K. and {Ishihara}, K. and {Ishino}, H. and {Itow}, Y. and {Kajita}, T. and {Kameda}, J. and {Kasuga}, S. and {Kobayashi}, K. and {Kobayashi}, Y. and {Koshio}, Y. and {Miura}, M. and {Nakahata}, M. and {Nakayama}, S. and {Okada}, A. and {Okumura}, K. and {Sakurai}, N. and {Shiozawa}, M. and {Suzuki}, Y. and {Takeuchi}, Y. and {Totsuka}, Y. and {Yamada}, S. and {Earl}, M. and {Habig}, A. and {Kearns}, E. and {Messier}, M.~D. and {Scholberg}, K. and {Stone}, J.~L. and {Sulak}, L.~R. and {Walter}, C.~W. and {Goldhaber}, M. and {Barszczxak}, T. and {Casper}, D. and {Gajewski}, W. and {Halverson}, P.~G. and {Hsu}, J. and {Kropp}, W.~R. and {Price}, L.~R. and {Reines}, F. and {Smy}, M. and {Sobel}, H.~W. and {Vagins}, M.~R. and {Ganezer}, K.~S. and {Keig}, W.~E. and {Ellsworth}, R.~W. and {Tasaka}, S. and {Flanagan}, J.~W. and {Kibayashi}, A. and {Learned}, J.~G. and {Matsuno}, S. and {Stenger}, V.~J. and {Takemori}, D. and {Ishii}, T. and {Kanzaki}, J. and {Kobayashi}, T. and {Mine}, S. and {Nakamura}, K. and {Nishikawa}, K. and {Oyama}, Y. and {Sakai}, A. and {Sakuda}, M. and {Sasaki}, O. and {Echigo}, S. and {Kohama}, M. and {Suzuki}, A.~T. and {Haines}, T.~J. and {Blaufuss}, E. and {Kim}, B.~K. and {Sanford}, R. and {Svoboda}, R. and {Chen}, M.~L. and {Conner}, Z. and {Goodman}, J.~A. and {Sullivan}, G.~W. and {Hill}, J. and {Jung}, C.~K. and {Martens}, K. and {Mauger}, C. and {McGrew}, C. and {Sharkey}, E. and {Viren}, B. and {Yanagisawa}, C. and {Doki}, W. and {Miyano}, K. and {Okazawa}, H. and {Saji}, C. and {Takahata}, M. and {Nagashima}, Y. and {Takita}, M. and {Yamaguchi}, T. and {Yoshida}, M. and {Kim}, S.~B. and {Etoh}, M. and {Fujita}, K. and {Hasegawa}, A. and {Hasegawa}, T. and {Hatakeyama}, S. and {Iwamoto}, T. and {Koga}, M. and {Maruyama}, T. and {Ogawa}, H. and {Shirai}, J. and {Suzuki}, A. and {Tsushima}, F. and {Koshiba}, M. and {Nemoto}, M. and {Nishijima}, K. and {Futagami}, T. and {Hayato}, Y. and {Kanaya}, Y. and {Kaneyuki}, K. and {Watanabe}, Y. and {Kielczewska}, D. and {Doyle}, R.~A. and {George}, J.~S. and {Stachyra}, A.~L. and {Wai}, L.~L. and {Wilkes}, R.~J. and {Young}, K.~K.},
        title = "{Evidence for Oscillation of Atmospheric Neutrinos}",
      journal = {Physical Review Letters},
         year = 1998,
        month = aug,
       volume = {81},
       number = {8},
        pages = {1562-1567},
          doi = {10.1103/PhysRevLett.81.1562},
archivePrefix = {arXiv},
       eprint = {hep-ex/9807003},
 primaryClass = {hep-ex},
       adsurl = {https://ui.adsabs.harvard.edu/abs/1998PhRvL..81.1562F}
}

@article{eguchi2003first,
author = {{Eguchi}, K. and {Enomoto}, S. and {Furuno}, K. and {Goldman}, J. and {Hanada}, H. and {Ikeda}, H. and {Ikeda}, K. and {Inoue}, K. and {Ishihara}, K. and {Itoh}, W. and {Iwamoto}, T. and {Kawaguchi}, T. and {Kawashima}, T. and {Kinoshita}, H. and {Kishimoto}, Y. and {Koga}, M. and {Koseki}, Y. and {Maeda}, T. and {Mitsui}, T. and {Motoki}, M. and {Nakajima}, K. and {Nakajima}, M. and {Nakajima}, T. and {Ogawa}, H. and {Owada}, K. and {Sakabe}, T. and {Shimizu}, I. and {Shirai}, J. and {Suekane}, F. and {Suzuki}, A. and {Tada}, K. and {Tajima}, O. and {Takayama}, T. and {Tamae}, K. and {Watanabe}, H. and {Busenitz}, J. and {Djurcic}, Z. and {McKinny}, K. and {Mei}, D. -M. and {Piepke}, A. and {Yakushev}, E. and {Berger}, B.~E. and {Chan}, Y.~D. and {Decowski}, M.~P. and {Dwyer}, D.~A. and {Freedman}, S.~J. and {Fu}, Y. and {Fujikawa}, B.~K. and {Heeger}, K.~M. and {Lesko}, K.~T. and {Luk}, K. -B. and {Murayama}, H. and {Nygren}, D.~R. and {Okada}, C.~E. and {Poon}, A.~W. and {Steiner}, H.~M. and {Winslow}, L.~A. and {Horton-Smith}, G.~A. and {McKeown}, R.~D. and {Ritter}, J. and {Tipton}, B. and {Vogel}, P. and {Lane}, C.~E. and {Miletic}, T. and {Gorham}, P.~W. and {Guillian}, G. and {Learned}, J.~G. and {Maricic}, J. and {Matsuno}, S. and {Pakvasa}, S. and {Dazeley}, S. and {Hatakeyama}, S. and {Murakami}, M. and {Svoboda}, R.~C. and {Dieterle}, B.~D. and {Dimauro}, M. and {Detwiler}, J. and {Gratta}, G. and {Ishii}, K. and {Tolich}, N. and {Uchida}, Y. and {Batygov}, M. and {Bugg}, W. and {Cohn}, H. and {Efremenko}, Y. and {Kamyshkov}, Y. and {Kozlov}, A. and {Nakamura}, Y. and {de Braeckeleer}, L. and {Gould}, C.~R. and {Karwowski}, H.~J. and {Markoff}, D.~M. and {Messimore}, J.~A. and {Nakamura}, K. and {Rohm}, R.~M. and {Tornow}, W. and {Young}, A.~R. and {Wang}, Y. -F.},
        title = "{First Results from KamLAND: Evidence for Reactor Antineutrino Disappearance}",
      journal = {Physical Review Letters},
         year = 2003,
        month = jan,
       volume = {90},
       number = {2},
          eid = {021802},
        pages = {021802},
          doi = {10.1103/PhysRevLett.90.021802},
archivePrefix = {arXiv},
       eprint = {hep-ex/0212021},
 primaryClass = {hep-ex},
       adsurl = {https://ui.adsabs.harvard.edu/abs/2003PhRvL..90b1802E}
}

@article{ahn2006measurement,
  author = {{Ahn}, M.~H. and {Aliu}, E. and {Andringa}, S. and {Aoki}, S. and {Aoyama}, Y. and {Argyriades}, J. and {Asakura}, K. and {Ashie}, R. and {Berghaus}, F. and {Berns}, H.~G. and {Bhang}, H. and {Blondel}, A. and {Borghi}, S. and {Bouchez}, J. and {Boyd}, S.~C. and {Burguet-Castell}, J. and {Casper}, D. and {Catala}, J. and {Cavata}, C. and {Cervera}, A. and {Chen}, S.~M. and {Cho}, K.~O. and {Choi}, J.~H. and {Dore}, U. and {Echigo}, S. and {Espinal}, X. and {Fechner}, M. and {Fernandez}, E. and {Fujii}, K. and {Fujii}, Y. and {Fukuda}, S. and {Fukuda}, Y. and {Gomez-Cadenas}, J. and {Gran}, R. and {Hara}, T. and {Hasegawa}, M. and {Hasegawa}, T. and {Hayashi}, K. and {Hayato}, Y. and {Helmer}, R.~L. and {Higuchi}, I. and {Hill}, J. and {Hiraide}, K. and {Hirose}, E. and {Hosaka}, J. and {Ichikawa}, A.~K. and {Ieiri}, M. and {Iinuma}, M. and {Ikeda}, A. and {Inagaki}, T. and {Ishida}, T. and {Ishihara}, K. and {Ishii}, H. and {Ishii}, T. and {Ishino}, H. and {Ishitsuka}, M. and {Itow}, Y. and {Iwashita}, T. and {Jang}, H.~I. and {Jang}, J.~S. and {Jeon}, E.~J. and {Jeong}, I.~S. and {Joo}, K.~K. and {Jover}, G. and {Jung}, C.~K. and {Kajita}, T. and {Kameda}, J. and {Kaneyuki}, K. and {Kang}, B.~H. and {Kato}, I. and {Kato}, Y. and {Kearns}, E. and {Kerr}, D. and {Kim}, C.~O. and {Khabibullin}, M. and {Khotjantsev}, A. and {Kielczewska}, D. and {Kim}, B.~J. and {Kim}, H.~I. and {Kim}, J.~H. and {Kim}, J.~Y. and {Kim}, S.~B. and {Kitamura}, M. and {Kitching}, P. and {Kobayashi}, K. and {Kobayashi}, T. and {Kohama}, M. and {Konaka}, A. and {Koshio}, Y. and {Kropp}, W. and {Kubota}, J. and {Kudenko}, Yu. and {Kume}, G. and {Kuno}, Y. and {Kurimoto}, Y. and {Kutter}, T. and {Learned}, J. and {Likhoded}, S. and {Lim}, I.~T. and {Lim}, S.~H. and {Loverre}, P.~F. and {Ludovici}, L. and {Maesaka}, H. and {Mallet}, J. and {Mariani}, C. and {Martens}, K. and {Maruyama}, T. and {Matsuno}, S. and {Matveev}, V. and {Mauger}, C. and {McConnel Mahn}, K.~B. and {McGrew}, C. and {Mikheyev}, S. and {Minakawa}, M. and {Minamino}, A. and {Mine}, S. and {Mineev}, O. and {Mitsuda}, C. and {Mitsuka}, G. and {Miura}, M. and {Moriguchi}, Y. and {Morita}, T. and {Moriyama}, S. and {Nakadaira}, T. and {Nakahata}, M. and {Nakamura}, K. and {Nakano}, I. and {Nakata}, F. and {Nakaya}, T. and {Nakayama}, S. and {Namba}, T. and {Nambu}, R. and {Nawang}, S. and {Nishikawa}, K. and {Nishino}, H. and {Nishiyama}, S. and {Nitta}, K. and {Noda}, S. and {Noumi}, H. and {Nova}, F. and {Novella}, P. and {Obayashi}, Y. and {Okada}, A. and {Okumura}, K. and {Okumura}, M. and {Onchi}, M. and {Ooyabu}, T. and {Oser}, S.~M. and {Otaki}, T. and {Oyama}, Y. and {Pac}, M.~Y. and {Park}, H. and {Pierre}, F. and {Rodriguez}, A. and {Saji}, C. and {Sakai}, A. and {Sakuda}, M. and {Sakurai}, N. and {Sanchez}, F. and {Sarrat}, A. and {Sasaki}, T. and {Sato}, H. and {Sato}, K. and {Scholberg}, K. and {Schroeter}, R. and {Sekiguchi}, M. and {Seo}, E. and {Sharkey}, E. and {Shima}, A. and {Shiozawa}, M. and {Shiraishi}, K. and {Sitjes}, G. and {Smy}, M. and {So}, H. and {Sobel}, H. and {Sorel}, M. and {Stone}, J. and {Sulak}, L. and {Suga}, Y. and {Suzuki}, A. and {Suzuki}, Y. and {Suzuki}, Y. and {Tada}, M. and {Takahashi}, T. and {Takasaki}, M. and {Takatsuki}, M. and {Takenaga}, Y. and {Takenaka}, K. and {Takeuchi}, H. and {Takeuchi}, Y. and {Taki}, K. and {Takubo}, Y. and {Tamura}, N. and {Tanaka}, H. and {Tanaka}, K. and {Tanaka}, M. and {Tanaka}, Y. and {Tashiro}, K. and {Terri}, R. and {T'Jampens}, S. and {Tornero-Lopez}, A. and {Toshito}, T. and {Totsuka}, Y. and {Ueda}, S. and {Vagins}, M. and {Whitehead}, L. and {Walter}, C.~W. and {Wang}, W. and {Wilkes}, R.~J. and {Yamada}, S. and {Yamada}, Y. and {Yamamoto}, S. and {Yamanoi}, Y. and {Yanagisawa}, C. and {Yershov}, N. and {Yokoyama}, H. and {Yokoyama}, M. and {Yoo}, J. and {Yoshida}, M. and {Zalipska}, J.},
        title = "{Measurement of neutrino oscillation by the K2K experiment}",
      journal = {Physical Review D},
         year = 2006,
        month = oct,
       volume = {74},
       number = {7},
          eid = {072003},
        pages = {072003},
          doi = {10.1103/PhysRevD.74.072003},
archivePrefix = {arXiv},
       eprint = {hep-ex/0606032},
 primaryClass = {hep-ex},
       adsurl = {https://ui.adsabs.harvard.edu/abs/2006PhRvD..74g2003A}
}

@article{michael2006observation,
  author = {{Michael}, D.~G. and {Adamson}, P. and {Alexopoulos}, T. and {Allison}, W.~W.~M. and {Alner}, G.~J. and {Anderson}, K. and {Andreopoulos}, C. and {Andrews}, M. and {Andrews}, R. and {Arms}, K.~E. and {Armstrong}, R. and {Arroyo}, C. and {Auty}, D.~J. and {Avvakumov}, S. and {Ayres}, D.~S. and {Baller}, B. and {Barish}, B. and {Barker}, M.~A. and {Barnes}, P.~D., Jr. and {Barr}, G. and {Barrett}, W.~L. and {Beall}, E. and {Becker}, B.~R. and {Belias}, A. and {Bergfeld}, T. and {Bernstein}, R.~H. and {Bhattacharya}, D. and {Bishai}, M. and {Blake}, A. and {Bocean}, V. and {Bock}, B. and {Bock}, G.~J. and {Boehm}, J. and {Boehnlein}, D.~J. and {Bogert}, D. and {Border}, P.~M. and {Bower}, C. and {Boyd}, S. and {Buckley-Geer}, E. and {Bungau}, C. and {Byon-Wagner}, A. and {Cabrera}, A. and {Chapman}, J.~D. and {Chase}, T.~R. and {Cherdack}, D. and {Chernichenko}, S.~K. and {Childress}, S. and {Choudhary}, B.~C. and {Cobb}, J.~H. and {Cossairt}, J.~D. and {Courant}, H. and {Crane}, D.~A. and {Culling}, A.~J. and {Dawson}, J.~W. and {de Jong}, J.~K. and {Demuth}, D.~M. and {de Santo}, A. and {Dierckxsens}, M. and {Diwan}, M.~V. and {Dorman}, M. and {Drake}, G. and {Drakoulakos}, D. and {Ducar}, R. and {Durkin}, T. and {Erwin}, A.~R. and {Escobar}, C.~O. and {Evans}, J.~J. and {Fackler}, O.~D. and {Falk Harris}, E. and {Feldman}, G.~J. and {Felt}, N. and {Fields}, T.~H. and {Ford}, R. and {Frohne}, M.~V. and {Gallagher}, H.~R. and {Gebhard}, M. and {Giurgiu}, G.~A. and {Godley}, A. and {Gogos}, J. and {Goodman}, M.~C. and {Gornushkin}, Yu. and {Gouffon}, P. and {Gran}, R. and {Grashorn}, E. and {Grossman}, N. and {Grudzinski}, J.~J. and {Grzelak}, K. and {Guarino}, V. and {Habig}, A. and {Halsall}, R. and {Hanson}, J. and {Harris}, D. and {Harris}, P.~G. and {Hartnell}, J. and {Hartouni}, E.~P. and {Hatcher}, R. and {Heller}, K. and {Hill}, N. and {Ho}, Y. and {Holin}, A. and {Howcroft}, C. and {Hylen}, J. and {Ignatenko}, M. and {Indurthy}, D. and {Irwin}, G.~M. and {Ishitsuka}, M. and {Jaffe}, D.~E. and {James}, C. and {Jenner}, L. and {Jensen}, D. and {Joffe-Minor}, T. and {Kafka}, T. and {Kang}, H.~J. and {Kasahara}, S.~M.~S. and {Kilmer}, J. and {Kim}, H. and {Kim}, M.~S. and {Koizumi}, G. and {Kopp}, S. and {Kordosky}, M. and {Koskinen}, D.~J. and {Kostin}, M. and {Kotelnikov}, S.~K. and {Krakauer}, D.~A. and {Kreymer}, A. and {Kumaratunga}, S. and {Ladran}, A.~S. and {Lang}, K. and {Laughton}, C. and {Lebedev}, A. and {Lee}, R. and {Lee}, W.~Y. and {Libkind}, M.~A. and {Ling}, J. and {Liu}, J. and {Litchfield}, P.~J. and {Litchfield}, R.~P. and {Longley}, N.~P. and {Lucas}, P. and {Luebke}, W. and {Madani}, S. and {Maher}, E. and {Makeev}, V. and {Mann}, W.~A. and {Marchionni}, A. and {Marino}, A.~D. and {Marshak}, M.~L. and {Marshall}, J.~S. and {Mayer}, N. and {McDonald}, J. and {McGowan}, A.~M. and {Meier}, J.~R. and {Merzon}, G.~I. and {Messier}, M.~D. and {Milburn}, R.~H. and {Miller}, J.~L. and {Miller}, W.~H. and {Mishra}, S.~R. and {Mislivec}, A. and {Miyagawa}, P.~S. and {Moore}, C.~D. and {Morf{\'\i}n}, J. and {Morse}, R. and {Mualem}, L. and {Mufson}, S. and {Murgia}, S. and {Murtagh}, M.~J. and {Musser}, J. and {Naples}, D. and {Nelson}, C. and {Nelson}, J.~K. and {Newman}, H.~B. and {Nezrick}, F. and {Nichol}, R.~J. and {Nicholls}, T.~C. and {Ochoa-Ricoux}, J.~P. and {Oliver}, J. and {Oliver}, W.~P. and {Onuchin}, V.~A. and {Osiecki}, T. and {Ospanov}, R. and {Paley}, J. and {Paolone}, V. and {Para}, A. and {Patzak}, T. and {Pavlovi{\'c}}, {\v{Z}}. and {Pearce}, G.~F. and {Pearson}, N. and {Peck}, C.~W. and {Perry}, C. and {Peterson}, E.~A. and {Petyt}, D.~A. and {Ping}, H. and {Piteira}, R. and {Pittam}, R. and {Pla-Dalmau}, A. and {Plunkett}, R.~K. and {Price}, L.~E. and {Proga}, M. and {Pushka}, D.~R. and {Rahman}, D. and {Rameika}, R.~A. and {Raufer}, T.~M. and {Read}, A.~L. and {Rebel}, B. and {Reichenbacher}, J. and {Reyna}, D.~E. and {Rosenfeld}, C. and {Rubin}, H.~A. and {Ruddick}, K. and {Ryabov}, V.~A. and {Saakyan}, R. and {Sanchez}, M.~C. and {Saoulidou}, N. and {Schneps}, J. and {Schoessow}, P.~V. and {Schreiner}, P. and {Schwienhorst}, R. and {Semenov}, V.~K. and {Seun}, S. -M. and {Shanahan}, P. and {Shield}, P.~D. and {Smart}, W. and {Smirnitsky}, V. and {Smith}, C. and {Smith}, P.~N. and {Sousa}, A. and {Speakman}, B. and {Stamoulis}, P. and {Stefanik}, A. and {Sullivan}, P. and {Swan}, J.~M. and {Symes}, P.~A. and {Tagg}, N. and {Talaga}, R.~L. and {Terekhov}, A. and {Tetteh-Lartey}, E. and {Thomas}, J. and {Thompson}, J. and {Thomson}, M.~A. and {Thron}, J.~L. and {Tinti}, G. and {Trendler}, R. and {Trevor}, J. and {Trostin}, I. and {Tsarev}, V.~A. and {Tzanakos}, G. and {Urheim}, J. and {Vahle}, P. and {Vakili}, M. and {Vaziri}, K. and {Velissaris}, C. and {Verebryusov}, V. and {Viren}, B. and {Wai}, L. and {Ward}, C.~P. and {Ward}, D.~R. and {Watabe}, M. and {Weber}, A. and {Webb}, R.~C. and {Wehmann}, A. and {West}, N. and {White}, C. and {White}, R.~F. and {Wojcicki}, S.~G. and {Wright}, D.~M. and {Wu}, Q.~K. and {Yan}, W.~G. and {Yang}, T. and {Yumiceva}, F.~X. and {Yun}, J.~C. and {Zheng}, H. and {Zois}, M. and {Zwaska}, R.},
        title = "{Observation of Muon Neutrino Disappearance with the MINOS Detectors in the NuMI Neutrino Beam}",
      journal = {Physical Review Letters},
         year = 2006,
        month = nov,
       volume = {97},
       number = {19},
          eid = {191801},
        pages = {191801},
          doi = {10.1103/PhysRevLett.97.191801},
archivePrefix = {arXiv},
       eprint = {hep-ex/0607088},
 primaryClass = {hep-ex},
       adsurl = {https://ui.adsabs.harvard.edu/abs/2006PhRvL..97s1801M}
}

@article{sh2009opera,
author = {{Acquafredda}, R. and {Adam}, T. and {Agafonova}, N. and {Alvarez Sanchez}, P. and {Ambrosio}, M. and {Anokhina}, A. and {Aoki}, S. and {Ariga}, A. and {Ariga}, T. and {Arrabito}, L. and {Aufranc}, C. and {Autiero}, D. and {Badertscher}, A. and {Bagulya}, A. and {Baussan}, E. and {Bergnoli}, A. and {Bersani Greggio}, F. and {Bertolin}, A. and {Besnier}, M. and {Biar{\'e}}, D. and {Bick}, D. and {Blin}, S. and {Borer}, K. and {Boucrot}, J. and {Boutigny}, D. and {Boyarkin}, V. and {Bozza}, C. and {Brugi{\`e}re}, T. and {Brugnera}, R. and {Brunetti}, G. and {Buontempo}, S. and {Campagne}, J.~E. and {Carlus}, B. and {Carrara}, E. and {Cazes}, A. and {Chaussard}, L. and {Chernyavsky}, M. and {Chiarella}, V. and {Chon-Sen}, N. and {Chukanov}, A. and {Ciesielski}, R. and {Consiglio}, L. and {Cozzi}, M. and {D'Amato}, G. and {Dal Corso}, F. and {D'Ambrosio}, N. and {Damet}, J. and {de La Taille}, C. and {DeLellis}, G. and {D{\'e}clais}, Y. and {Descombes}, T. and {DeSerio}, M. and {Di Capua}, F. and {Di Ferdinando}, D. and {Di Giovanni}, A. and {Di Marco}, N. and {Di Troia}, C. and {Dick}, N. and {Dmitrievski}, S. and {Dominjon}, A. and {Dracos}, M. and {Duchesneau}, D. and {Dulach}, B. and {Dusini}, S. and {Ebert}, J. and {Efthymiopoulos}, I. and {Egorov}, O. and {Elsener}, K. and {Enikeev}, R. and {Ereditato}, A. and {Esposito}, L.~S. and {Fanin}, C. and {Favier}, J. and {Felici}, G. and {Ferber}, T. and {Fini}, R. and {Fournier}, L. and {Franceschi}, A. and {Frekers}, D. and {Fukuda}, T. and {Fukushima}, C. and {Galkin}, V.~I. and {Galkin}, V.~A. and {Gallet}, R. and {Gardien}, S. and {Garfagnini}, A. and {Gaudiot}, G. and {Giacomelli}, G. and {Giorgini}, M. and {Girerd}, C. and {Goellnitz}, C. and {Goeltzenlichter}, T. and {Goldberg}, J. and {Golubkov}, D. and {Gornushkin}, Y. and {Grapton}, J. -N. and {Grella}, G. and {Grianti}, F. and {Gschwendtner}, E. and {Guerin}, C. and {Guler}, M. and {Gustavino}, C. and {Guyonnet}, J. -L. and {Hagner}, C. and {Hamane}, T. and {Hara}, T. and {Hauger}, M. and {Hess}, M. and {Hierholzer}, M. and {Hoshino}, K. and {Ieva}, M. and {Incurvati}, M. and {Jakovcic}, K. and {Janicsko Csathy}, J. and {Janutta}, B. and {Jollet}, C. and {Juget}, F. and {Kazuyama}, M. and {Kim}, S.~H. and {Khovansky}, N. and {Kimura}, M. and {Klicek}, B. and {Knuesel}, J. and {Kodama}, K. and {Kolev}, D. and {Komatsu}, M. and {Kose}, U. and {Krasnoperov}, A. and {Kreslo}, I. and {Krumstein}, Z. and {Kutsenov}, V.~V. and {Kuznetsov}, V.~A. and {Laktineh}, I. and {Lavy}, M. and {Lazzaro}, C. and {Le}, T.~D. and {LeFlour}, T. and {Lenkeit}, J. and {Lewis}, J. and {Lieunard}, S. and {Ljubicic}, A. and {Longhin}, A. and {Lutter}, G. and {Malgin}, A. and {Manai}, K. and {Mandrioli}, G. and {Marotta}, A. and {Marteau}, J. and {Martin-Chassard}, G. and {Matveev}, V. and {Mauri}, N. and {Meddahi}, M. and {Meisel}, F. and {Meregaglia}, A. and {Meschini}, A. and {Messina}, M. and {Migliozzi}, P. and {Monacelli}, P. and {Monteiro}, I. and {Moreau}, F. and {Morishima}, K. and {Moser}, U. and {Muciaccia}, M.~T. and {Mugnier}, P. and {Naganawa}, N. and {Nakamura}, M. and {Nakano}, T. and {Napolitano}, T. and {Nikitina}, V. and {Niwa}, K. and {Nonoyama}, Y. and {Nozdrin}, A. and {Ogawa}, S. and {Olchevski}, A. and {Orlandi}, D. and {Orlova}, G. and {Osedlo}, V. and {Ossetski}, D. and {Paniccia}, M. and {Paoloni}, A. and {Park}, B.~D. and {Park}, I.~G. and {Pastore}, A. and {Patrizii}, L. and {Pellegrino}, L. and {Pennacchio}, E. and {Pessard}, H. and {Pilipenko}, V. and {Pistillo}, C. and {Polukhina}, N. and {Pozzato}, M. and {Pretzl}, K. and {Publichenko}, P. and {Pupilli}, F. and {Raux}, L. and {Repellin}, J.~P. and {Rescigno}, R. and {Rizhikov}, D. and {Roganova}, T. and {Romano}, G. and {Rosa}, G. and {Rostovtseva}, I. and {Rubbia}, A. and {Russo}, A. and {Ryasny}, V. and {Ryazhskaya}, O. and {Sadovski}, A. and {Sanelli}, C. and {Sato}, O. and {Sato}, Y. and {Saveliev}, V. and {Sazhina}, G. and {Schembri}, A. and {Schmidt Parzefall}, W. and {Schroeder}, H. and {Sch{\"u}tz}, H.~U. and {Schuler}, J. and {Scotto Lavina}, L. and {Serrano}, J. and {Shibuya}, H. and {Simone}, S. and {Sioli}, M. and {Sirignano}, C. and {Sirri}, G. and {Song}, J.~S. and {Spinetti}, M. and {Stanco}, L. and {Starkov}, N. and {Stipcevic}, M. and {Strauss}, T. and {Strolin}, P. and {Sugonyaev}, V. and {Takahashi}, S. and {Talochkin}, V. and {Tenti}, M. and {Tereschenko}, V. and {Terranova}, F. and {Tezuka}, I. and {Tioukov}, V. and {Tolun}, P. and {Tsarev}, V. and {Tsenov}, R. and {Tufanli}, S. and {Ugolino}, U. and {Ushida}, N. and {Van Beek}, G. and {Verguilov}, V. and {Viant}, T. and {Vilain}, P. and {Vladimirov}, M. and {Votano}, L. and {Vuilleumier}, J.~L. and {Waelchli}, T. and {Weber}, M. and {Wilquet}, G. and {Wonsak}, B. and {Wurtz}, J. and {Yakushev}, V. and {Yoon}, C.~S. and {Zaitsev}, Y. and {Zghiche}, A. and {Zimmermann}, R.},
        title = "{The OPERA experiment in the CERN to Gran Sasso neutrino beam}",
      journal = {Journal of Instrumentation},
         year = 2009,
        month = apr,
       volume = {04},
       number = {4},
        pages = {04018},
          doi = {10.1088/1748-0221/4/04/P04018},
       adsurl = {https://ui.adsabs.harvard.edu/abs/2009JInst...4.4018A}
}

@article{an2012observation,
  author = {{An}, F.~P. and {Bai}, J.~Z. and {Balantekin}, A.~B. and {Band}, H.~R. and {Beavis}, D. and {Beriguete}, W. and {Bishai}, M. and {Blyth}, S. and {Boddy}, K. and {Brown}, R.~L. and {Cai}, B. and {Cao}, G.~F. and {Cao}, J. and {Carr}, R. and {Chan}, W.~T. and {Chang}, J.~F. and {Chang}, Y. and {Chasman}, C. and {Chen}, H.~S. and {Chen}, H.~Y. and {Chen}, S.~J. and {Chen}, S.~M. and {Chen}, X.~C. and {Chen}, X.~H. and {Chen}, X.~S. and {Chen}, Y. and {Chen}, Y.~X. and {Cherwinka}, J.~J. and {Chu}, M.~C. and {Cummings}, J.~P. and {Deng}, Z.~Y. and {Ding}, Y.~Y. and {Diwan}, M.~V. and {Dong}, L. and {Draeger}, E. and {Du}, X.~F. and {Dwyer}, D.~A. and {Edwards}, W.~R. and {Ely}, S.~R. and {Fang}, S.~D. and {Fu}, J.~Y. and {Fu}, Z.~W. and {Ge}, L.~Q. and {Ghazikhanian}, V. and {Gill}, R.~L. and {Goett}, J. and {Gonchar}, M. and {Gong}, G.~H. and {Gong}, H. and {Gornushkin}, Y.~A. and {Greenler}, L.~S. and {Gu}, W.~Q. and {Guan}, M.~Y. and {Guo}, X.~H. and {Hackenburg}, R.~W. and {Hahn}, R.~L. and {Hans}, S. and {He}, M. and {He}, Q. and {He}, W.~S. and {Heeger}, K.~M. and {Heng}, Y.~K. and {Hinrichs}, P. and {Ho}, T.~H. and {Hor}, Y.~K. and {Hsiung}, Y.~B. and {Hu}, B.~Z. and {Hu}, T. and {Hu}, T. and {Huang}, H.~X. and {Huang}, H.~Z. and {Huang}, P.~W. and {Huang}, X. and {Huang}, X.~T. and {Huber}, P. and {Isvan}, Z. and {Jaffe}, D.~E. and {Jetter}, S. and {Ji}, X.~L. and {Ji}, X.~P. and {Jiang}, H.~J. and {Jiang}, W.~Q. and {Jiao}, J.~B. and {Johnson}, R.~A. and {Kang}, L. and {Kettell}, S.~H. and {Kramer}, M. and {Kwan}, K.~K. and {Kwok}, M.~W. and {Kwok}, T. and {Lai}, C.~Y. and {Lai}, W.~C. and {Lai}, W.~H. and {Lau}, K. and {Lebanowski}, L. and {Lee}, J. and {Lee}, M.~K.~P. and {Leitner}, R. and {Leung}, J.~K.~C. and {Leung}, K.~Y. and {Lewis}, C.~A. and {Li}, B. and {Li}, F. and {Li}, G.~S. and {Li}, J. and {Li}, Q.~J. and {Li}, S.~F. and {Li}, W.~D. and {Li}, X.~B. and {Li}, X.~N. and {Li}, X.~Q. and {Li}, Y. and {Li}, Z.~B. and {Liang}, H. and {Liang}, J. and {Lin}, C.~J. and {Lin}, G.~L. and {Lin}, S.~K. and {Lin}, S.~X. and {Lin}, Y.~C. and {Ling}, J.~J. and {Link}, J.~M. and {Littenberg}, L. and {Littlejohn}, B.~R. and {Liu}, B.~J. and {Liu}, C. and {Liu}, D.~W. and {Liu}, H. and {Liu}, J.~C. and {Liu}, J.~L. and {Liu}, S. and {Liu}, X. and {Liu}, Y.~B. and {Lu}, C. and {Lu}, H.~Q. and {Luk}, A. and {Luk}, K.~B. and {Luo}, T. and {Luo}, X.~L. and {Ma}, L.~H. and {Ma}, Q.~M. and {Ma}, X.~B. and {Ma}, X.~Y. and {Ma}, Y.~Q. and {Mayes}, B. and {McDonald}, K.~T. and {McFarlane}, M.~C. and {McKeown}, R.~D. and {Meng}, Y. and {Mohapatra}, D. and {Morgan}, J.~E. and {Nakajima}, Y. and {Napolitano}, J. and {Naumov}, D. and {Nemchenok}, I. and {Newsom}, C. and {Ngai}, H.~Y. and {Ngai}, W.~K. and {Nie}, Y.~B. and {Ning}, Z. and {Ochoa-Ricoux}, J.~P. and {Oh}, D. and {Olshevski}, A. and {Pagac}, A. and {Patton}, S. and {Pearson}, C. and {Pec}, V. and {Peng}, J.~C. and {Piilonen}, L.~E. and {Pinsky}, L. and {Pun}, C.~S.~J. and {Qi}, F.~Z. and {Qi}, M. and {Qian}, X. and {Raper}, N. and {Rosero}, R. and {Roskovec}, B. and {Ruan}, X.~C. and {Seilhan}, B. and {Shao}, B.~B. and {Shih}, K. and {Steiner}, H. and {Stoler}, P. and {Sun}, G.~X. and {Sun}, J.~L. and {Tam}, Y.~H. and {Tanaka}, H.~K. and {Tang}, X. and {Themann}, H. and {Torun}, Y. and {Trentalange}, S. and {Tsai}, O. and {Tsang}, K.~V. and {Tsang}, R.~H.~M. and {Tull}, C. and {Viren}, B. and {Virostek}, S. and {Vorobel}, V. and {Wang}, C.~H. and {Wang}, L.~S. and {Wang}, L.~Y. and {Wang}, L.~Z. and {Wang}, M. and {Wang}, N.~Y. and {Wang}, R.~G. and {Wang}, T. and {Wang}, W. and {Wang}, X. and {Wang}, X. and {Wang}, Y.~F. and {Wang}, Z. and {Wang}, Z. and {Wang}, Z.~M. and {Webber}, D.~M. and {Wei}, Y.~D. and {Wen}, L.~J. and {Wenman}, D.~L. and {Whisnant}, K. and {White}, C.~G. and {Whitehead}, L. and {Whitten}, C.~A., Jr. and {Wilhelmi}, J. and {Wise}, T. and {Wong}, H.~C. and {Wong}, H.~L.~H. and {Wong}, J. and {Worcester}, E.~T. and {Wu}, F.~F. and {Wu}, Q. and {Xia}, D.~M. and {Xiang}, S.~T. and {Xiao}, Q. and {Xing}, Z.~Z. and {Xu}, G. and {Xu}, J. and {Xu}, J. and {Xu}, J.~L. and {Xu}, W. and {Xu}, Y. and {Xue}, T. and {Yang}, C.~G. and {Yang}, L. and {Ye}, M. and {Yeh}, M. and {Yeh}, Y.~S. and {Yip}, K. and {Young}, B.~L. and {Yu}, Z.~Y. and {Zhan}, L. and {Zhang}, C. and {Zhang}, F.~H. and {Zhang}, J.~W. and {Zhang}, Q.~M. and {Zhang}, K. and {Zhang}, Q.~X. and {Zhang}, S.~H. and {Zhang}, Y.~C. and {Zhang}, Y.~H. and {Zhang}, Y.~X. and {Zhang}, Z.~J. and {Zhang}, Z.~P. and {Zhang}, Z.~Y. and {Zhao}, J. and {Zhao}, Q.~W. and {Zhao}, Y.~B. and {Zheng}, L. and {Zhong}, W.~L. and {Zhou}, L. and {Zhou}, Z.~Y. and {Zhuang}, H.~L. and {Zou}, J.~H.},
        title = "{Observation of Electron-Antineutrino Disappearance at Daya Bay}",
      journal = {Physical Review Letters},
         year = 2012,
        month = apr,
       volume = {108},
       number = {17},
          eid = {171803},
        pages = {171803},
          doi = {10.1103/PhysRevLett.108.171803},
archivePrefix = {arXiv},
       eprint = {1203.1669},
 primaryClass = {hep-ex},
       adsurl = {https://ui.adsabs.harvard.edu/abs/2012PhRvL.108q1803A}
}

@article{an2015new,
  author = {{An}, F.~P. and {Balantekin}, A.~B. and {Band}, H.~R. and {Bishai}, M. and {Blyth}, S. and {Butorov}, I. and {Cao}, G.~F. and {Cao}, J. and {Cen}, W.~R. and {Chan}, Y.~L. and {Chang}, J.~F. and {Chang}, L.~C. and {Chang}, Y. and {Chen}, H.~S. and {Chen}, Q.~Y. and {Chen}, S.~M. and {Chen}, Y.~X. and {Chen}, Y. and {Cheng}, J.~H. and {Cheng}, J. and {Cheng}, Y.~P. and {Cherwinka}, J.~J. and {Chu}, M.~C. and {Cummings}, J.~P. and {de Arcos}, J. and {Deng}, Z.~Y. and {Ding}, X.~F. and {Ding}, Y.~Y. and {Diwan}, M.~V. and {Draeger}, E. and {Dwyer}, D.~A. and {Edwards}, W.~R. and {Ely}, S.~R. and {Gill}, R. and {Gonchar}, M. and {Gong}, G.~H. and {Gong}, H. and {Grassi}, M. and {Gu}, W.~Q. and {Guan}, M.~Y. and {Guo}, L. and {Guo}, X.~H. and {Hackenburg}, R.~W. and {Han}, R. and {Hans}, S. and {He}, M. and {Heeger}, K.~M. and {Heng}, Y.~K. and {Higuera}, A. and {Hor}, Y.~K. and {Hsiung}, Y.~B. and {Hu}, B.~Z. and {Hu}, L.~M. and {Hu}, L.~J. and {Hu}, T. and {Hu}, W. and {Huang}, E.~C. and {Huang}, H.~X. and {Huang}, X.~T. and {Huber}, P. and {Hussain}, G. and {Jaffe}, D.~E. and {Jaffke}, P. and {Jen}, K.~L. and {Jetter}, S. and {Ji}, X.~P. and {Ji}, X.~L. and {Jiao}, J.~B. and {Johnson}, R.~A. and {Kang}, L. and {Kettell}, S.~H. and {Kramer}, M. and {Kwan}, K.~K. and {Kwok}, M.~W. and {Kwok}, T. and {Langford}, T.~J. and {Lau}, K. and {Lebanowski}, L. and {Lee}, J. and {Lei}, R.~T. and {Leitner}, R. and {Leung}, K.~Y. and {Leung}, J.~K.~C. and {Lewis}, C.~A. and {Li}, D.~J. and {Li}, F. and {Li}, G.~S. and {Li}, Q.~J. and {Li}, S.~C. and {Li}, W.~D. and {Li}, X.~N. and {Li}, X.~Q. and {Li}, Y.~F. and {Li}, Z.~B. and {Liang}, H. and {Lin}, C.~J. and {Lin}, G.~L. and {Lin}, P.~Y. and {Lin}, S.~K. and {Ling}, J.~J. and {Link}, J.~M. and {Littenberg}, L. and {Littlejohn}, B.~R. and {Liu}, D.~W. and {Liu}, H. and {Liu}, J.~L. and {Liu}, J.~C. and {Liu}, S.~S. and {Lu}, C. and {Lu}, H.~Q. and {Lu}, J.~S. and {Luk}, K.~B. and {Ma}, Q.~M. and {Ma}, X.~Y. and {Ma}, X.~B. and {Ma}, Y.~Q. and {Martinez Caicedo}, D.~A. and {McDonald}, K.~T. and {McKeown}, R.~D. and {Meng}, Y. and {Mitchell}, I. and {Monari Kebwaro}, J. and {Nakajima}, Y. and {Napolitano}, J. and {Naumov}, D. and {Naumova}, E. and {Ngai}, H.~Y. and {Ning}, Z. and {Ochoa-Ricoux}, J.~P. and {Olshevski}, A. and {Park}, J. and {Patton}, S. and {Pec}, V. and {Peng}, J.~C. and {Piilonen}, L.~E. and {Pinsky}, L. and {Pun}, C.~S.~J. and {Qi}, F.~Z. and {Qi}, M. and {Qian}, X. and {Raper}, N. and {Ren}, B. and {Ren}, J. and {Rosero}, R. and {Roskovec}, B. and {Ruan}, X.~C. and {Shao}, B.~B. and {Steiner}, H. and {Sun}, G.~X. and {Sun}, J.~L. and {Tang}, W. and {Taychenachev}, D. and {Themann}, H. and {Tsang}, K.~V. and {Tull}, C.~E. and {Tung}, Y.~C. and {Viaux}, N. and {Viren}, B. and {Vorobel}, V. and {Wang}, C.~H. and {Wang}, M. and {Wang}, N.~Y. and {Wang}, R.~G. and {Wang}, W. and {Wang}, W.~W. and {Wang}, X. and {Wang}, Y.~F. and {Wang}, Z. and {Wang}, Z. and {Wang}, Z.~M. and {Wei}, H.~Y. and {Wen}, L.~J. and {Whisnant}, K. and {White}, C.~G. and {Whitehead}, L. and {Wise}, T. and {Wong}, H.~L.~H. and {Wong}, S.~C.~F. and {Worcester}, E. and {Wu}, Q. and {Xia}, D.~M. and {Xia}, J.~K. and {Xia}, X. and {Xing}, Z.~Z. and {Xu}, J.~Y. and {Xu}, J.~L. and {Xu}, J. and {Xu}, Y. and {Xue}, T. and {Yan}, J. and {Yang}, C.~G. and {Yang}, L. and {Yang}, M.~S. and {Yang}, M.~T. and {Ye}, M. and {Yeh}, M. and {Yeh}, Y.~S. and {Young}, B.~L. and {Yu}, G.~Y. and {Yu}, Z.~Y. and {Zang}, S.~L. and {Zhan}, L. and {Zhang}, C. and {Zhang}, H.~H. and {Zhang}, J.~W. and {Zhang}, Q.~M. and {Zhang}, Y.~M. and {Zhang}, Y.~X. and {Zhang}, Y.~M. and {Zhang}, Z.~J. and {Zhang}, Z.~Y. and {Zhang}, Z.~P. and {Zhao}, J. and {Zhao}, Q.~W. and {Zhao}, Y.~F. and {Zhao}, Y.~B. and {Zheng}, L. and {Zhong}, W.~L. and {Zhou}, L. and {Zhou}, N. and {Zhuang}, H.~L. and {Zou}, J.~H. and {Daya Bay Collaboration}},
        title = "{New Measurement of Antineutrino Oscillation with the Full Detector Configuration at Daya Bay}",
      journal = {Physical Review Letters},
         year = 2015,
        month = sep,
       volume = {115},
       number = {11},
          eid = {111802},
        pages = {111802},
          doi = {10.1103/PhysRevLett.115.111802},
       adsurl = {https://ui.adsabs.harvard.edu/abs/2015PhRvL.115k1802A}
}

@article{vien2016delta,
  author = {{Vien}, V.~V. and {C{\'a}rcamo Hern{\'a}ndez}, A.~E. and {Long}, H.~N.},
        title = "{The {\ensuremath{\Delta}}(27) flavor 3-3-1 model with neutral leptons}",
      journal = {Nuclear Physics B},
         year = 2016,
        month = dec,
       volume = {913},
        pages = {792-814},
          doi = {10.1016/j.nuclphysb.2016.10.010},
archivePrefix = {arXiv},
       eprint = {1601.03300},
 primaryClass = {hep-ph},
       adsurl = {https://ui.adsabs.harvard.edu/abs/2016NuPhB.913..792V}
}

@article{hirsch2009a4,
  author = {{Hirsch}, M. and {Morisi}, S. and {Valle}, J.~W.~F.},
        title = "{A$_4$-based tri-bimaximal mixing within inverse and linear seesaw schemes}",
      journal = {Physics Letters B},
         year = 2009,
        month = sep,
       volume = {679},
       number = {5},
        pages = {454-459},
          doi = {10.1016/j.physletb.2009.08.003},
archivePrefix = {arXiv},
       eprint = {0905.3056},
 primaryClass = {hep-ph},
       adsurl = {https://ui.adsabs.harvard.edu/abs/2009PhLB..679..454H}
}

@article{carcamo2021controlled,
  author = {{C{\'a}rcamo Hern{\'a}ndez}, A.~E. and {de Medeiros Varzielas}, Ivo and {L{\'o}pez-Ib{\'a}{\~n}ez}, M.~L. and {Melis}, Aurora},
        title = "{Controlled fermion mixing and FCNCs in a $\Delta$(27) 3+1 Higgs Doublet Model}",
      journal = {Journal of High Energy Physics},
         year = 2021,
        month = may,
       volume = {2021},
       number = {5},
          eid = {215},
        pages = {215},
          doi = {10.1007/JHEP05(2021)215},
archivePrefix = {arXiv},
       eprint = {2102.05658},
 primaryClass = {hep-ph},
       adsurl = {https://ui.adsabs.harvard.edu/abs/2021JHEP...05..215C}
}

@article{sruthilaya2018a_4,
  author = {{Sruthilaya}, M. and {Mohanta}, Rukmani and {Patra}, Sudhanwa},
        title = "{A$_4$ realization of linear seesaw and neutrino phenomenology}",
      journal = {European Physical Journal C},
         year = 2018,
        month = sep,
       volume = {78},
       number = {9},
          eid = {719},
        pages = {719},
          doi = {10.1140/epjc/s10052-018-6181-6},
archivePrefix = {arXiv},
       eprint = {1709.01737},
 primaryClass = {hep-ph},
       adsurl = {https://ui.adsabs.harvard.edu/abs/2018EPJC...78..719S}
}

@article{borah2019linear,
  author = {{Borah}, Debasish and {Karmakar}, Biswajit},
        title = "{Linear seesaw for Dirac neutrinos with A$_{4}$ flavour symmetry}",
      journal = {Physics Letters B},
         year = 2019,
        month = feb,
       volume = {789},
        pages = {59-70},
          doi = {10.1016/j.physletb.2018.12.006},
archivePrefix = {arXiv},
       eprint = {1806.10685},
 primaryClass = {hep-ph},
       adsurl = {https://ui.adsabs.harvard.edu/abs/2019PhLB..789...59B}
}

@article{T71,
  author = {{Luhn}, Christoph and {Nasri}, Salah and {Ramond}, Pierre},
        title = "{Tri-bimaximal neutrino mixing and the family symmetry Z$_{7}${\ensuremath{\rtimes}}Z$_{3}$}",
      journal = {Physics Letters B},
         year = 2007,
        month = aug,
       volume = {652},
       number = {1},
        pages = {27-33},
          doi = {10.1016/j.physletb.2007.06.059},
archivePrefix = {arXiv},
       eprint = {0706.2341},
 primaryClass = {hep-ph},
       adsurl = {https://ui.adsabs.harvard.edu/abs/2007PhLB..652...27L}
}

@article{T72,
  author = {{Hagedorn}, Claudia and {Schmidt}, Michael A. and {Smirnov}, Alexei Yu.},
        title = "{Lepton mixing and cancellation of the Dirac mass hierarchy in SO(10) GUTs with flavor symmetries T$_{7}$ and {\ensuremath{\Sigma}}(81)}",
      journal = {Physical Review D},
         year = 2009,
        month = feb,
       volume = {79},
       number = {3},
          eid = {036002},
        pages = {036002},
          doi = {10.1103/PhysRevD.79.036002},
archivePrefix = {arXiv},
       eprint = {0811.2955},
 primaryClass = {hep-ph},
       adsurl = {https://ui.adsabs.harvard.edu/abs/2009PhRvD..79c6002H}
}

@article{T73,
  author = {{Cao}, Qing-Hong and {Khalil}, Shaaban and {Ma}, Ernest and {Okada}, Hiroshi},
        title = "{Observable T$_{7}$ Lepton Flavor Symmetry at the Large Hadron Collider}",
      journal = {Physical Review Letters},
         year = 2011,
        month = apr,
       volume = {106},
       number = {13},
          eid = {131801},
        pages = {131801},
          doi = {10.1103/PhysRevLett.106.131801},
archivePrefix = {arXiv},
       eprint = {1009.5415},
 primaryClass = {hep-ph},
       adsurl = {https://ui.adsabs.harvard.edu/abs/2011PhRvL.106m1801C}
}

@article{T74,
  author = {{Luhn}, Christoph and {Parattu}, Krishna Mohan and {Wingerter}, Ak{\i}n},
        title = "{A minimal model of neutrino flavor}",
      journal = {Journal of High Energy Physics},
         year = 2012,
        month = dec,
       volume = {2012},
          eid = {96},
        pages = {96},
          doi = {10.1007/JHEP12(2012)096},
archivePrefix = {arXiv},
       eprint = {1210.1197},
 primaryClass = {hep-ph},
       adsurl = {https://ui.adsabs.harvard.edu/abs/2012JHEP...12..096L}
}

@article{T75,
  author = {{Kajiyama}, Yuji and {Okada}, Hiroshi and {Yagyu}, Kei},
        title = "{T$_{7}$ flavor model in three loop seesaw and Higgs phenomenology}",
      journal = {Journal of High Energy Physics},
         year = 2013,
        month = oct,
       volume = {2013},
          eid = {196},
        pages = {196},
          doi = {10.1007/JHEP10(2013)196},
archivePrefix = {arXiv},
       eprint = {1307.0480},
 primaryClass = {hep-ph},
       adsurl = {https://ui.adsabs.harvard.edu/abs/2013JHEP...10..196K}
}

@article{T76,
  author = {{Bonilla}, Cesar and {Morisi}, Stefano and {Peinado}, Eduardo and {Valle}, J.~W.~F.},
        title = "{Relating quarks and leptons with the T$_{7}$ flavour group}",
      journal = {Physics Letters B},
         year = 2015,
        month = mar,
       volume = {742},
        pages = {99-106},
          doi = {10.1016/j.physletb.2015.01.017},
archivePrefix = {arXiv},
       eprint = {1411.4883},
 primaryClass = {hep-ph},
       adsurl = {https://ui.adsabs.harvard.edu/abs/2015PhLB..742...99B}
}

@article{T77,
  author = {{Vien}, V.~V. and {Long}, H.~N.},
        title = "{The T$_{7}$ flavor symmetry in 3-3-1 model with neutral leptons}",
      journal = {Journal of High Energy Physics},
         year = 2014,
        month = apr,
       volume = {2014},
          eid = {133},
        pages = {133},
          doi = {10.1007/JHEP04(2014)133},
archivePrefix = {arXiv},
       eprint = {1402.1256},
 primaryClass = {hep-ph},
       adsurl = {https://ui.adsabs.harvard.edu/abs/2014JHEP...04..133V}
}

@article{T78,
  author = {{Vien}, V.~V.},
        title = "{T$_{7}$ flavor symmetry scheme for understanding neutrino mass and mixing in 3-3-1 model with neutral leptons}",
      journal = {Modern Physics Letters A},
         year = 2014,
        month = sep,
       volume = {29},
       number = {28},
          eid = {1450139},
        pages = {1450139},
          doi = {10.1142/S0217732314501399},
archivePrefix = {arXiv},
       eprint = {1508.02585},
 primaryClass = {hep-ph},
       adsurl = {https://ui.adsabs.harvard.edu/abs/2014MPLA...2950139V}
}

@article{T79,
  author = {{C{\'a}rcamo Hern{\'a}ndez}, A.~E. and {Martinez}, R.},
        title = "{Fermion mass and mixing pattern in a minimal T$_7$ flavor 331 model}",
      journal = {Journal of Physics G: Nuclear and Particle Physics},
         year = 2016,
        month = feb,
     volume = {43},
       number = {4},
        pages = {45003},
          doi = {10.1088/0954-3899/43/4/045003},
archivePrefix = {arXiv},
       eprint = {1511.07997},
 primaryClass = {hep-ph},
       adsurl = {https://ui.adsabs.harvard.edu/abs/2015arXiv151107997C}
}

@article{T710,
  author = {{Arbel{\'a}ez}, Carolina and {C{\'a}rcamo Hern{\'a}ndez}, A.~E. and {Kovalenko}, Sergey and {Schmidt}, Iv{\'a}n},
        title = "{Adjoint SU(5) GUT model with T$_{7}$ flavor symmetry}",
      journal = {Physical Review D},
         year = 2015,
        month = dec,
       volume = {92},
       number = {11},
          eid = {115015},
        pages = {115015},
          doi = {10.1103/PhysRevD.92.115015},
archivePrefix = {arXiv},
       eprint = {1507.03852},
 primaryClass = {hep-ph},
       adsurl = {https://ui.adsabs.harvard.edu/abs/2015PhRvD..92k5015A}
}

@article{T713,
  author = {{Vien}, V.~V. and {Long}, H.~N.},
        title = "{Fermion Mass and Mixing in a Simple Extension of the Standard Model Based on T$_{7}$ Flavor Symmetry}",
      journal = {Physics of Atomic Nuclei},
         year = 2019,
        month = mar,
       volume = {82},
       number = {2},
        pages = {168-182},
          doi = {10.1134/S1063778819020133},
       adsurl = {https://ui.adsabs.harvard.edu/abs/2019PAN....82..168V}
}

@article{stacey2003particle,
  title="{Particle swarm optimization with mutation}",
  author={Stacey, Andrew and Jancic, Mirjana and Grundy, Ian},
  booktitle={The 2003 Congress on Evolutionary Computation, 2003. CEC'03.},
  volume={2},
  pages={1425--1430},
  year={2003},
  organization={IEEE}
}

@article{eberhart1995new,
  title="{A new optimizer using particle swarm theory}",
  author={Eberhart, Russell and Kennedy, James},
  booktitle={MHS'95. Proceedings of the sixth international symposium on micro machine and human science},
  pages={39--43},
  year={1995},
  organization={Ieee}
}

@article{esmin2015review,
  title="{A review on particle swarm optimization algorithm and its variants to clustering high-dimensional data}",
  author={Esmin, Ahmed AA and Coelho, Rodrigo A and Matwin, Stan},
  journal={Artificial Intelligence Review},
  volume={44},
  pages={23--45},
  year={2015},
  publisher={Springer}
}

@article{rana2011review,
  title="{A review on particle swarm optimization algorithms and their applications to data clustering}",
  author={Rana, Sandeep and Jasola, Sanjay and Kumar, Rajesh},
  journal={Artificial Intelligence Review},
  volume={35},
  pages={211--222},
  year={2011},
  publisher={Springer}
}

@article{he2007parameter,
  title="{Parameter estimation for chaotic systems by particle swarm optimization}",
  author={He, Qie and Wang, Ling and Liu, Bo},
  journal={Chaos, Solitons \& Fractals},
  volume={34},
  number={2},
  pages={654--661},
  year={2007},
  publisher={Elsevier}
}

@article{alatas2009chaos,
  title="{Chaos embedded particle swarm optimization algorithms}",
  author={Alatas, Bilal and Akin, Erhan and Ozer, A Bedri},
  journal={Chaos, Solitons \& Fractals},
  volume={40},
  number={4},
  pages={1715--1734},
  year={2009},
  publisher={Elsevier}
}

@article{babazadeh2009optimization,
  title="{Optimization of fuel core loading pattern design in a VVER nuclear power reactors using Particle Swarm Optimization (PSO)}",
  author={Babazadeh, Davood and Boroushaki, Mehrdad and Lucas, Caro},
  journal={Annals of nuclear energy},
  volume={36},
  number={7},
  pages={923--930},
  year={2009},
  publisher={Elsevier}
}

@article{ibrahim2019hybridization,
  title="{A hybridization of cuckoo search and particle swarm optimization for solving nonlinear systems}",
  author={Ibrahim, Abdelmonem M and Tawhid, Mohamed A},
  journal={Evolutionary Intelligence},
  volume={12},
  pages={541--561},
  year={2019},
  publisher={Springer}
}

@article{subbaraj2010hybrid,
  title="{Hybrid particle swarm optimization based optimal reactive power dispatch}",
  author={Subbaraj, P and Rajnarayanan, PN},
  journal={International Journal of Computer Applications},
  volume={1},
  number={5},
  pages={65--70},
  year={2010},
  publisher={Citeseer}
}

@article{jiang2020multilayer,
  title="{Multilayer optical thin film design with deep Q learning}",
  author={Jiang, Anqing and Osamu, Yoshie and Chen, Liangyao},
  journal={Scientific reports},
  volume={10},
  number={1},
  pages={12780},
  year={2020},
  publisher={Nature Publishing Group UK London}
}

@article{yue2019determination,
  title="{Determination of thin metal film’s thickness and optical constants based on SPR phase detection by simulated annealing particle swarm optimization}",
  author={Yue, Chong and Qin, Zirui and Lang, Yaopu and Liu, Qinggang},
  journal={Optics Communications},
  volume={430},
  pages={238--245},
  year={2019},
  publisher={Elsevier}
}

@article{rabady2014global,
  title="{Global optimal design of optical multilayer thin-film filters using particle swarm optimization}",
  author={Rabady, Rabi I and Ababneh, Almahdi},
  journal={Optik},
  volume={125},
  number={1},
  pages={548--553},
  year={2014},
  publisher={Elsevier}
}

@article{ruan2016determination,
  title="{Determination of optical properties and thickness of optical thin film using stochastic particle swarm optimization}",
  author={Ruan, Zhao-Hui and Yuan, Yuan and Zhang, Xiao-Xian and Shuai, Yong and Tan, He-Ping},
  journal={Solar Energy},
  volume={127},
  pages={147--158},
  year={2016},
  publisher={Elsevier}
}

@article{mehmood2019nature,
  title="{Nature-inspired heuristic paradigms for parameter estimation of control autoregressive moving average systems}",
  author={Mehmood, Ammara and Zameer, Aneela and Raja, Muhammad Asif Zahoor and Bibi, Rabia and Chaudhary, Naveed Ishtiaq and Aslam, Muhammad Saeed},
  journal={Neural Computing and Applications},
  volume={31},
  pages={5819--5842},
  year={2019},
  publisher={Springer}
}

@article{yetis2014forecasting,
  title="{Forecasting of Turkey's electricity consumption using Artificial Neural Network}",
  author={Yetis, Yunus and Jamshidi, Mo},
  booktitle={2014 World Automation Congress (WAC)},
  pages={723--728},
  year={2014},
  organization={IEEE}
}

@article{yassin2016binary,
  title="{Binary particle swarm optimization structure selection of nonlinear autoregressive moving average with exogenous inputs (NARMAX) model of a flexible robot arm}",
  author={Yassin, Ihsan Mohd and Zabidi, Azlee and Amin Megat Ali, Megat Syahirul and Md Tahir, Nooritawati and Zainol Abidin, Husna and Rizman, Zairi Ismael},
  journal={International Journal on Advanced Science, Engineering and Information Technology},
  volume={6},
  number={5},
  pages={630--637},
  year={2016}
}

@article{akbar2019novel,
  title="{Novel application of FO-DPSO for 2-D parameter estimation of electromagnetic plane waves}",
  author={Akbar, Sadiq and Zaman, Fawad and Asif, Muhammad and Rehman, Ata Ur and Raja, Muhammad Asif Zahoor},
  journal={Neural Computing and Applications},
  volume={31},
  pages={3681--3690},
  year={2019},
  publisher={Springer}
}

@article{guedria2016improved,
  title="{Improved accelerated PSO algorithm for mechanical engineering optimization problems}",
  author={Guedria, Najeh Ben},
  journal={Applied Soft Computing},
  volume={40},
  pages={455--467},
  year={2016},
  publisher={Elsevier}
}

@article{kaveh2013engineering,
  title={Engineering design optimization using a hybrid PSO and HS algorithm},
  author={Kaveh, A and NASR, ELAHI A},
  year={2013},
  publisher={ASIAN JOURNAL OF CIVIL ENGINEERING (BUILDING AND HOUSING)}
}

@article{kumar2021design,
  title="{Design of an efficient hybridized CS-PSO algorithm and its applications for solving constrained and bound constrained structural engineering design problems}",
  author={Kumar, Nirmal and Mahato, Sanat Kumar and Bhunia, Asoke Kumar},
  journal={Results in Control and Optimization},
  volume={5},
  pages={100064},
  year={2021},
  publisher={Elsevier}
}

@article{djemame2019solving,
  title="{Solving reverse emergence with quantum PSO application to image processing}",
  author={Djemame, Safia and Batouche, Mohamed and Oulhadj, Hamouche and Siarry, Patrick},
  journal={Soft Computing},
  volume={23},
  number={16},
  pages={6921--6935},
  year={2019},
  publisher={Springer}
}

@inproceedings{singh2014image,
  title="{Image contrast enhancement using GA and PSO: a survey}",
  author={Singh, Ravindra Pal and Dixit, Manish and Silakari, Sanjay},
  booktitle={2014 International Conference on Computational Intelligence and Communication Networks},
  pages={186--189},
  year={2014},
  organization={IEEE}
}

@article{pramanik2015image,
  title="{Image registration using PSO and APSO: a comparative analysis}",
  author={Pramanik, Jitendra and Dalai, Sunita and Rana, Debaraj},
  journal={International Journal of Computer Applications},
  volume={116},
  number={21},
  year={2015},
  publisher={Citeseer}
}

@inproceedings{azayite2019financial,
  title="{Financial Early Warning System Model Based on Neural Networks, PSO and SA Algorithms}",
  author={Azayite, Fatima Zahra and Achchab, Said},
  booktitle={EngOpt 2018 Proceedings of the 6th International Conference on Engineering Optimization},
  pages={970--982},
  year={2019},
  organization={Springer}
}

@article{chiam2009memetic,
  title="{A memetic model of evolutionary PSO for computational finance applications}",
  author={Chiam, Swee Chiang and Tan, Kay Chen and Mamun, A Al},
  journal={Expert Systems with Applications},
  volume={36},
  number={2},
  pages={3695--3711},
  year={2009},
  publisher={Elsevier}
}

@article{pan2022design,
  title="{Design of financial management model using the forward neural network based on particle swarm optimization algorithm}",
  author={Pan, Yilin and others},
  journal={Computational Intelligence and Neuroscience},
  volume={2022},
  year={2022},
  publisher={Hindawi}
}

@article{marinakis2009ant,
  title="{Ant colony and particle swarm optimization for financial classification problems}",
  author={Marinakis, Yannis and Marinaki, Magdalene and Doumpos, Michael and Zopounidis, Constantin},
  journal={Expert Systems with Applications},
  volume={36},
  number={7},
  pages={10604--10611},
  year={2009},
  publisher={Elsevier}
}

@article{meissner2006optimized,
  title="{Optimized Particle Swarm Optimization (OPSO) and its application to artificial neural network training}",
  author={Meissner, Michael and Schmuker, Michael and Schneider, Gisbert},
  journal={BMC bioinformatics},
  volume={7},
  number={1},
  pages={1--11},
  year={2006},
  publisher={BioMed Central}
}

@inproceedings{rauf2018training,
  title="{Training of artificial neural network using pso with novel initialization technique}",
  author={Rauf, Hafiz Tayyab and Bangyal, Waqas Haider and Ahmad, Jamil and Bangyal, Saad Abdullah},
  booktitle={2018 international conference on innovation and intelligence for informatics, computing, and technologies (3ICT)},
  pages={1--8},
  year={2018},
  organization={IEEE}
}

@article{rodejohann2011neutrino,
  author = {{Rodejohann}, Werner},
        title = "{Neutrino-Less Double Beta Decay and Particle Physics}",
      journal = {International Journal of Modern Physics E},
         year = 2011,
        month = jan,
       volume = {20},
       number = {9},
        pages = {1833-1930},
          doi = {10.1142/S0218301311020186},
archivePrefix = {arXiv},
       eprint = {1106.1334},
 primaryClass = {hep-ph},
       adsurl = {https://ui.adsabs.harvard.edu/abs/2011IJMPE..20.1833R}
}

@article{mitra2012neutrinoless,
  author = {{Mitra}, Manimala and {Senjanovi{\'c}}, Goran and {Vissani}, Francesco},
        title = "{Neutrinoless double beta decay and heavy sterile neutrinos}",
      journal = {Nuclear Physics B},
         year = 2012,
        month = mar,
       volume = {856},
       number = {1},
        pages = {26-73},
          doi = {10.1016/j.nuclphysb.2011.10.035},
archivePrefix = {arXiv},
       eprint = {1108.0004},
 primaryClass = {hep-ph},
       adsurl = {https://ui.adsabs.harvard.edu/abs/2012NuPhB.856...26M}
}

@article{bilenky2012neutrinoless,
  author = {{Bilenky}, S.~M. and {Giunti}, C.},
        title = "{Neutrinoless Double-Beta Decay:. a Brief Review}",
      journal = {Modern Physics Letters A},
         year = 2012,
        month = jan,
       volume = {27},
       number = {13},
          eid = {1230015-1-1230015-22},
        pages = {1230015-1-1230015-22},
          doi = {10.1142/S0217732312300157},
archivePrefix = {arXiv},
       eprint = {1203.5250},
 primaryClass = {hep-ph},
       adsurl = {https://ui.adsabs.harvard.edu/abs/2012MPLA...2730015B}
}

@article{rodejohann2012neutrinolessssw,
  author = {{Rodejohann}, Werner},
        title = "{Neutrinoless double-beta decay and neutrino physics}",
      journal = {Journal of Physics G Nuclear Physics},
         year = 2012,
        month = dec,
       volume = {39},
       number = {12},
          eid = {124008},
        pages = {124008},
          doi = {10.1088/0954-3899/39/12/124008},
archivePrefix = {arXiv},
       eprint = {1206.2560},
 primaryClass = {hep-ph},
       adsurl = {https://ui.adsabs.harvard.edu/abs/2012JPhG...39l4008R}
}

@article{vergados2012theory,
  author = {{Vergados}, J.~D. and {Ejiri}, H. and {{\v{S}}imkovic}, F.},
        title = "{Theory of neutrinoless double-beta decay}",
      journal = {Reports on Progress in Physics},
         year = 2012,
        month = oct,
       volume = {75},
       number = {10},
          eid = {106301},
        pages = {106301},
          doi = {10.1088/0034-4885/75/10/106301},
archivePrefix = {arXiv},
       eprint = {1205.0649},
 primaryClass = {hep-ph},
       adsurl = {https://ui.adsabs.harvard.edu/abs/2012RPPh...75j6301V}
}

@article{R1,
  author = "He, Xiao-Gang and Keum, Yong-Yeon and Volkas, Raymond R.",
    title = "{A(4) flavor symmetry breaking scheme for understanding quark and neutrino mixing angles}",
    eprint = "hep-ph/0601001",
    archivePrefix = "arXiv",
    doi = "10.1088/1126-6708/2006/04/039",
    journal = "JHEP",
    volume = "04",
    pages = "039",
    year = "2006"
}

@article{R2,
  author = "Ma, Ernest",
    title = "{Dark Scalar Doublets and Neutrino Tribimaximal Mixing from A(4) Symmetry}",
    eprint = "0808.1729",
    archivePrefix = "arXiv",
    primaryClass = "hep-ph",
    reportNumber = "UCRHEP-T453",
    doi = "10.1016/j.physletb.2008.12.038",
    journal = "Phys. Lett. B",
    volume = "671",
    pages = "366--368",
    year = "2009"
}

@article{R3,
  author = "Altarelli, Guido and Meloni, Davide",
    title = "{A Simplest $A_4$ Model for Tri-Bimaximal Neutrino Mixing}",
    eprint = "0905.0620",
    archivePrefix = "arXiv",
    primaryClass = "hep-ph",
    reportNumber = "RM3-TH-09-8, CERN-PH-TH-2009-046",
    doi = "10.1088/0954-3899/36/8/085005",
    journal = "J. Phys. G",
    volume = "36",
    pages = "085005",
    year = "2009"
}

@article{R4,
  author = "Lin, Yin",
    title = "{A Predictive A(4) model, Charged Lepton Hierarchy and Tri-bimaximal Sum Rule}",
    eprint = "0804.2867",
    archivePrefix = "arXiv",
    primaryClass = "hep-ph",
    doi = "10.1016/j.nuclphysb.2008.12.025",
    journal = "Nucl. Phys. B",
    volume = "813",
    pages = "91--105",
    year = "2009"
}

@article{R5,
  author = "Ahn, Y. H. and Chen, Chian-Shu",
    title = "{Non-zero $U_{e3}$ and TeV-Leptogenesis through $A_4$ symmetry breaking}",
    eprint = "1001.2869",
    archivePrefix = "arXiv",
    primaryClass = "hep-ph",
    doi = "10.1103/PhysRevD.81.105013",
    journal = "Phys. Rev. D",
    volume = "81",
    pages = "105013",
    year = "2010"
}

@article{R6,
  author = "Barry, James and Rodejohann, Werner",
    title = "{Deviations from tribimaximal mixing due to the vacuum expectation value misalignment in $A_4$ models}",
    eprint = "1003.2385",
    archivePrefix = "arXiv",
    primaryClass = "hep-ph",
    doi = "10.1103/PhysRevD.81.119901",
    journal = "Phys. Rev. D",
    volume = "81",
    pages = "093002",
    year = "2010",
    note = "[Erratum: Phys.Rev.D 81, 119901 (2010)]"
}

@article{R8,
  author = "Vien, Vo Van and Long, Hoang Ngoc",
    title = "{Neutrino mixing with nonzero $\theta_{13}$ and CP violation in the 3-3-1 model based on $A_4$ flavor symmetry}",
    eprint = "1405.4665",
    archivePrefix = "arXiv",
    primaryClass = "hep-ph",
    doi = "10.1142/S0217751X15501171",
    journal = "Int. J. Mod. Phys. A",
    volume = "30",
    number = "21",
    pages = "1550117",
    year = "2015"
}

@article{R10,
    author = "Vien, V. V.",
    title = "{Cobimaximal neutrino mixing in the $U(1)_{B-L}$ extension with $A_4$ symmetry}",
    doi = "10.1142/S0217732320503113",
    journal = "Mod. Phys. Lett. A",
    volume = "35",
    number = "38",
    pages = "2050311",
    year = "2020"
}

@article{R11,
    author = "Ding, Gui-Jun and Lu, Jun-Nan and Valle, Jos\'e W. F.",
    title = "{Trimaximal neutrino mixing from scotogenic $A_4$ family symmetry}",
    eprint = "2009.04750",
    archivePrefix = "arXiv",
    primaryClass = "hep-ph",
    reportNumber = "USTC-ICTS/PCFT-20-27, IFIC/20-XXX",
    doi = "10.1016/j.physletb.2021.136122",
    journal = "Phys. Lett. B",
    volume = "815",
    pages = "136122",
    year = "2021"
}

@article{R12,
  author = "Dey, Manash and Chakraborty, Pralay and Roy, Subhankar",
    title = "{The \ensuremath{\mu}-\ensuremath{\tau} mixed symmetry and neutrino mass matrix}",
    eprint = "2211.01314",
    archivePrefix = "arXiv",
    primaryClass = "hep-ph",
    doi = "10.1016/j.physletb.2023.137767",
    journal = "Phys. Lett. B",
    volume = "839",
    pages = "137767",
    year = "2023",
    note = "[Erratum: Phys.Lett.B 852, 138602 (2024)]"
}

@article{Fawzi2019EffectiveMB,
  title="{Effective methodology based on neural network optimizer for extracting model parameters of PEM fuel cells}",
  author={Mohamed A. Fawzi and Attia A. El‐Fergany and Hany M. Hasanien},
  journal={International Journal of Energy Research},
  year={2019},
  volume={43},
  pages={8136 - 8147},
  url={https://api.semanticscholar.org/CorpusID:202035751}
}

@article{AbouOmar2022ObserverbasedIT,
  title="{Observer-based interval type-2 fuzzy PID controller for PEMFC air feeding system using novel hybrid neural network algorithm-differential evolution optimizer}",
  author={Mahmoud S. AbouOmar and Yi Xin Su and Huajun Zhang and Binghua Shi and Li-li Wan},
  journal={Alexandria Engineering Journal},
  year={2022},
  url={https://api.semanticscholar.org/CorpusID:245994750}
}

@article{Aslam2024NeurocomputingSF,
  title="{Neuro-computing solution for Lorenz differential equations through artificial neural networks integrated with PSO-NNA hybrid meta-heuristic algorithms: a comparative study}",
  author={Muhammad Naeem Aslam and Muhammad Waheed Aslam and Muhammad Sarmad Arshad and Zeeshan Afzal and Murad Khan Hassani and Ahmed M. Zidan and Ali Akg{\"u}l},
  journal={Scientific Reports},
  year={2024},
  volume={14},
  url={https://api.semanticscholar.org/CorpusID:268751273}
}

@article{JasimShaban2023NNAAA,
  title="{NNA and Activation Equation-Based Prediction of New COVID-19 Infections}",
  author={Faris Ali Jasim Shaban},
  journal={2023 5th International Congress on Human-Computer Interaction, Optimization and Robotic Applications (HORA)},
  year={2023},
  pages={1-6},
  url={https://api.semanticscholar.org/CorpusID:259279562}
}

@article{Zhang2020HybridTO,
  title="{Hybrid teaching-learning-based optimization and neural network algorithm for engineering design optimization problems}",
  author={Yiying Zhang and Zhigang Jin and Ye Chen},
  journal={Knowl. Based Syst.},
  year={2020},
  volume={187},
  url={https://api.semanticscholar.org/CorpusID:199009221}
}

@article{Zhang2021NeuralNA,
  title="{Neural Network Algorithm With Reinforcement Learning for Parameters Extraction of Photovoltaic Models}",
  author={Yiying Zhang},
  journal={IEEE Transactions on Neural Networks and Learning Systems},
  year={2021},
  volume={34},
  pages={2806-2816},
  url={https://api.semanticscholar.org/CorpusID:237516372}
}

@article{Qadeer2022NeuralNP,
  title="{Neural network-inspired performance enhancement of synthetic natural gas liquefaction plant with different minimum approach temperatures}",
  author={Kinza Qadeer and Ashfaq Ahmad and Ahmad Naquash and Muhammad Abdul Qyyum and Khaliq Majeed and Zhongming Zhou and Tianbiao He and A. S. Nizami and Moonyong Lee},
  journal={Fuel},
  year={2022},
  url={https://api.semanticscholar.org/CorpusID:240583364}
}

@article{Zhang2022MultipleLN,
  title="{Multiple Learning Neural Network Algorithm for Parameter Estimation of Proton Exchange Membrane Fuel Cell Models}",
  author={Yiying Zhang and Chao Huang and Hailong Huang and Jingda Wu},
  journal={Green Energy and Intelligent Transportation},
  year={2022},
  url={https://api.semanticscholar.org/CorpusID:253076457}
}

@article{Elsisi2021AnIN,
  title="{An Improved Neural Network Algorithm to Efficiently Track Various Trajectories of Robot Manipulator Arms}",
  author={Mahmoud Elsisi and Karar Mahmoud and Matti Lehtonen and Mohamed M. F. Darwish},
  journal={IEEE Access},
  year={2021},
  volume={9},
  pages={11911-11920},
  url={https://api.semanticscholar.org/CorpusID:231714779}
}

@article{OsornioRos2017IdentificationOP,
  title="{Identification of Positioning System for Industrial Applications using Neural Network}",
  author={Roque Alfredo Osornio-R{\'i}os},
  journal={Journal of Scientific \& Industrial Research},
  year={2017},
  volume={76},
  pages={144},
  url={https://api.semanticscholar.org/CorpusID:37974345}
}

@article{boger2000sudbury,
  author = "Boger, J. and others",
    collaboration = "SNO",
    title = "{The Sudbury neutrino observatory}",
    eprint = "nucl-ex/9910016",
    archivePrefix = "arXiv",
    reportNumber = "SNO-STR-99-025",
    doi = "10.1016/S0168-9002(99)01469-2",
    journal = "Nucl. Instrum. Meth. A",
    volume = "449",
    pages = "172--207",
    year = "2000"
}

@article{aslam2025particle,
    author = "Aslam, Muhammad Waheed and Zafar, Abrar Ahmad and Aslam, Muhammad Naeem and Bhatti, Abdul Aziz and Hussain, Talab and Iqbal, Munawar and Abdo, Mohammed S.",
    title = "{Particle swarm optimization based analysis to unlocking the neutrino mass puzzle using $SU(2)_L \times U(1)_Y \times A_{4}\times S_2\times Z_{10} \times Z_{3}$ flavor symmetry}",
    eprint = "2404.14917",
    archivePrefix = "arXiv",
    primaryClass = "hep-ph",
    doi = "10.1038/s41598-024-81791-3",
    journal = "Sci. Rep.",
    volume = "15",
    number = "1",
    pages = "5129",
    year = "2025"
}

@article{esteban2025nufit,
  title="{NuFit-6.0: Updated global analysis of three-flavor neutrino oscillations}",
  author={Esteban, Ivan and Gonzalez-Garcia, MC and Maltoni, Michele and Martinez-Soler, Ivan and Pinheiro, Jo{\~a}o Paulo and Schwetz, Thomas},
  journal={Journal of High Energy Physics},
  volume={2024},
  number={12},
  pages={1--33},
  year={2025},
  publisher={Springer}
}

@article{CentellesChulia:2016fxr,
    author = "Centelles Chuli\'a, Salvador and Srivastava, Rahul and Valle, Jos\'e W. F.",
    title = "{CP violation from flavor symmetry in a lepton quarticity dark matter model}",
    eprint = "1606.06904",
    archivePrefix = "arXiv",
    primaryClass = "hep-ph",
    reportNumber = "REPORT-N-IFIC-16-43",
    doi = "10.1016/j.physletb.2016.08.028",
    journal = "Phys. Lett. B",
    volume = "761",
    pages = "431--436",
    year = "2016"
}

@article{Branco:1983tn,
    author = "Branco, G. C. and Gerard, J. M. and Grimus, W.",
    title = "{GEOMETRICAL T VIOLATION}",
    reportNumber = "CFMC E-9/83",
    doi = "10.1016/0370-2693(84)92024-0",
    journal = "Phys. Lett. B",
    volume = "136",
    pages = "383--386",
    year = "1984"
}

@article{Ma:2007wu,
    author = "Ma, Ernest",
    title = "{Near tribimaximal neutrino mixing with Delta(27) symmetry}",
    eprint = "0709.0507",
    archivePrefix = "arXiv",
    primaryClass = "hep-ph",
    reportNumber = "UCRHEP-T439",
    doi = "10.1016/j.physletb.2007.12.060",
    journal = "Phys. Lett. B",
    volume = "660",
    pages = "505--507",
    year = "2008"
}

@article{Abbas:2014ewa,
    author = "Abbas, Mohammed and Khalil, Shaaban",
    title = "{Fermion masses and mixing in $\Delta(27)$ flavour model}",
    eprint = "1406.6716",
    archivePrefix = "arXiv",
    primaryClass = "hep-ph",
    doi = "10.1103/PhysRevD.91.053003",
    journal = "Phys. Rev. D",
    volume = "91",
    number = "5",
    pages = "053003",
    year = "2015"
}

@article{Chen:2015jta,
    author = "Chen, Peng and Ding, Gui-Jun and Rojas, Alma. D. and Vaquera-Araujo, C. A. and Valle, J. W. F.",
    title = "{Warped flavor symmetry predictions for neutrino physics}",
    eprint = "1509.06683",
    archivePrefix = "arXiv",
    primaryClass = "hep-ph",
    reportNumber = "IFIC-15-XX",
    doi = "10.1007/JHEP01(2016)007",
    journal = "JHEP",
    volume = "01",
    pages = "007",
    year = "2016"
}

@article{Vien:2020hzy,
    author = "Vien, V. V. and Khoi, D. P.",
    title="{U(1) B-L extension based on $\Delta$ (27) symmetry for lepton masses and mixings}",
    doi = "10.1142/S0217732320501813",
    journal = "Mod. Phys. Lett. A",
    volume = "35",
    number = "22",
    pages = "2050181",
    year = "2020"
}

@article{Ma:2001dn,
    author = "Ma, Ernest and Rajasekaran, G.",
    title = "{Softly broken A(4) symmetry for nearly degenerate neutrino masses}",
    eprint = "hep-ph/0106291",
    archivePrefix = "arXiv",
    reportNumber = "UCRHEP-T308",
    doi = "10.1103/PhysRevD.64.113012",
    journal = "Phys. Rev. D",
    volume = "64",
    pages = "113012",
    year = "2001"
}

@article{Brown:1984dk,
    author = "Brown, T. and Pakvasa, S. and Sugawara, H. and Yamanaka, Y.",
    title = "{Neutrino Masses, Mixing and Oscillations in S(4) Model of Permutation Symmetry}",
    reportNumber = "UH-511-523-84",
    doi = "10.1103/PhysRevD.30.255",
    journal = "Phys. Rev. D",
    volume = "30",
    pages = "255",
    year = "1984"
}

@article{Chamoun:2023vnn,
    author = "Chamoun, N. and Lashin, E. I.",
    title = "{Textures of neutrino mass matrix from S$_{4}$-flavor symmetry}",
    eprint = "2308.10985",
    archivePrefix = "arXiv",
    primaryClass = "hep-ph",
    doi = "10.1007/JHEP01(2025)016",
    journal = "JHEP",
    volume = "01",
    pages = "016",
    year = "2025"
}

@article{Ding:2024inn,
    author = "Ding, Gui-Jun and Lu, Jun-Nan and Petcov, S. T. and Qu, Bu-Yao",
    title = "{Non-holomorphic modular S$_{4}$ lepton flavour models}",
    eprint = "2408.15988",
    archivePrefix = "arXiv",
    primaryClass = "hep-ph",
    reportNumber = "SISSA 17/2024/FISI",
    doi = "10.1007/JHEP01(2025)191",
    journal = "JHEP",
    volume = "01",
    pages = "191",
    year = "2025"
}

@article{Penedo:2018nmg,
    author = "Penedo, J. T. and Petcov, S. T.",
    title = "{Lepton Masses and Mixing from Modular $S_4$ Symmetry}",
    eprint = "1806.11040",
    archivePrefix = "arXiv",
    primaryClass = "hep-ph",
    reportNumber = "SISSA 25/2018/FISI, IPMU18-0121, SISSA-25-2018-FISI",
    doi = "10.1016/j.nuclphysb.2018.12.016",
    journal = "Nucl. Phys. B",
    volume = "939",
    pages = "292--307",
    year = "2019"
}

@article{Kobayashi:2019xvz,
    author = "Kobayashi, Tatsuo and Shimizu, Yusuke and Takagi, Kenta and Tanimoto, Morimitsu and Tatsuishi, Takuya H.",
    title = "{$A_4$ lepton flavor model and modulus stabilization from $S_4$ modular symmetry}",
    eprint = "1909.05139",
    archivePrefix = "arXiv",
    primaryClass = "hep-ph",
    reportNumber = "EPHOU-19-012, HUPD1912",
    doi = "10.1103/PhysRevD.100.115045",
    journal = "Phys. Rev. D",
    volume = "100",
    number = "11",
    pages = "115045",
    year = "2019",
    note = "[Erratum: Phys.Rev.D 101, 039904 (2020)]"
}

@article{Novichkov:2018ovf,
    author = "Novichkov, P. P. and Penedo, J. T. and Petcov, S. T. and Titov, A. V.",
    title = "{Modular S$_{4}$ models of lepton masses and mixing}",
    eprint = "1811.04933",
    archivePrefix = "arXiv",
    primaryClass = "hep-ph",
    reportNumber = "SISSA 47/2018/FISI, IPMU18-0187, IPPP/18/98",
    doi = "10.1007/JHEP04(2019)005",
    journal = "JHEP",
    volume = "04",
    pages = "005",
    year = "2019"
}

@article{Liu:2020akv,
    author = "Liu, Xiang-Gan and Yao, Chang-Yuan and Ding, Gui-Jun",
    title = "{Modular invariant quark and lepton models in double covering of $S_4$ modular group}",
    eprint = "2006.10722",
    archivePrefix = "arXiv",
    primaryClass = "hep-ph",
    reportNumber = "USTC-ICTS/PCFT-20-18",
    doi = "10.1103/PhysRevD.103.056013",
    journal = "Phys. Rev. D",
    volume = "103",
    number = "5",
    pages = "056013",
    year = "2021"
}

@article{deMedeirosVarzielas:2023crv,
    author = "de Medeiros Varzielas, I. and Levy, M. and Penedo, J. T. and Petcov, S. T.",
    title = "{Quarks at the modular S$_{4}$ cusp}",
    eprint = "2307.14410",
    archivePrefix = "arXiv",
    primaryClass = "hep-ph",
    reportNumber = "SISSA 11/2023/FISI, CFTP/23-002",
    doi = "10.1007/JHEP09(2023)196",
    journal = "JHEP",
    volume = "09",
    pages = "196",
    year = "2023"
}

@article{Aslam:2025vvp,
    author = "Aslam, Muhammad Waheed and Zafar, Abrar Ahmad",
    title = "{Optimizing Neutrino Mass Predictions with Neural Network Algorithm and $SU(2)_L \times U(1)_Y \times T_7 \times Z_{10}$ Symmetry}",
    doi = "10.1007/s10773-025-05993-9",
    journal = "Int. J. Theor. Phys.",
    volume = "64",
    number = "5",
    pages = "124",
    year = "2025"
}

@article{doi:10.1142/S0217732325501068,
author = {Aslam, Muhammad Waheed and Zafar, Abrar Ahmad and Aslam, Muhammad Naeem and Saleem, S.},
title = "{Optimizing neutrino mass predictions with competitive chaotic learning neural network algorithm and $SU(2)_L \times A_4 \times Z_3 \times Z_2$ symmetry}",
journal = {Modern Physics Letters A},
volume = {40},
number = {27},
pages = {2550106},
year = {2025},
doi ={10.1142/S0217732325501068},
URL = {https://doi.org/10.1142/S0217732325501068
    
    

},
eprint = { 
    
        https://doi.org/10.1142/S0217732325501068
    
    

}
}

@article{burrows1990neutrinos,
  author = "Burrows, A.",
    title = "{Neutrinos From Supernova Explosions}",
    reportNumber = "STEWARD-925, ATAP-90-09",
    doi = "10.1146/annurev.ns.40.120190.001145",
    journal = "Ann. Rev. Nucl. Part. Sci.",
    volume = "40",
    pages = "181--212",
    year = "1990"
}

@article{tamborra2018neutrinos,
  author = "Tamborra, Irene and Murase, Kohta",
    title = "{Neutrinos from Supernovae}",
    doi = "10.1007/s11214-018-0468-7",
    journal = "Space Sci. Rev.",
    volume = "214",
    number = "1",
    pages = "31",
    year = "2018"
}

@article{burrows2000neutrinos,
  author = "Burrows, A. and Young, T.",
    title = "{Neutrinos and supernova theory}",
    doi = "10.1016/S0370-1573(00)00016-8",
    journal = "Phys. Rept.",
    volume = "333",
    pages = "63--75",
    year = "2000"
}

@article{cooperstein1988neutrinos,
  author = "Cooperstein, J.",
    title = "{Neutrinos in Supernovae}",
    doi = "10.1016/0370-1573(88)90038-5",
    journal = "Phys. Rept.",
    volume = "163",
    pages = "95--126",
    year = "1988"
}

@article{herant1997neutrinos,
  title="{Neutrinos and supernovae}",
  author={Herant, Marc and Colgate, Stirling A and Benz, Willy and Fryer, Chris},
  journal={Los Alamos Science},
  volume={25},
  pages={64--78},
  year={1997}
}

@article{sarkar2003neutrinos,
 author = {{Sarkar}, Subir},
        title = "{Neutrinos from the Big Bang}",
      journal = {arXiv e-prints},
         year = 2003,
        month = feb,
          eid = {hep-ph/0302175},
        pages = {hep-ph/0302175},
          doi = {10.48550/arXiv.hep-ph/0302175},
archivePrefix = {arXiv},
       eprint = {hep-ph/0302175},
 primaryClass = {hep-ph},
       adsurl = {https://ui.adsabs.harvard.edu/abs/2003hep.ph....2175S}
}

@article{steigman2005neutrinos,
  author = {{Steigman}, Gary},
        title = "{Neutrinos and Big Bang Nucleosynthesis}",
      journal = {Physica Scripta Volume T},
         year = 2005,
        month = jan,
       volume = {121},
        pages = {142-146},
          doi = {10.1088/0031-8949/2005/T121/021},
archivePrefix = {arXiv},
       eprint = {hep-ph/0501100},
 primaryClass = {hep-ph},
       adsurl = {https://ui.adsabs.harvard.edu/abs/2005PhST..121..142S}
}

@article{blennow2013probing,
 author = {{Blennow}, Mattias and {Carrigan}, Marcus and {Fernandez Martinez}, Enrique},
        title = "{Probing the Dark Matter mass and nature with neutrinos}",
      journal = {Journal of Cosmology and Astroparticle Physics},
         year = 2013,
        month = jun,
       volume = {2013},
       number = {6},
          eid = {038},
        pages = {038},
          doi = {10.1088/1475-7516/2013/06/038},
archivePrefix = {arXiv},
       eprint = {1303.4530},
 primaryClass = {hep-ph},
       adsurl = {https://ui.adsabs.harvard.edu/abs/2013JCAP...06..038B}
}

@article{agarwalla2011neutrino,
  author = "Agarwalla, Sanjib Kumar and Blennow, Mattias and Fernandez Martinez, Enrique and Mena, Olga",
    title = "{Neutrino Probes of the Nature of Light Dark Matter}",
    eprint = "1105.4077",
    archivePrefix = "arXiv",
    primaryClass = "hep-ph",
    reportNumber = "CERN-PH-TH-2011-118, EURONU-WP6-11-35, IFIC-11-24, MPP-2011-56",
    doi = "10.1088/1475-7516/2011/09/004",
    journal = "JCAP",
    volume = "09",
    pages = "004",
    year = "2011"
}

@article{dodelson1994sterile,
  author = {{Dodelson}, Scott and {Widrow}, Lawrence M.},
        title = "{Sterile neutrinos as dark matter}",
      journal = {Physical Review Letters},
         year = 1994,
        month = jan,
       volume = {72},
       number = {1},
        pages = {17-20},
          doi = {10.1103/PhysRevLett.72.17},
archivePrefix = {arXiv},
       eprint = {hep-ph/9303287},
 primaryClass = {hep-ph},
       adsurl = {https://ui.adsabs.harvard.edu/abs/1994PhRvL..72...17D}
}

@article{bertone2018new,
  author = "Bertone, Gianfranco and Tait, M. P., Tim",
    title = "{A new era in the search for dark matter}",
    eprint = "1810.01668",
    archivePrefix = "arXiv",
    primaryClass = "astro-ph.CO",
    doi = "10.1038/s41586-018-0542-z",
    journal = "Nature",
    volume = "562",
    number = "7725",
    pages = "51--56",
    year = "2018"
}

@article{hooper2008strategies,
  author = "Hooper, Dan and Baltz, Edward A.",
    title = "{Strategies for Determining the Nature of Dark Matter}",
    eprint = "0802.0702",
    archivePrefix = "arXiv",
    primaryClass = "hep-ph",
    reportNumber = "SLAC-PUB-13146, FERMILAB-PUB-08-026-A",
    doi = "10.1146/annurev.nucl.58.110707.171217",
    journal = "Ann. Rev. Nucl. Part. Sci.",
    volume = "58",
    pages = "293--314",
    year = "2008"
}

@article{t2k2020constraint,
  author = "Abe, K. and others",
    collaboration = "T2K",
    title = "{Constraint on the matter\textendash{}antimatter symmetry-violating phase in neutrino oscillations}",
    eprint = "1910.03887",
    archivePrefix = "arXiv",
    primaryClass = "hep-ex",
    doi = "10.1038/s41586-020-2177-0",
    journal = "Nature",
    volume = "580",
    number = "7803",
    pages = "339--344",
    year = "2020",
    note = "[Erratum: Nature 583, E16 (2020)]"
}

@article{weinheimer2013neutrino,
  author = "Weinheimer, Christian and Zuber, Kai",
    title = "{Neutrino Masses}",
    eprint = "1307.3518",
    archivePrefix = "arXiv",
    primaryClass = "hep-ex",
    doi = "10.1002/andp.201300063",
    journal = "Annalen Phys.",
    volume = "525",
    number = "8-9",
    pages = "565--575",
    year = "2013"
}

@article{chakraborty2014higgs,
  author = {{Chakraborty}, Sabyasachi and {Roy}, Sourov},
        title = "{Higgs boson mass, neutrino masses and mixing and keV dark matter in an U(1)$_{ R }$ - lepton number model}",
      journal = {Journal of High Energy Physics},
         year = 2014,
        month = jan,
       volume = {2014},
          eid = {101},
        pages = {101},
          doi = {10.1007/JHEP01(2014)101},
archivePrefix = {arXiv},
       eprint = {1309.6538},
 primaryClass = {hep-ph},
       adsurl = {https://ui.adsabs.harvard.edu/abs/2014JHEP...01..101C}
}

@article{raffelt2008neutrinos,
  title={Neutrinos and Stars},
  author={Raffelt, Georg G},
  booktitle={Neutrinos in Particle Physics, Astrophysics and Cosmology},
  pages={185--194},
  year={2008},
archivePrefix = {arXiv},
       eprint = {1201.1637},
 primaryClass = {astro-ph.SR},
       adsurl = {https://ui.adsabs.harvard.edu/abs/2012arXiv1201.1637R}
}

@article{leontaris1998modular,
     author = "Leontaris, G. K. and Tracas, N. D.",
    title = "{Modular weights, U(1)'s and mass matrices}",
    eprint = "hep-ph/9709510",
    archivePrefix = "arXiv",
    reportNumber = "NTUA-65-97A, IOA-TH-97-005",
    doi = "10.1016/S0370-2693(97)01412-3",
    journal = "Phys. Lett. B",
    volume = "419",
    pages = "206--210",
    year = "1998"
}

@article{Kobayashi:2018vbk,
    author = "Kobayashi, Tatsuo and Tanaka, Kentaro and Tatsuishi, Takuya H.",
    title = "{Neutrino mixing from finite modular groups}",
    eprint = "1803.10391",
    archivePrefix = "arXiv",
    primaryClass = "hep-ph",
    reportNumber = "HPHOU-18-002",
    doi = "10.1103/PhysRevD.98.016004",
    journal = "Phys. Rev. D",
    volume = "98",
    number = "1",
    pages = "016004",
    year = "2018"
}

@article{Mishra:2020gxg,
    author = "Mishra, Subhasmita",
    title = "{Neutrino mixing and Leptogenesis with modular $S_3$ symmetry in the framework of type III seesaw}",
    eprint = "2008.02095",
    archivePrefix = "arXiv",
    primaryClass = "hep-ph",
    month = "8",
    year = "2020"
}
\end{document}